\documentclass[reprint,superscriptaddress,amsmath,aps,prd]{revtex4-2}

\usepackage{ifthen}  %
\usepackage{ifptex}  %
\ifthenelse{\boolean{ptex}}{
\usepackage[dvipdfmx,final]{graphicx}
\usepackage[dvipsnames]{xcolor}
}{
\usepackage{graphicx} 
\usepackage{epstopdf} 
\usepackage{xcolor} 
}
\usepackage{dcolumn} 
\usepackage{amssymb} 
\usepackage[utf8]{inputenc} 
\usepackage[english]{babel} 
\usepackage{blindtext} 
\usepackage{multirow}
\usepackage{here}
\usepackage{amsmath}
\usepackage{amsfonts}
\usepackage{siunitx}

\usepackage{adjustbox} 
\usepackage{tabularx} 

\ifthenelse{\boolean{ptex}}{
\usepackage[dvipdfmx]{hyperref} 
}{
\usepackage{hyperref} 
}

\usepackage[caption=false]{subfig}

\hypersetup{%
  setpagesize=false,
  bookmarksnumbered=true,
  bookmarksopen=true,
  colorlinks=true,
  linkcolor=teal,
  citecolor=teal,
  urlcolor=black,
}

\graphicspath{{figures/}} 

\input belle2sym.tex
\newcommand{\intlum} {\ensuremath{\lum_{\mathrm{int}}}\xspace}

\def\amm        {\ensuremath{a_{\mu}}\xspace}
\def\ammpppz    {\ensuremath{a_{\mu}^{3\pi}}\xspace}
\def\ammhvp     {\ensuremath{a^{\rm{HVP}}_{\mu}}\xspace}
\def\ammhvplo   {\ensuremath{a^{\rm{HVP,LO}}_{\mu}}\xspace}

\def\rrad        {\ensuremath{r_{\mathrm{rad}}}\xspace}

\def\chisqfctpg    {\ensuremath{\chi^{2}_{\mathrm{2\pi3\g}}}\xspace} 
\def\chisqfcfpg    {\ensuremath{\chi^{2}_{\mathrm{2\pi5\g}}}\xspace}

\def\chisqrec      {\ensuremath{\chi^{2}_{\mathrm{1C,2\pi\g}}}\xspace}

\def\pppz         {\ensuremath{\pi^{+}\pi^{-}\pi^{0}}\xspace}
\def\pppzg        {\ensuremath{\pi^{+}\pi^{-}\pi^{0}\gamma}\xspace}

\def\pppzpz       {\ensuremath{\pi^{+}\pi^{-}\pi^{0}\pi^{0}}\xspace}
\def\pppzpzg      {\ensuremath{\pi^{+}\pi^{-}\pi^{0}\pi^{0}\gamma}\xspace}
\def\ppg          {\ensuremath{\pi^{+}\pi^{-}\gamma}\xspace}
\def\kkpzg        {\ensuremath{K^{+}K^{-}\pi^{0}\gamma}\xspace}

\def\mmg       {\ensuremath{\mu^+\mu^-\gamma}\xspace}
\def\ksklg        {\ensuremath{K^0_{\scriptscriptstyle S}K^0_{\scriptscriptstyle L}\gamma}\xspace}

\def\mgg            {\ensuremath{M(\gaga)}\xspace}
\def\mpig           {\ensuremath{M(\pi^{\pm}\g)}\xspace}
\def\mggisr         {\ensuremath{M(\g_{\mathrm{ISR}}\g)}\xspace}
\def\mpppz          {\ensuremath{M(3\pi)}\xspace}
\def\mpppzrec       {\ensuremath{M(\pi^+\pi^-\pi^0_\mathrm{rec.})}\xspace}

\def\thetagcms      {\ensuremath{\theta^{*}_\g}\xspace}

\renewcommand{\d}{\mathrm{d}}

\newlength{\figcolumnwidth}
\newboolean{articletitles}
\setboolean{articletitles}{true} 

\usepackage{orcidlink}

\begin{document}

\title{%
\texorpdfstring{%
Measurement of the \boldmath{\epem \to \pppz} cross section in the energy range 0.62--3.50\,GeV at Belle~II%
}{%
Measurement of the e+e- -> pi+ pi- pi0 cross section in the energy range 0.62-3.50 GeV at Belle II%
}%
}

  \author{I.~Adachi\,\orcidlink{0000-0003-2287-0173}} 
  \author{L.~Aggarwal\,\orcidlink{0000-0002-0909-7537}} 
  \author{H.~Aihara\,\orcidlink{0000-0002-1907-5964}} 
  \author{N.~Akopov\,\orcidlink{0000-0002-4425-2096}} 
  \author{A.~Aloisio\,\orcidlink{0000-0002-3883-6693}} 
  \author{N.~Anh~Ky\,\orcidlink{0000-0003-0471-197X}} 
  \author{D.~M.~Asner\,\orcidlink{0000-0002-1586-5790}} 
  \author{H.~Atmacan\,\orcidlink{0000-0003-2435-501X}} 
  \author{V.~Aushev\,\orcidlink{0000-0002-8588-5308}} 
  \author{M.~Aversano\,\orcidlink{0000-0001-9980-0953}} 
% ...
\collaboration{The Belle II Collaboration}

\begin{abstract}
We report a measurement of the $\epem \to \pppz$ cross section in the energy range from 0.62 to 3.50\gev using an initial-state radiation technique. 
We use an $\epem$ data sample corresponding to 191\invfb of integrated luminosity, collected at a center-of-mass energy at or near the \Y4S resonance with the \belletwo detector at the SuperKEKB collider.
Signal yields are extracted by fitting the two-photon mass distribution in $\epem \to \pppzg$ events, which involve a $\piz \to \gaga$ decay and an energetic photon radiated from the initial state. Signal efficiency corrections with an accuracy of 1.6\% are obtained from several control data samples.
The uncertainty on the cross section at the $\omega$ and $\phi$ resonances is dominated by the systematic uncertainty of 2.2\%.
The resulting cross sections in the 0.62--1.80\gev energy range yield 
$
\ammpppz = [48.91 \pm 0.23~(\mathrm{stat}) \pm 1.07~(\mathrm{syst})] \times 10^{-10}
$
for the leading-order hadronic vacuum polarization contribution to the muon anomalous magnetic moment.
This result differs by $2.5$ standard deviations from the most precise current determination.
\end{abstract}

\maketitle

\def\prd{true}

\def\userFigureWidth{0.96\columnwidth}
\section{Introduction} 
%
%
The measured hadronic cross section for \epem annihilation
at center-of-mass~(c.m.)\ energies below 2\gev plays an important role in
obtaining the standard model~(SM)\ prediction for the hadronic vacuum polarization~(HVP)\ contribution to the muon anomalous magnetic moment 
$\amm \equiv (g_{\mu}-2)/2$ through dispersion relations, where $g_{\mu}$ is the muon gyromagnetic ratio.
A discrepancy of about five standard deviations~($\sigma$)\ between 
the world average of $\amm$ measured by experiments at BNL~\cite{Muong-2:2006rrc} and FNAL~\cite{Muong-2:2021ojo,Muong-2:2023cdq}, 
and SM predictions based on dispersion relations has been reported~\cite{Davier:2017zfy,Keshavarzi:2018mgv,Colangelo:2018mtw,Hoferichter:2019mqg,Davier:2019can,Keshavarzi:2019abf}
The HVP contribution is the main source of uncertainty in the SM prediction, accounting for more than 80\% of the current theoretical uncertainty.
Therefore, improving the precision of the HVP term is the key to clarifying the picture.
Discrepancies between measurements of the cross section have an impact on the uncertainty in the HVP contribution to \amm, \ammhvp.
In the \pipi final state, the long-standing difference in experimental values between \babar~\cite{BaBar:2012bdw} and KLOE~\cite{KLOE:2008fmq, KLOE:2010qei, KLOE:2012anl, KLOE-2:2017fda} contributes to the systematic uncertainty in \ammhvp. 
A result recently reported from CMD-3~\cite{CMD-3:2023alj,CMD-3:2023rfe} suggests a different value closer to the SM.
In addition, recent predictions based on lattice QCD calculations show 2--3$\sigma$ differences from values based on dispersion relations, i.e., a smaller difference from the measured \amm~\cite{Borsanyi:2020mff, Ce:2022kxy, ExtendedTwistedMass:2022jpw, RBC:2023pvn}.
Therefore, additional experimental measurements are important to clarify the situation.
%
%
%
\par
We report herein a new measurement of the cross section for $\epem \to \pppz$ in the energy range from 0.62 to 3.50\gev using a 191\invfb data sample from the Belle~II experiment and applying an initial-state radiation~(ISR) technique~\cite{Rodrigo:2002rz, Druzhinin:2011qd}.
The Belle~II experiment is an electron-positron-collider detector operating at c.m.~energies at and near the $\Y4S$ resonance, 10.58\gev.
Using ISR production and a technique that employs events in which a single energetic ISR photon is emitted, one can measure \epem cross sections to hadrons as a function of c.m.\ energy while operating the accelerator at a fixed c.m.\ energy.
The ISR technique is complementary to the competing method of measuring the hadronic cross section by varying the c.m.\ energy directly.
\par
We measure the cross section for $\epem \to \pppz$ in ISR production, in which the photon is emitted from ISR.
The $\epem \to \pppz$ cross section $\sigma_{3\pi} (\sqrt{s'})$, including vacuum-polarization contributions (``dressed'') as a function of energy $\sqrt{s'} = \mpppz$, is obtained from the relation $s'=s-2\sqrt{s}E^{*}_\g$.
Here, $\sqrt{s}$ and $E^{*}_\g$ are the energy of the \epem system and ISR photon in the c.m.~frame, respectively.
The double differential cross section of the ISR process is related to $\sigma_{3\pi}(\sqrt{s'})$ by
\begin{align}
    \frac{\d \sigma_{3\pi\g}}{\d \sqrt{s'} \d \cos{\thetagcms}} = \frac{2\sqrt{s'}}{s} W(s, s', \thetagcms) \sigma_{3\pi}(\sqrt{s'}),
\end{align}
where \thetagcms is the polar angle of the ISR photon momentum relative to the beam axis, in the c.m.~frame.
The radiator function $W(s, s', \thetagcms)$ describes the probability of the ISR photon emission, which is calculated by QED~\cite{Druzhinin:2011qd}.
At leading order and neglecting the term proportional to $m^2_e/s$, where $m_e$ is the electron mass, the radiator function is given by
\begin{align}\label{eq:radiator}
  W(s, s', \thetagcms) = \frac{\alpha}{\pi} \biggl( \frac{s^2 + s'^2}{s(s-s')} \frac{1}{\sin^2{\thetagcms}} - \frac{1}{2}\frac{s-s'}{s} \biggr),
\end{align}
where $\alpha$ is the fine-structure constant.
In ISR processes, the ISR photons are emitted predominantly almost collinearly to the beam direction.
This analysis includes a small fraction (about 10\%) of the ISR photons emitted at large polar angles but within the detector acceptance.
\par
The measured (visible)\ three-pion mass spectrum is given by the following relation:
\begin{equation}\label{eq:spectrum}
    \frac{\d N_{\mathrm{vis}}}{\d \sqrt{s'}} = \sigma_{3\pi}  \varepsilon  \frac{\d L_{\mathrm{eff}}}{\d\sqrt{s'}} \rrad,
\end{equation}
where
$\varepsilon$ is the signal efficiency,
$\d L_{\mathrm{eff}}/\d\sqrt{s'}$ is the so-called effective luminosity, and
\rrad is the radiative correction to the leading order ISR $\epem \to \pppzg$ cross section, to take higher-order ISR processes into account~\cite{Rodrigo:2001jr, Rodrigo:2001kf}.
The effective luminosity, given from the radiator function [Eq.~\eqref{eq:radiator}] for the ISR photon in the angular range, $\theta^{*}_{\mathrm{min}} < \theta^{*}_{\g} < 180 - \theta^{*}_{\mathrm{min}}$, is defined by
\begin{align} \notag 
    \frac{\d L_{\mathrm{eff}}}{\d\sqrt{s'}} 
    &=  \intlum \frac{2\sqrt{s'}}{s} \int^{\pi-\theta^{*}_{\mathrm{min}}}_{\theta^{*}_{\mathrm{min}}}{W(s,s',\theta') \sin{\theta'} \d \theta'}\\
    &=  \intlum \frac{2\sqrt{s'}}{s} \frac{\alpha}{\pi} \biggl( \frac{s^{2} + s'^{2}}{s (s - s')} \ln{\frac{1+C}{1-C}} -  \frac{s - s'}{s} C \biggr),
\end{align}
where \intlum is the integrated luminosity of the dataset, 
$\theta^{*}_{\mathrm{min}}$ is the minimum polar angle of an ISR photon in the c.m.~frame,
and $C$ is $\cos{\theta^{*}_{\mathrm{min}}}$.
\par
The three-pion process is the second largest contributor to \ammhvp after the \pipi final state.
The uncertainty in the three-pion contribution accounts for 15\% of the total uncertainty in \ammhvp~\cite{Aoyama:2020ynm}.
The $\epem \to \pppz$ cross section has been measured previously by several $\epem$ experiments.
These include the SND and CMD-2 experiments at the VEPP-2M collider~\cite{Achasov:2000am,Achasov:2002ud,Achasov:2003ir,Akhmetshin:1995vz,Akhmetshin:1998se,CMD-2:2003gqi,Akhmetshin:2006sc},
which measured the cross section up to 1.4\gev by scanning the beam energy.
In addition, SND at VEPP-2000 measured the cross section by scanning the energy range of 1.05--2.0\gev~\cite{Aulchenko:2015mwt}.
The \babar experiment at the PEP-II collider measured three-pion cross sections up to around 3.5\gev with a data sample recorded at a c.m.~energy of 10.58\gev using the ISR technique~\cite{BaBar:2004ytv, BABAR:2021cde}.
The $\epem \to \pppz$ contribution to \ammhvp, \ammpppz, is summarized in Ref~\cite{Aoyama:2020ynm}, which quotes an uncertainty of 3\%.
However, this average does not include the recent BABAR 2021 result~\cite{BABAR:2021cde}, which has 1.3\% precision.
A global fit including the BABAR 2021 result, achieves a precision of 1.2\%~\cite{Hoferichter:2023bjm}.
%
%
\par
Our analysis follows the \babar 2021 measurement in many aspects.
To obtain the mass spectrum $\d N_{\mathrm{vis}}/\d\sqrt{s'}$, we reconstruct $\epem \to \pppzg$ candidates after an initial event selection and extract the signal by fitting the two-photon mass distribution as a function of \pppz mass and determining the yield of $\piz \to \gaga$ decays.
The $3\pi$ mass spectrum is characterized by the $\omega(782)$ and $\phi(1020)$ resonances below 1.05\gevcc,
the $\omega(1420)$ and $\omega(1650)$ resonances in the range 1.1--1.8\gevcc, and the \jpsi resonance at 3.09\gevcc.
Residual background events are estimated using control samples and subtracted from the observed spectrum.
The spectrum is then unfolded to correct for the effect of detector resolution.
The signal efficiency $\varepsilon$ is obtained from simulation and corrected for possible differences between the data and simulation using various data control samples.
The cross section is obtained from Eq.~\eqref{eq:spectrum}, and the three-pion contribution to \ammhvp is determined based on the measured cross section.
Key aspects are signal selection, background estimation, and efficiency determination and their associated systematic uncertainties.
\par
This article is organized as follows. We first discuss the Belle~II detector and the Monte Carlo simulation program in Sec.~\ref{sec:setup}; event selection criteria are discussed in Sec.~\ref{sec:selection}.
Section~\ref{sec:bkg} describes the background estimation using several data control samples, selected to enhance each of the background components.
Section~\ref{sec:efficiency} presents the validation of efficiencies from various control samples.
The unfolding is discussed in Sec.~\ref{sec:unfold}.
After discussing the systematic uncertainty on the cross section measurement, we provide the cross section results in Sec.~\ref{sec:xsec}.
Using these results, we evaluate the three-pion contribution to the HVP term in $a_\mu$ in Sec.~\ref{sec:amm}.
Finally, we discuss differences from the \babar measurement in Sec.~\ref{sec:discuss}
and conclude in Sec.~\ref{sec:summary}.
%
%
%
%
%
%
%
\section{The Belle~II detector and simulation} \label{sec:setup}
%
%
We use a 191\invfb data sample collected from 2019 to 2021 at a c.m.~energy of 10.58\gev at the Belle~II experiment.
The Belle~II experiment is located at SuperKEKB, which collides 7-GeV electrons with 4-GeV positrons at energies at or near the $\Upsilon(4S)$ resonance~\cite{Akai:2018mbz}.
The Belle~II detector~\cite{Abe:2010gxa} has a cylindrical geometry.
Its coordinate system is defined with the $z$ axis in the laboratory frame, which is the symmetry axis of the superconducting solenoid.
The polar angle $\theta$ is defined with respect to the $+z$ axis.
The detectors that are used to reconstruct trajectories of charged particles are a two-layer silicon-pixel detector~(PXD)\ surrounded by a four-layer double-sided silicon-strip detector~(SVD)~\cite{Belle-IISVD:2022upf} and a 56-layer central drift chamber~(CDC).
The second layer of the PXD covered only one-sixth of the azimuth for the data analyzed in this paper.
Surrounding the CDC, which also provides ionization-energy-loss measurements, is a time-of-propagation counter~(TOP)~\cite{Kotchetkov:2018qzw} in the barrel region and an aerogel-based ring-imaging Cherenkov counter~(ARICH)\ in the forward region.
These detectors provide charged-particle identification.
Surrounding the TOP and ARICH is an electromagnetic calorimeter~(ECL)\ composed of CsI(Tl)\ crystals that primarily provide energy and timing measurements for photons and electrons.
Outside of the ECL is the superconducting solenoid magnet.
The magnet provides a 1.5~T axial magnetic field.
Its flux return is instrumented with resistive-plate chambers and plastic scintillator modules to detect muons, $K^0_L$ mesons, and neutrons~(KLM).
%
%
%
\par
Signal and background simulation data are used for the determination of the signal efficiency and background.
The detector geometry and response are simulated using \texttt{GEANT4}-based simulation framework~\cite{Agostinelli:2002hh,Kuhr:2018lps,basf2repo}.
The experimental and simulated samples are reconstructed using the Belle~II software~\cite{Kuhr:2018lps,basf2repo}.
Signal $\epem \to \pppzg$ events are modeled by \texttt{PHOKHARA}~\cite{Rodrigo:2001jr, Rodrigo:2001kf, Czyz:2005as} (version 9.1), where we limit the polar angle of the ISR photon to the range $20\degrees < \thetagcms < 160\degrees$.
The kinematic properties of the $3\pi$ system implemented in \texttt{PHOKHARA} are based on previous measurements in \epem annihilation~\cite{Czyz:2005as}.
The vacuum polarization corrections given in Ref.~\cite{F.Jegerlehner:2019} are used in the event generation.
In addition, the invariant mass of the hadronic system and the ISR photon at the generator level is required to be greater than 8\gevcc to suppress events with extra ISR photon emissions.
\par
The background processes $\epem \to \pppzpzg$ and $\epem \to \ppg$ are also generated with \texttt{PHOKHARA}.
The reaction $\epem \to \kkpzg$ is simulated by \texttt{PHOKHARA} for the production of the ISR photon, and \texttt{EvtGen}~\cite{Lange:2001uf} is used for the $\g^{*} \to K^{+}K^{-}\piz$ decay from the hadronic system at a given mass.
Backgrounds from $\epem \to \mmg$ and $\epem \to \qqbar$ processes, with $q$ indicating an $u$, $d$, $s$, or $c$ quark, are generated using \texttt{KKMC}~\cite{Jadach:1999vf}, with hadronization and hadronic decays simulated by \texttt{PYTHIA8}~\cite{Sjostrand:2014zea} and \texttt{EVTGEN}, respectively.
Hadronic $\epem \to \qqbar$ events with ISR emission are excluded from the samples, to avoid overlap with the \texttt{PHOKHARA} sample.
The $\epem \to \tautau$ process is generated with \texttt{KKMC}, and $\tau$ decays are simulated with \texttt{TAUOLA}~\cite{Jadach:1990mz}.
Electron-related backgrounds including $\epem \to \epem(\g)$ and $\epem \to \gaga$ are simulated using the \texttt{BABAYAGA@NLO} generator~\cite{CarloniCalame:2003yt}.
Beam backgrounds are simulated separately and overlaid on each Monte Carlo sample~\cite{Lewis:2018ayu}.
\par
To minimize experimental bias, the event selection and analysis workflow are optimized using simulated events before examining data.
Several data control samples are used to determine the dominant backgrounds as well as signal efficiency correction factors.
%
%
%
%
%
\section{Event selection} \label{sec:selection}
%
%
%
\subsection{Baseline selection}
Events including an energetic ISR photon are selected with a hardware-based ECL trigger.
Events are accepted if energy deposits exceeding 2\gev of c.m.~energy are detected in the barrel region of the ECL.
For processes with $3\pi$ masses below 3.5\gevcc, which are measured in this paper, the ISR photon has an energy greater than 4\gev, and satisfies this trigger condition.
\par
In the offline analysis, signal $\epem \to \pppzg$ candidates are reconstructed from two oppositely charged particles and three photons.
Charged particles are reconstructed from tracks using PXD, SVD, and CDC information.
To suppress misreconstructed and beam-induced background tracks, each charged particle is required to have a transverse momentum greater than 0.2\gevc,
have more than 20 CDC wire hits,
and originate from the interaction point; we require the transverse and longitudinal projections of the distance between the interaction point and the track to be smaller than 0.5\cm and 2.0\cm, respectively.
Photons are selected from clusters, sets of adjacent crystals detected in the ECL that do not match the extrapolation of CDC tracks.
The clusters are required to have energies greater than $100\mev$ and polar angles within the CDC acceptance in the laboratory frame of $17\degrees<\theta^{~\mathrm{lab.}}<150\degrees$.
Events with two oppositely charged particles and at least three candidates passing the photon requirements are selected.
The ISR photon candidate is required to have an energy greater than $2\gev$ in the c.m.~frame and have a polar angle in the barrel region $37.3\degrees<\theta^{~\mathrm{lab.}}_{\g}<123.7\degrees$,
which corresponds approximately to the angular range 48--135\degrees in the c.m.~system.
We require the invariant mass of the $ \pi^+\pi^- \gaga\g$ system to be greater than 8.0\gevcc to suppress extra ISR emission.
\par
\begin{figure}
  \centering
  \includegraphics[width=\userFigureWidth]{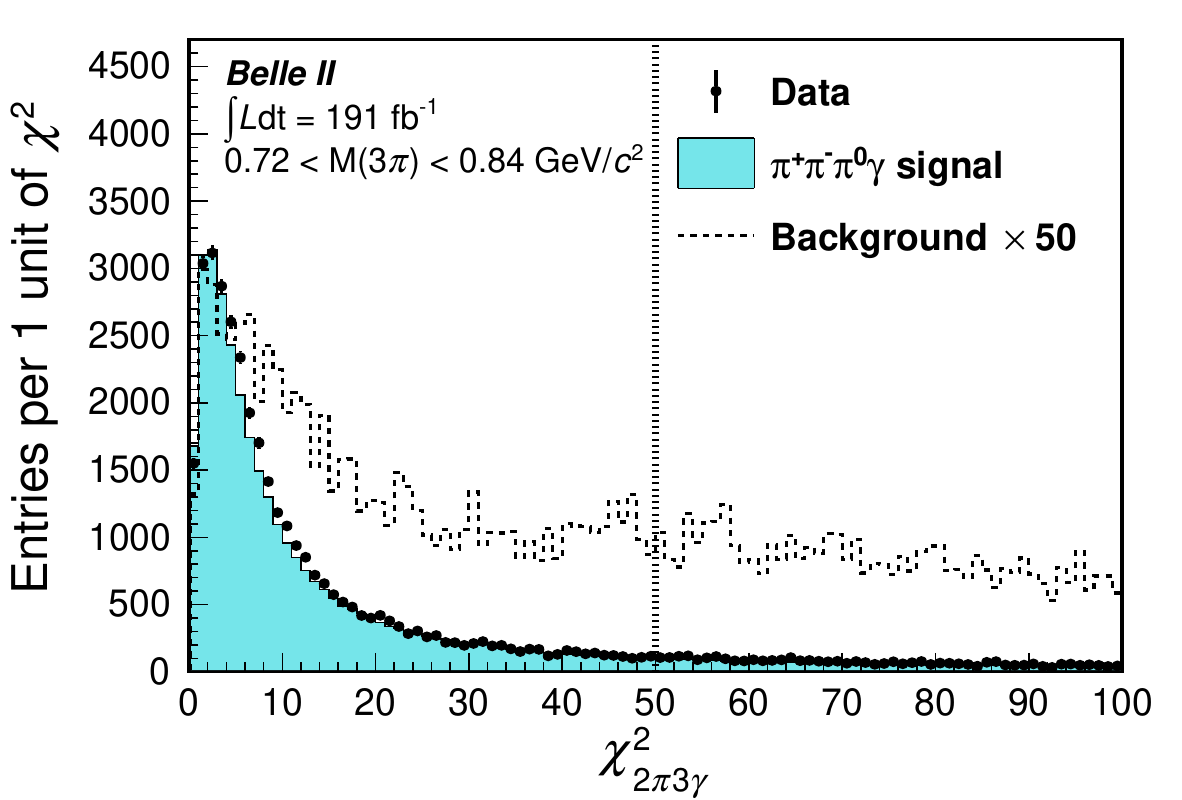}
    \caption{
        Distribution of \chisqfctpg for events in the range $0.72 < \mpppz < 0.84\gevcc$.
        The points with error bars show the data after the baseline selection.
        The filled and dashed histograms are the simulations for signal and background, respectively.
        The simulated signal is normalized to the integrated luminosity of data.
        The background simulation is scaled by a factor of 50.
        The dotted vertical line represents the upper bound of the \chisqfctpg selection.
  }
  \label{fig:spectrum_chi2}
\end{figure}

We kinematically fit the four-momentum of the $\pipi\gaga\g$ final state to match the four-momentum of the \epem beams (4C fit).
In order to extract signal yields from fits to diphoton invariant mass, the \piz mass constraint is not imposed in the 4C fit.
A 4C kinematic fit is performed for all $\pipi\gaga\g$ combinations if one photon satisfies the ISR criterion and the invariant mass of the remaining two photons (\mgg)\ is less than 0.3\gevcc.
We use a relatively wide \piz sideband to determine the level of combinatorial \gaga background.
If there are more than three photons in an event, we take all three-photon combinations.
Figure~\ref{fig:spectrum_chi2} shows the distribution of \chisqfctpg, the chi-squared value obtained from the 4C fit.
In Fig.~\ref{fig:spectrum_chi2}, the background simulation is scaled by a factor of 50.
We require $\chisqfctpg$ to be less than 50.
After applying the kinematic fit, \piz candidates are reconstructed from two-photon combinations.
Each \piz candidate is required to have a \gaga opening angle in the laboratory frame less than 1.4\rad
and the angle between the transverse momenta of the \gaga, $|\Delta \varphi_{\gaga}|$, less than 1.5\rad,
where $\varphi$ is the azimuthal angle around the magnetic field axis.
%
%
%
%
\subsection{Background suppression criteria}\label{sec:sel_bkgsup}
To reduce the systematic uncertainty due to background subtraction, further selections are imposed to suppress possible backgrounds.
We reject any charged particle identified as an electron or a kaon to reduce the contribution from the $\epem \to \epem(\g)$, $\epem \to \gaga$, and $\epem \to \Kp\Km\piz\g$ processes.
The likelihood for pion-kaon identification is obtained by combining information from all sub-detectors except for the PXD.
Information from the CDC, ECL, ARICH, and KLM is used for pion-electron identification.
We require the square of the mass recoiling against the \pipi pair to be greater than 4\gevccsq to reduce $\epem \to \ppg$ and $\epem \to \mmg$ backgrounds.
%
%
%
\par
One of the largest backgrounds is from the $\epem \to \pppzpzg$ process.
For events with two charged particles and at least five photons, we reconstruct the \pppzpzg candidate and reject events containing a candidate that meets the following criteria.
We apply the same selection criteria for charged particles and an ISR photon as those used in the \pppzg signal selection.
Two \piz's are reconstructed with different selections.
The minimum photon energy requirement is 100\mev for one \piz and 50\mev for the other.
The invariant mass selection $0.110 < \mgg < 0.158 \gevcc$ is applied to both \piz's.
A kinematic fit is then applied to the \pppzpzg candidate imposing four-momentum conservation.
Events containing a reconstructed \pppzpzg candidate having a chi-squared of the 4C kinematic fit $\chisqfcfpg < 30$ are rejected.
This criterion reduces 43\% of the \pppzpzg background, retaining 97\% of the signal.
%
%
%
\par
\begin{figure}
  \centering
  \includegraphics[width=\userFigureWidth]{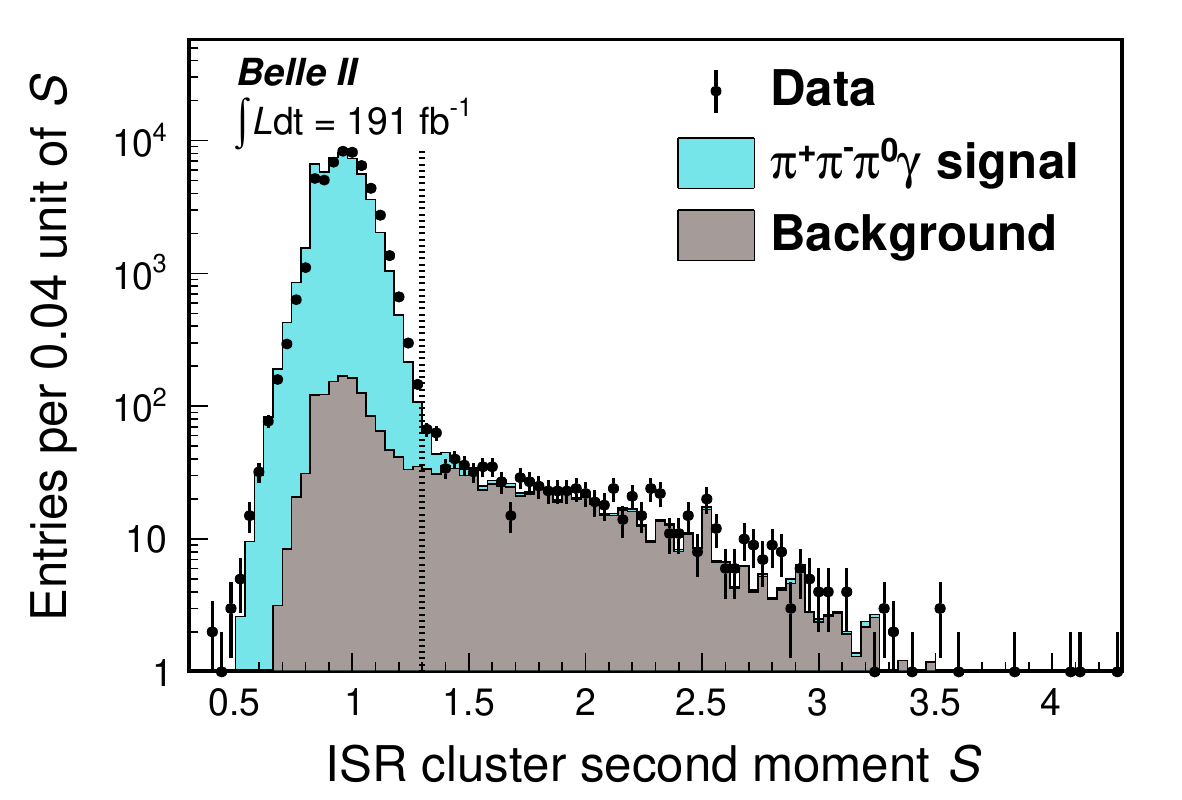}
    \caption{
        Distribution of ISR cluster second moment $S$ for events with all background-suppression requirements applied except for those on $S$.
        The points with error bars show the data.
        The stacked histograms are the simulations for signal (cyan) and background (grey).
        The simulated signal is normalized to the integrated luminosity of data.
        Events to the right of the dotted vertical line are rejected.
  }
  \label{fig:csm_sr}
\end{figure}

A selection on the invariant mass of a charged pion and the most energetic \g in the c.m.~frame, $\mpig> 2\gevcc$, is applied to reduce background from $\epem \to \tautau$ and non-ISR $\epem \to \qqbar$ backgrounds that include a high momentum $\rho^{\pm}\to\pi^{\pm}\pi^{0}$ decay.
Most of the remaining non-ISR \qqbar backgrounds are from events in which one or both of the photons from the \piz have high energy and can be reconstructed as an ISR photon.
To reduce this background, a selection on the invariant mass of the ISR photon candidate and any additional photon, \mggisr, is applied; we reject events containing a photon satisfying $0.10 < \mggisr < 0.17\gevcc$.
Two photons from a high momentum \piz occasionally produce a single cluster (a merged cluster)\ in the ECL.
These merged clusters can be distinguished from single photons using their ECL energy distributions.
We use the second moment of an ECL cluster, which is defined by%
\begin{align}
  S \propto \frac{\sum^{n}_{i} E_{i} r^{2}_{i}}{\sum^n_{i} E_{i}},
\end{align}
where $n$ is the number of ECL crystals associated with the cluster, $i$ is the index of the ECL crystal, $E_{i}$ is the energy of crystal $i$, and $r_{i}$ is the distance between the center of the crystal and the cluster axis.
The distribution of the cluster second moment is shown in Fig.~\ref{fig:csm_sr}.
The cluster's second moment peaks around 1.0 for a typical ISR photon cluster, whereas it has a wider distribution in the range of 1.0--3.0 for merged single clusters.
Candidate ISR photons with $S$ values greater than 1.3 are rejected. %
The signal efficiency for the background suppression criteria is about 91\%, while the signal purity increases from 91\% to 98\% in the \mpppz
region below 1.05\gevcc. 
At this stage, 92\% of events have a single signal candidate.
The candidate with the minimum \chisqfctpg is selected in events with multiple candidates.
\subsection{Signal extraction}\label{sec:signal_extraction}
\quad  
To select the final signal sample, we require that the \piz candidates have two-photon invariant mass in the range  $0.123 < \mgg < 0.147\gevcc$. 
We use this sample to study the general features of signal events unless otherwise specified.
\par
\begin{figure*}[t]
  \centering
  \includegraphics[width=0.48\textwidth]{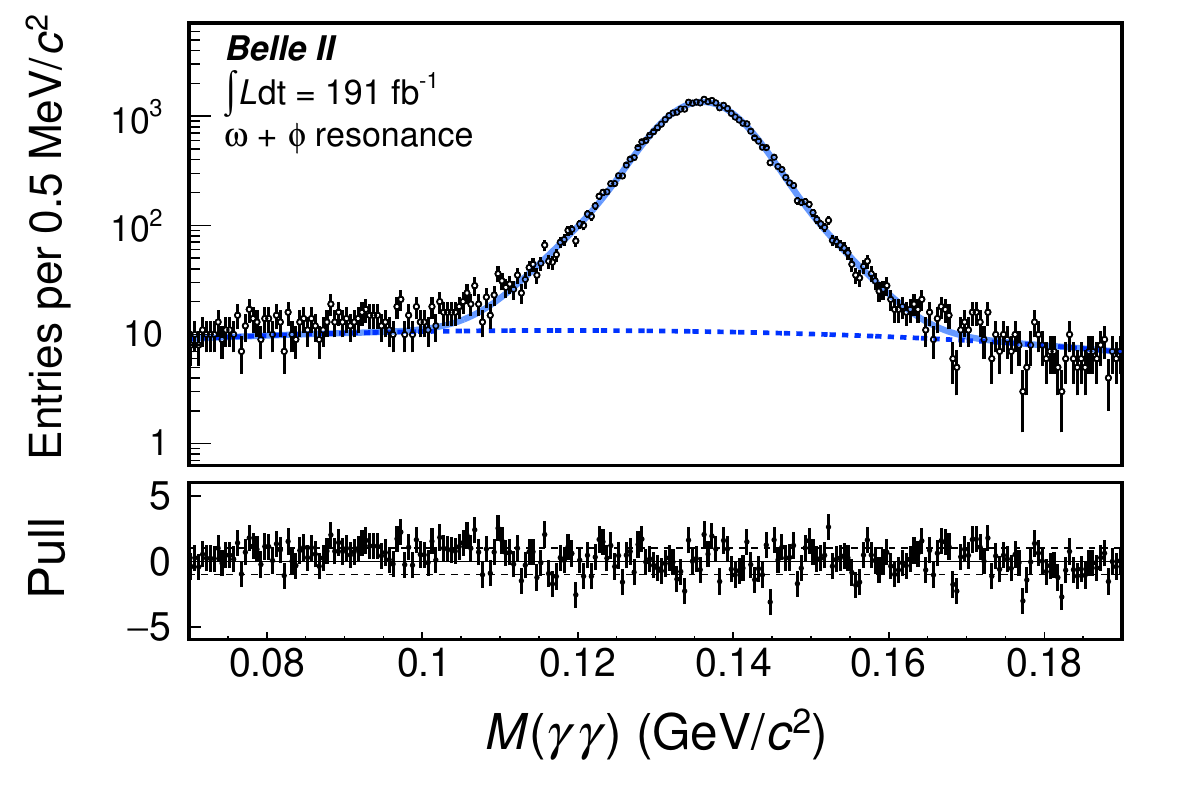}
  \includegraphics[width=0.48\textwidth]{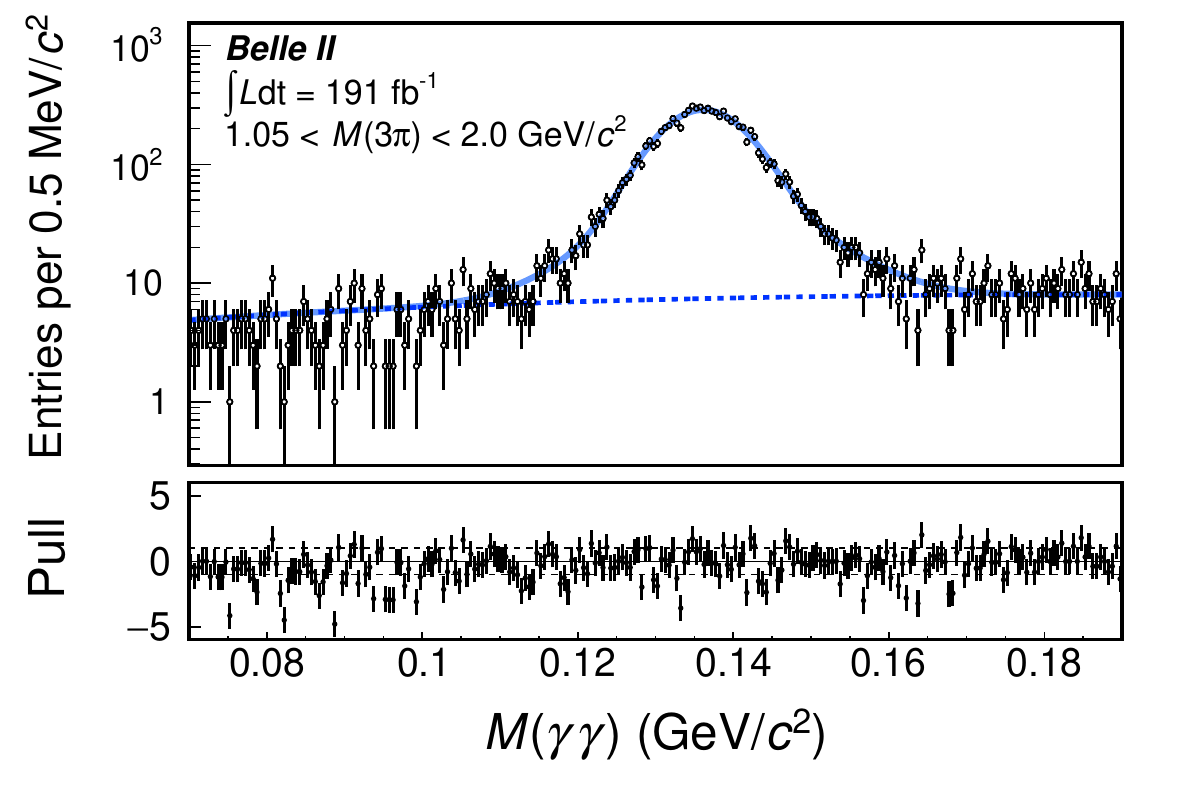}
    \caption{
      Diphoton mass distributions for (left) events in the $3\pi$ mass region below 1.05\gevcc and (right) in the 1.05--2.00\gevcc range with the fit projection overlaid.
	   The points with error bars represent the data passing all event selections.
	   The solid and dashed curves show the total and background contributions to the fit result.
      The differences between data and fit results divided by the data uncertainties (pull) are shown in the bottom panel.
    }
  \label{fig:final_sample_pi0mass}
\end{figure*}

The \piz signals selected in this way, however, include both true \piz signals and combinatorial background in which the invariant mass of random combinations of two photons has a value near the mass of the \piz meson.
To extract a three pion mass \mpppz distribution from the true \piz signal without the combinatorial background, we fit to the \mgg distribution in each three pion mass \mpppz range.
The \mgg distributions for the data, in the low and high \mpppz mass regions are shown
in Fig.~\ref{fig:final_sample_pi0mass}.
A clear \piz signal is seen above a linear background. 
We model the signal PDF with the sum of a Novosibirsk function~\cite{Belle:1999bhb} and a Gaussian function, the combinatorial background is modeled with 1st- or 2nd-order polynomials.
The fit is carried out in the \mgg range 0.05--0.23\gevcc for the $3\pi$ mass range 0.5--1.05\gevcc, 0.05--0.29\gevcc for the $3\pi$ mass range 1.05--2\gevcc, and 0.07--0.29\gevcc for the $3\pi$ mass range 2.0--3.5\gevcc.
In each $3\pi$ mass region, a wide sideband region is selected so that the background event model correctly determines the non-\piz background component.
\par
\begin{figure*} 
  \centering
    \includegraphics[width=0.48\textwidth]{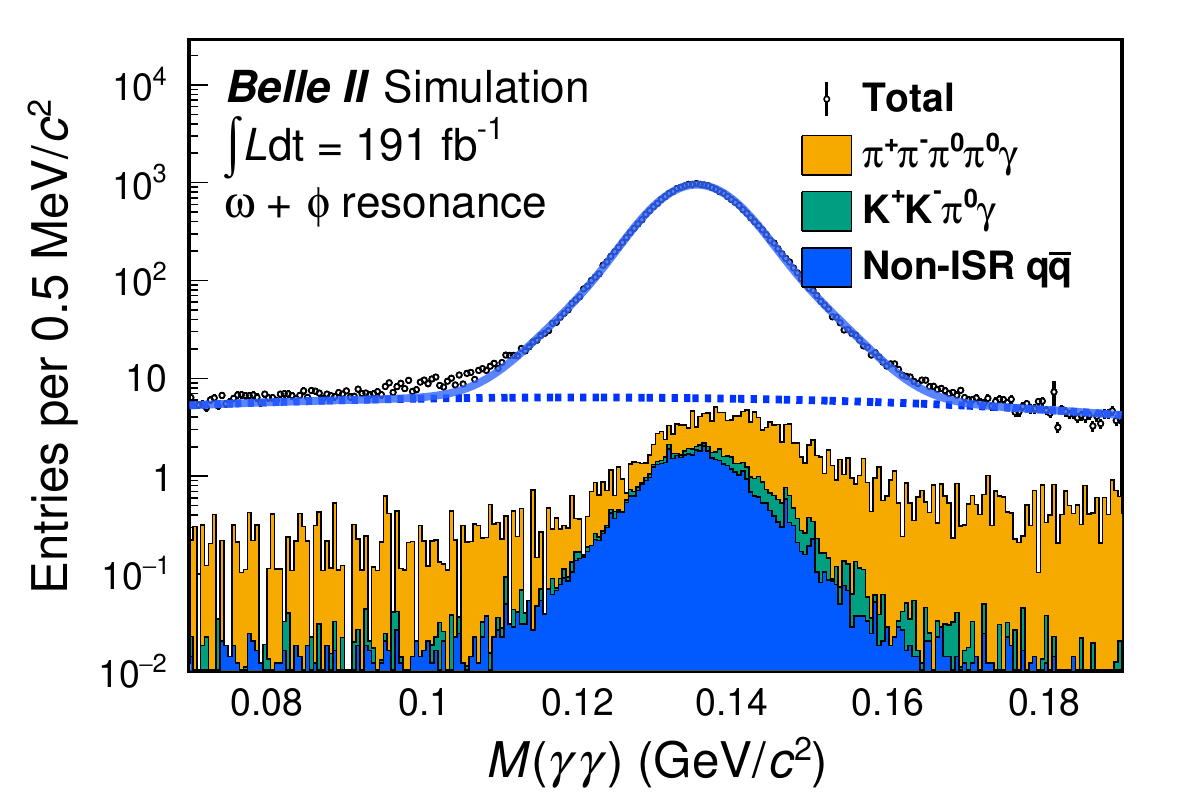}
    \includegraphics[width=0.48\textwidth]{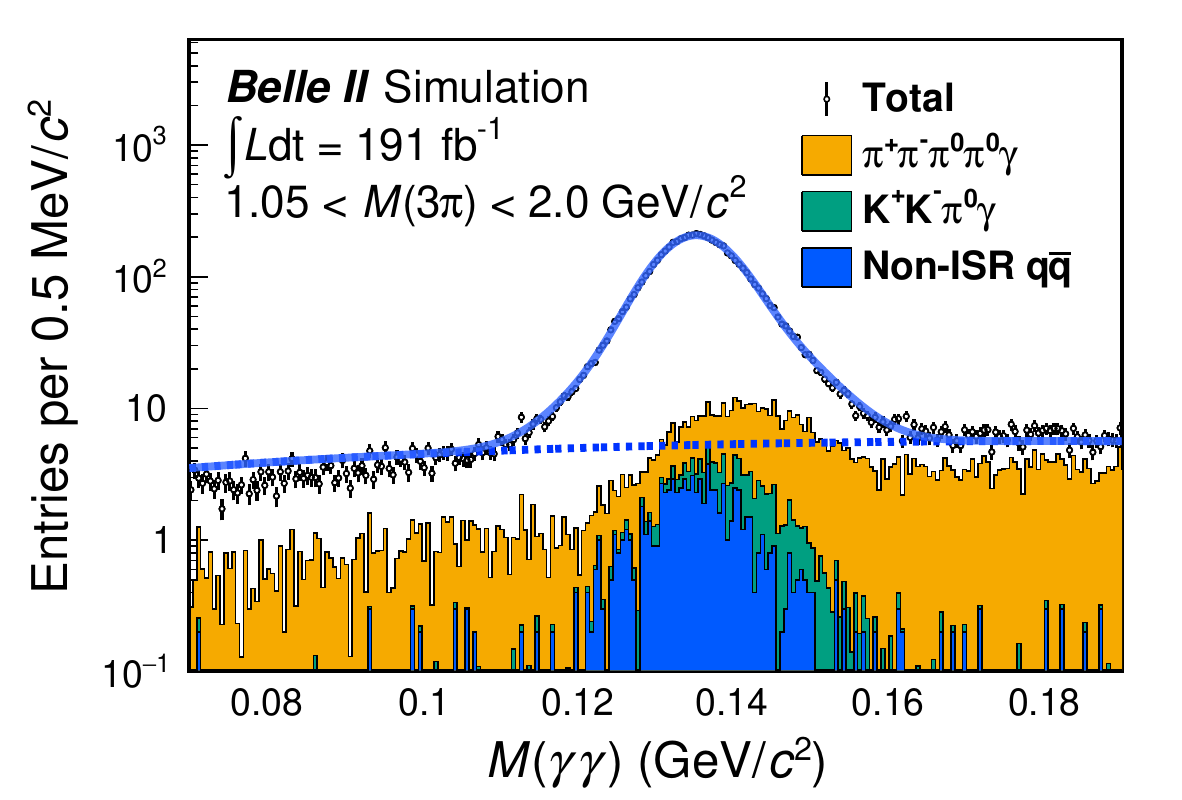}
    \caption{
      Diphoton mass distributions for simulated events (left) in the $3\pi$ mass region below 1.05\gevcc and (right) in the 1.05--2.00\gevcc range with the fit projection overlaid.
      The simulated signal is normalized to the integrated luminosity of data.
	    The points with error bars represent the simulated events passing all event selections.
	    The solid and dashed curves show the total and background contributions to the fit result.
      The filled stacked histograms are the estimated contributions of residual backgrounds.
    }
  \label{fig:final_sample_pi0mass_mc}
\end{figure*}

Figure~\ref{fig:final_sample_pi0mass_mc} shows the simulated \mgg distributions in the $\omega$ and $\phi$ resonance regions.
The values of the parameters used to describe the \piz line shape agree within errors between the data and the simulation except for the line width.
The width of the \piz signal in data is about 0.8\mevcc wider than that of the simulation. 
The consequences of this discrepancy between data and simulation on signal efficiency and the associated systematic errors are evaluated and discussed in detail in Sec.~VI~A.
\par
The distribution of the background processes exhibits a tail on the higher side in contrast to the signal, due to the 4C fit being carried out assuming the signal process, even though this assumption is incorrect for the background.
However, simulation studies demonstrate that these background events do not affect the overall line shape, as the fraction of these backgrounds is small (0.6\% for $\omega$ region and 8\% for the higher mass region).
The uncertainty due to the small difference in the line shape is negligible compared to the uncertainty in the \piz detection efficiency from other sources discussed in Sec.~V~C.
\par
To obtain the \piz yield, we integrate over the signal function in the range $0.123 < \mgg < 0.147\gevcc$, so that the \piz yields from both methods, counting and fitting, agree when the combinatorial background is negligible.
The \mgg range is optimized for the fraction of the combinatorial background especially in the $3\pi$ mass region below 1.05\gevcc.
We use the fitting method to obtain the \mpppz distribution for both the data and the simulation in the later sections.
More details of the fitting procedures are described in Sec.~\ref{sec:spectrum}.

%
\subsection{Comparisons of data and Monte Carlo simulation}
\begin{figure*}
  \centering
    \includegraphics[width=\userFigureWidth]{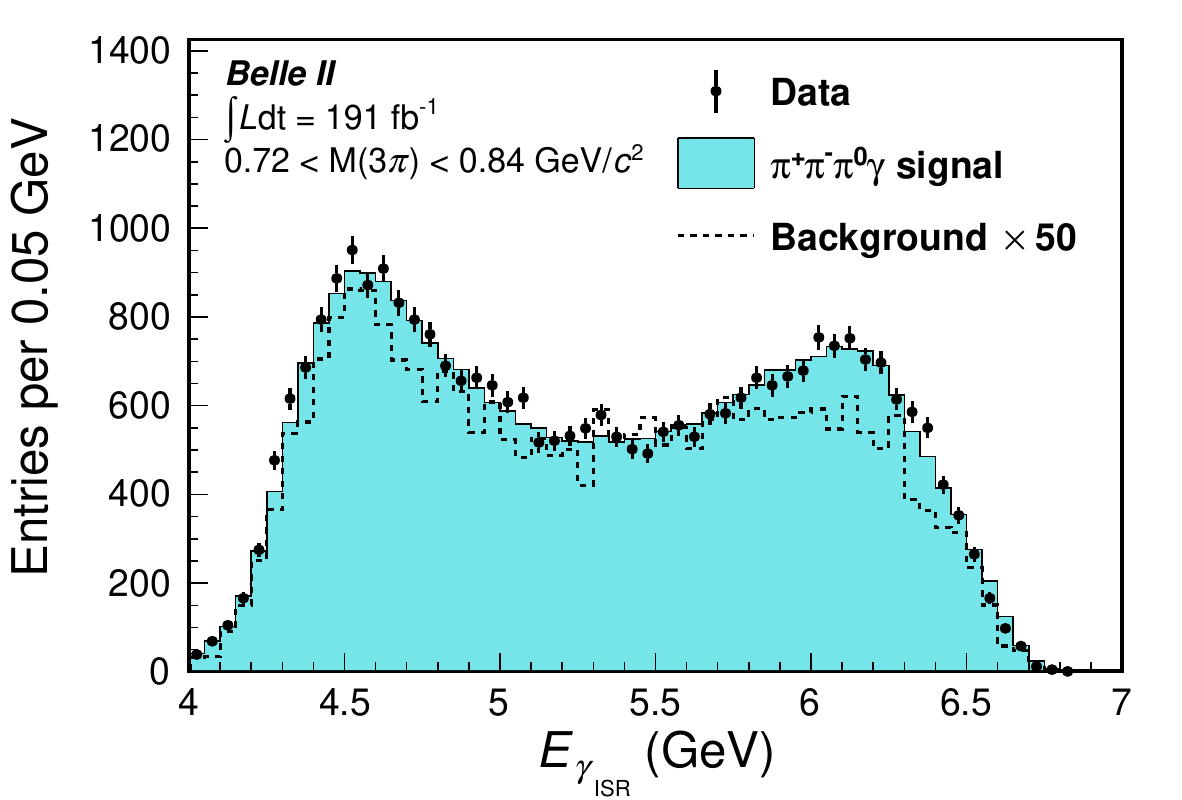}\qquad
    \includegraphics[width=\userFigureWidth]{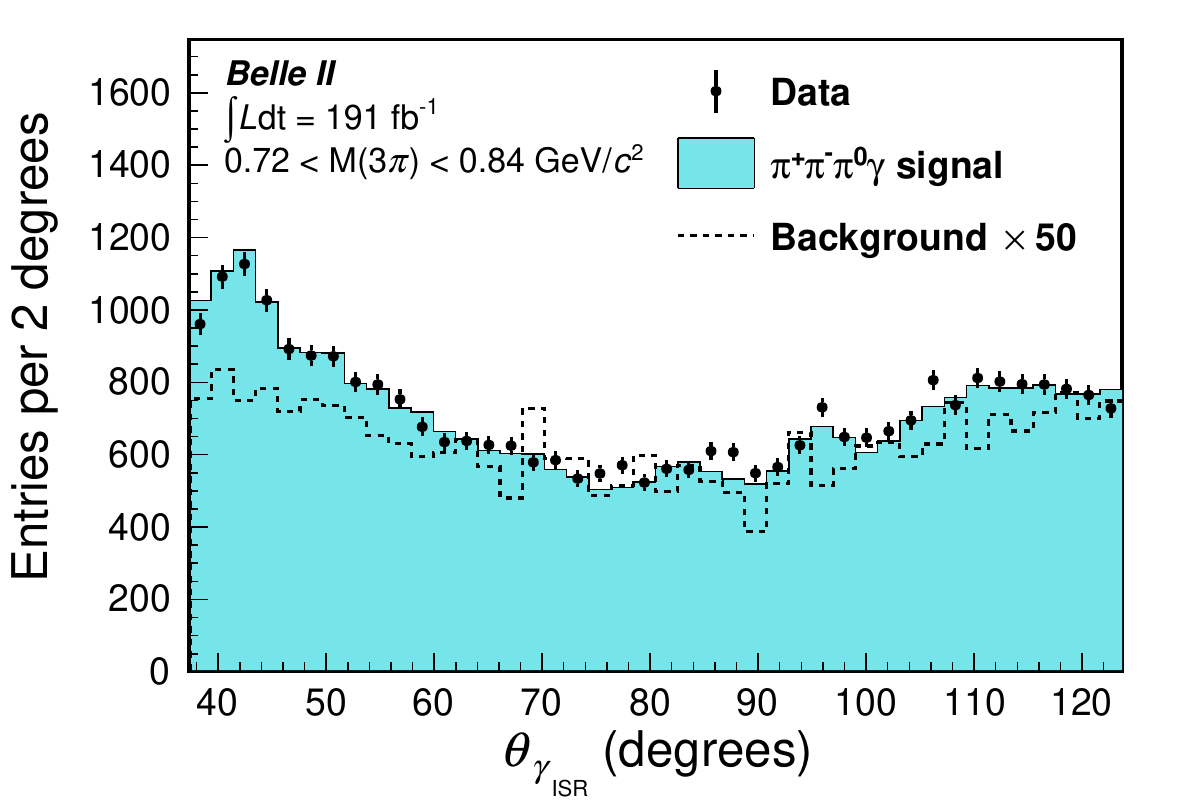}
    \caption{
        Distribution of (left) ISR energy and (right) polar angle in the laboratory frame.
        The points with error bars show the data after all signal selections.
        The filled and dashed histograms are the simulations for the signal and background, respectively.
        The simulated signal is normalized to the integrated luminosity of the data.
        The background simulation is scaled by a factor of 50.
    }
  \label{fig:variable_isr}
\end{figure*}
\begin{figure*}
  \centering
  \includegraphics[width=\columnwidth]{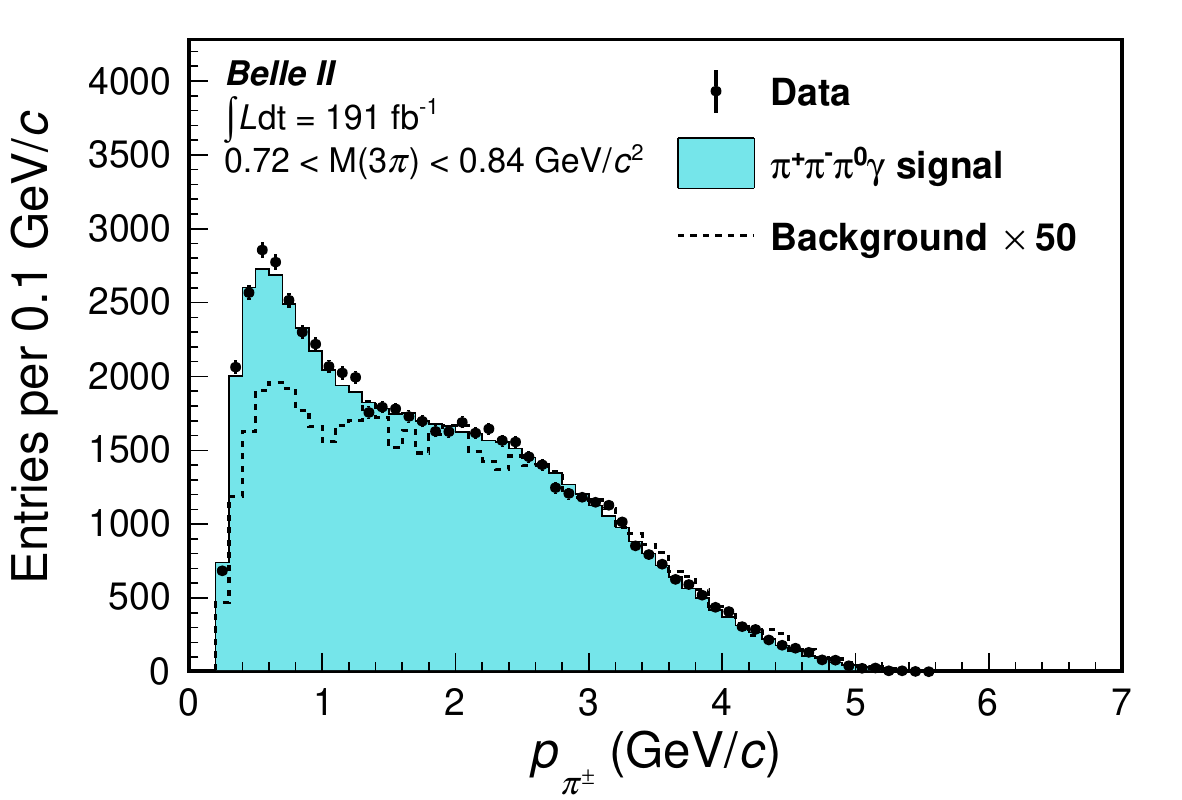}
  \includegraphics[width=\columnwidth]{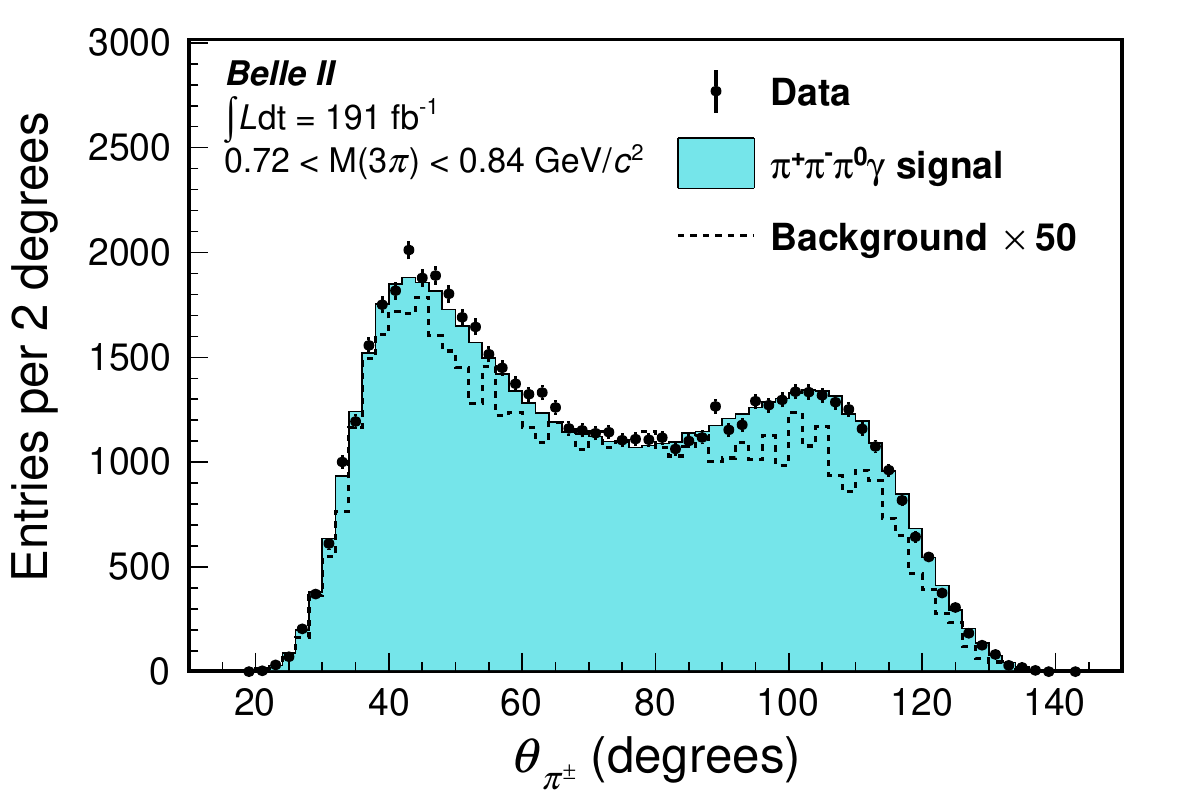}
  \caption{
        Distribution of (left) charged-pion momentum and (right) polar angle in the laboratory frame.
        The convention in the figure is the same as in Fig.~\ref{fig:variable_isr}.
  } 
  \label{fig:variable_pi}
\end{figure*}
\begin{figure*}
  \centering
    \includegraphics[width=\columnwidth]{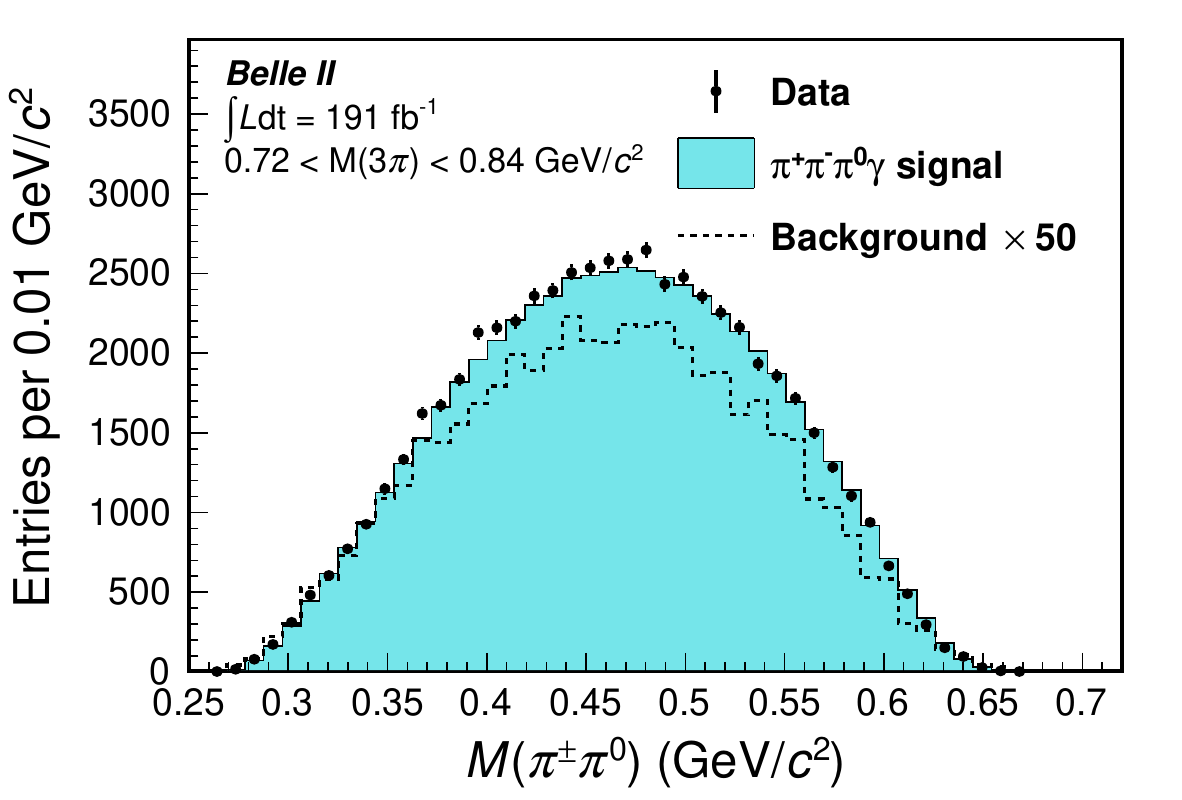}
    \includegraphics[width=\columnwidth]{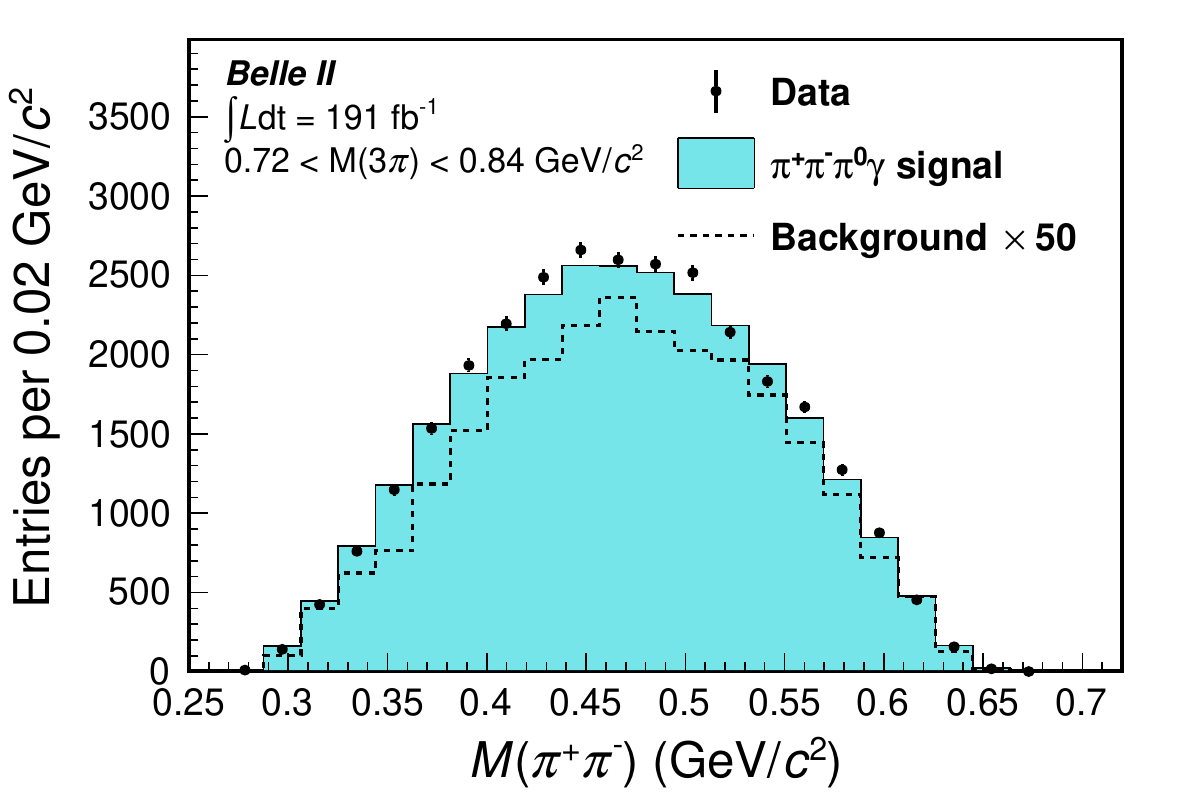}
    \caption{
        Distribution of (left) charged and (right) neutral dipion mass.
        The convention in the figure is the same as in Fig.~\ref{fig:variable_isr}.
  }
  \label{fig:spectrum_pipimass}
\end{figure*}

As a consistency check, we now compare detailed distributions in data and Monte Carlo simulation after all selection requirements are applied.
\par
\begin{figure}
  \centering
    \includegraphics[width=\userFigureWidth]{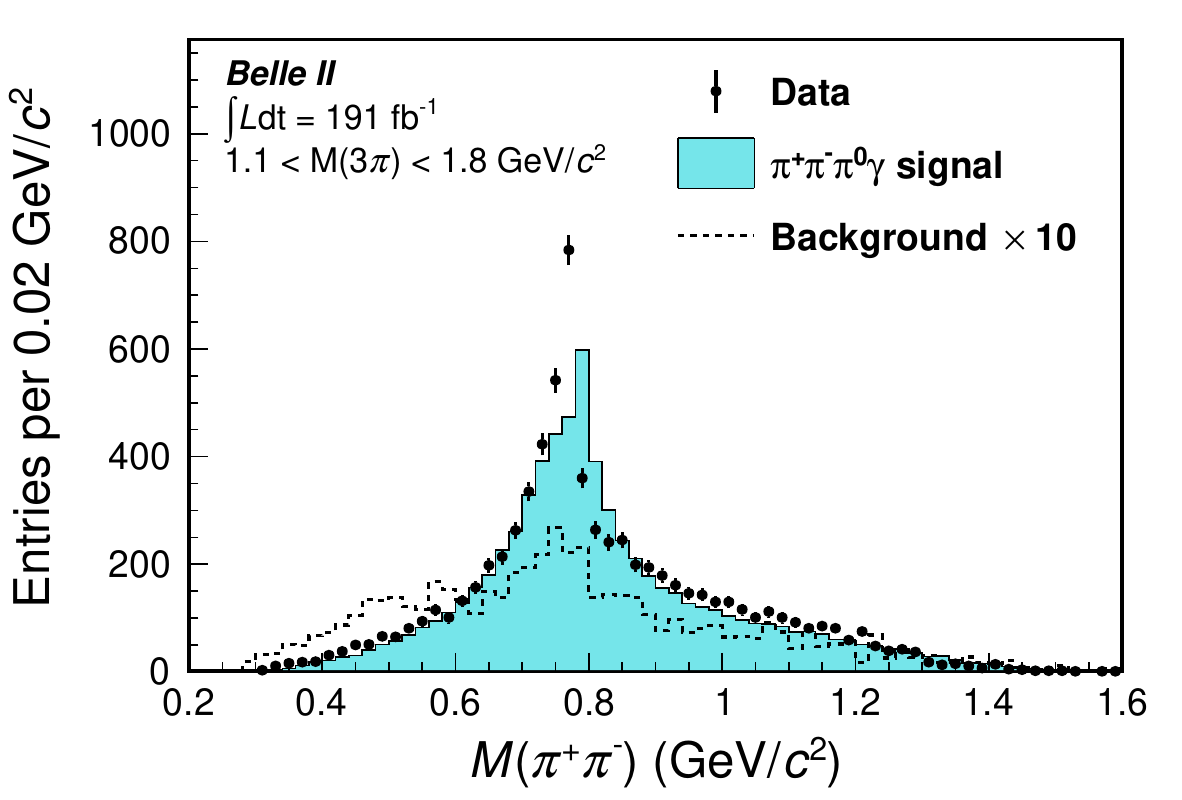}
    \caption{
        Distribution of $\pipi$ invariant mass in the $3\pi$ mass range of 1.1--1.8\gevcc.
        The convention in the figure is the same as in Fig.~\ref{fig:variable_isr}.
        The background simulation is scaled by a factor of 10.
  }
  \label{fig:spectrum_pipimass_2}
\end{figure}

The energy and polar angle distributions of ISR photons in data are compared with simulation in Fig.~\ref{fig:variable_isr}.
Figure~\ref{fig:variable_pi} shows the momentum and polar angle distributions of charged pions.
Figure~\ref{fig:spectrum_pipimass} shows the invariant mass distribution of the two pion system
when the invariant mass of the \pppz system is in the $\omega$ region, $0.72 < \mpppz < 0.84 \gevcc$.
In Figs.~{\ref{fig:variable_isr}}--{\ref{fig:spectrum_pipimass}}, the background simulation is scaled up by a factor of 50.
A good agreement in Figs.~\ref{fig:variable_isr}, \ref{fig:variable_pi}, and \ref{fig:spectrum_pipimass} confirms that the Monte Carlo event generator and detector simulation reproduce the basic variables well.
There are slight discrepancies in the charged dipion invariant mass distribution for the $3\pi$ mass region 1.1--1.8\gevcc, as shown in Fig.~\ref{fig:spectrum_pipimass_2}.
In Figs.~\ref{fig:spectrum_pipimass_2}, the background simulation is scaled up by a factor of 10.
\babar in Ref.~\cite{Czyz:2005as} also observed a similar $\rho$-$\omega$ interference pattern near $m_\omega$, which is not modeled by \texttt{PHOKHARA}.
Although the model implemented in \texttt{PHOKHARA} is not perfect, this difference does not affect our cross section measurement as a function of \mpppz.
We confirm that the efficiency does not depend on the dipion mass $M(\pipi)$.
%
%
%
%
\\
\section{Background estimation}~\label{sec:bkg}
\begin{figure}
  \centering
  \includegraphics[width=\userFigureWidth]{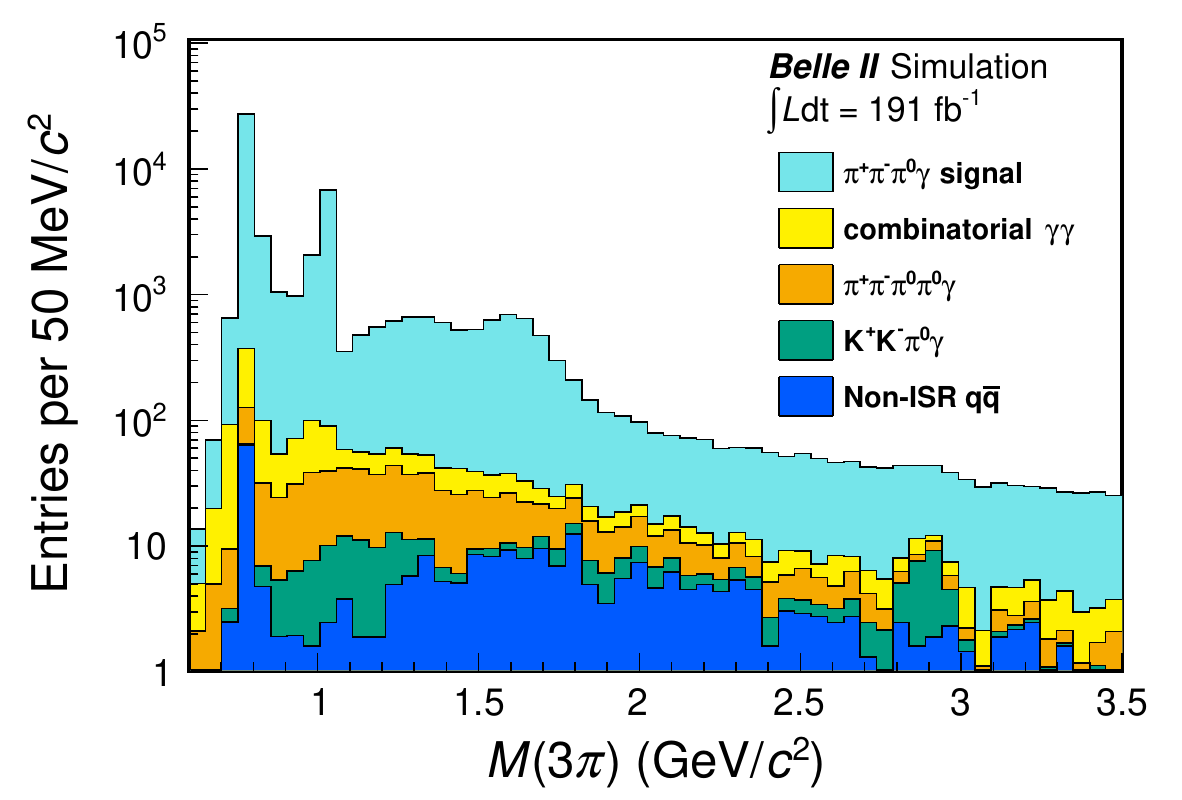}
  \caption{
      Distribution of $3\pi$ mass spectrum in Monte Carlo simulation after all background suppression requirements.
      The filled stacked histograms are the individual contributions:
      signal $\epem \to \pppzg$ (cyan),
      combinatorial \gaga background (yellow),
      $\epem \to \pppzpzg$ (orange),
      $\epem \to \kkpzg$ (green),
      and non-ISR \qqbar (blue).
      The samples are normalized to the integrated luminosity of data.
  }
  \label{fig:bkg_sim}
\end{figure}

We use Monte Carlo simulation normalized to the integrated luminosity of the data as an approximate description of the sample composition.
The simulated signal and backgrounds after applying all selection requirements are shown in Fig.~\ref{fig:bkg_sim}.
The overall background level is less than 1\% in the $\omega$ and $\phi$ resonance regions and is about 10\% above 1.05\gevcc.
The main backgrounds remaining in the sample are $\epem \to \pppzpzg$, $\epem \to \kkpzg$, $\epem \to \qqbar$, and combinatorial \gaga background.
The last one refers to a background in which one or both photons in the \piz candidate do not originate from a real \piz.
This background dominates in the $\omega$ and $\phi$ resonance regions but does not have a signal-like peak at the \piz mass, so it is not expected to bias our signal extraction, which is based on determining the \piz yield.
Since the processes $\epem \to \ppg$ and $\epem \to \mmg$ contribute only to the combinatorial \gaga background (i.e. the \mgg distribution for the background).
Since there are no contributions in the \piz peak, we do not discuss the $\epem \to \ppg$ and $\epem \to \mmg$ backgrounds further in this section.
\par
To confirm the background expectation from the simulation, we examine several data control samples, each enhanced in a specific background source.
%
%
%
%
%
\par
\begin{figure*}
\centering
  \subfloat[][]{\includegraphics[width=0.48\textwidth]{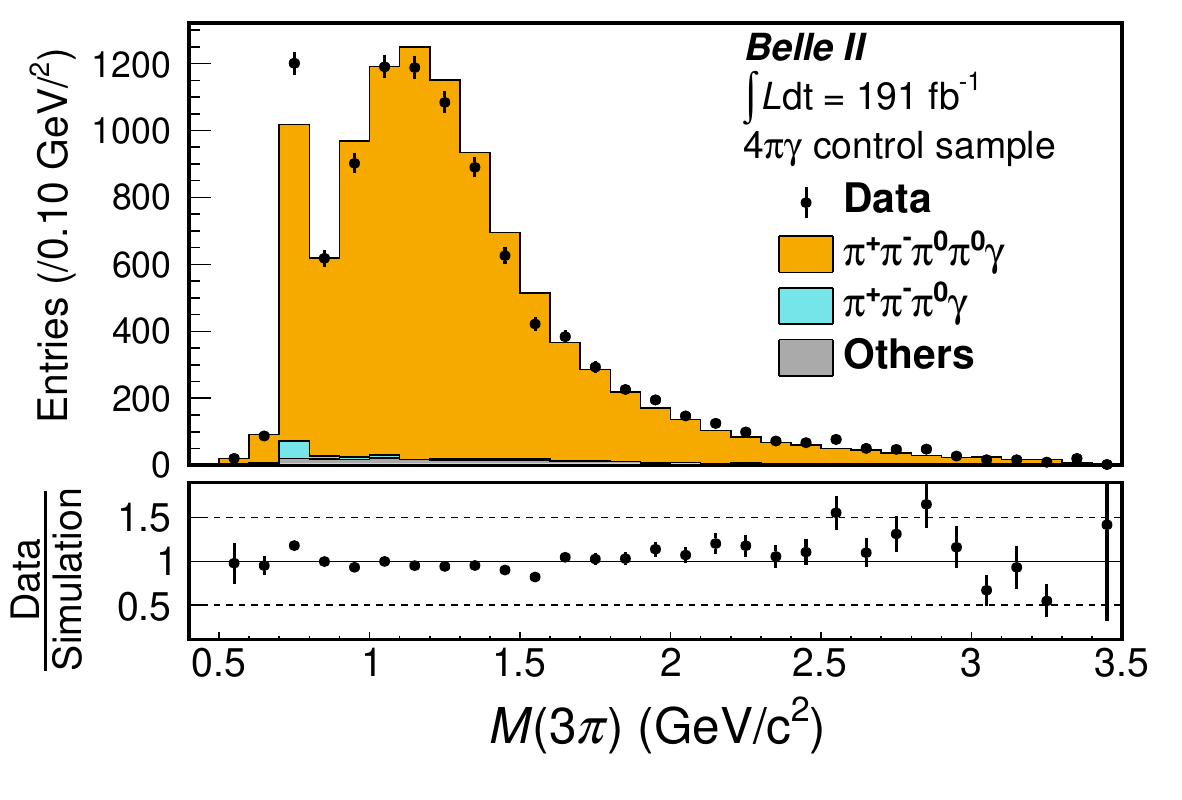}\label{fig:bkgest_isr4pi_cr_a}}
  \subfloat[][]{\includegraphics[width=0.48\textwidth]{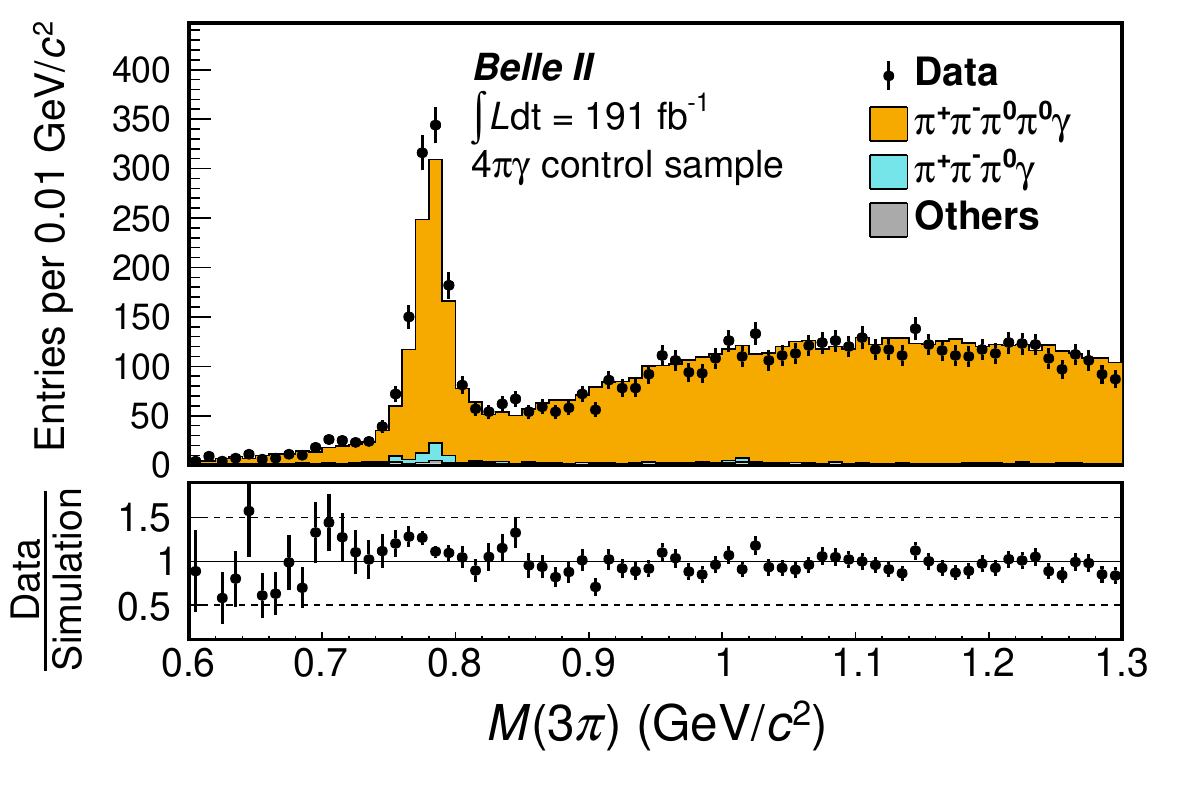}\label{fig:bkgest_isr4pi_cr_b}}
    \caption{
        Distribution of $3\pi$ mass in the $\epem \to \pppzpzg$ control sample
				(a) for the $3\pi$ mass range 0.4--3.5\gevcc in 100-\txtmevcc-wide bins and
				(b) for the $3\pi$ mass range 0.6--1.3\gevcc in 10-\txtmevcc-wide bins.
        Events are selected with the requirements $100 < \chisqfctpg < 1000$ and $\chisqfcfpg < 100$.
        Points with error bars show the data and the stacked histograms show the simulations.
        Bottom panels show data-to-simulation ratios.
  }
  \label{fig:bkgest_isr4pi_cr}
\end{figure*}

A control sample that enhances the $\epem \to \pppzpzg$ background is selected by requiring $\chisqfcfpg<100$ and $100 < \chisqfctpg <1000$.
The first condition selects $\epem \to \pppzpzg$ candidates while the second reduces the contamination from $\epem \to \pppzg$.
The resulting \pppzpzg sample has 95\% purity.
Figures~\ref{fig:bkgest_isr4pi_cr} show data-simulation comparisons in the \mpppz distribution of the \pppzpzg sample for \subref{fig:bkgest_isr4pi_cr_a} the full mass region and \subref{fig:bkgest_isr4pi_cr_b} the region below $1.3\gevcc$.
The $\epem \to \pppzpzg$ model implemented in \texttt{PHOKHARA} reproduces the \mpppz distribution well, except for the normalization of the $\omega$ signal.
After subtracting the small contamination from other processes, a scale factor as a function of $3\pi$ mass is determined from Fig.~\ref{fig:bkgest_isr4pi_cr} by taking the bin-by-bin ratio between the data and the simulation.
For the $3\pi$ mass region below 1.05\gevcc, there is a clear $\omega$ signal from $\epem \to \omega\piz\g \to \pppzpzg$.
Since the $\omega$ resonance is narrow, we fit the $3\pi$ mass shape with an $\omega$ signal and a smooth component for the data and the simulation,
and obtain a scale factor from the ratio between the data and the simulation.
The \mpppz distribution is modeled using a Voigt function, which is a convolution of a Breit-Wigner function with a Gaussian function, for the $\omega$ resonance peak and a Gaussian function for the non-$\omega$ component up to 1.05\gevcc.
The scale factor is 1.3 at the $\omega$ resonance and varies from 0.8 to 1.2 in the neighboring regions.
The uncertainty in the \pppzpzg level is 10\%--20\%, which includes the statistical uncertainty of the simulation and the uncertainty on the efficiency associated with the additional \piz.
\par
So far we have determined the data-to-simulation scale factors for the \pppzpzg background without separating $4\pi$ mass ($M(4\pi)$) regions.
To test the $M(4\pi)$ dependence of the corrections, we have prepared separate data-to-simulation scale factors as a function of \mpppz in six $M(4\pi)$ regions.
We apply these corrections to the simulated \pppzpzg sample.
We find the difference in the resulting \mpppz distributions after applying the scale factors before and after separating the $M(4\pi)$ regions is negligibly small.
%
%
%
%
%
\par
\begin{figure}
  \centering
  \includegraphics[width=\userFigureWidth]{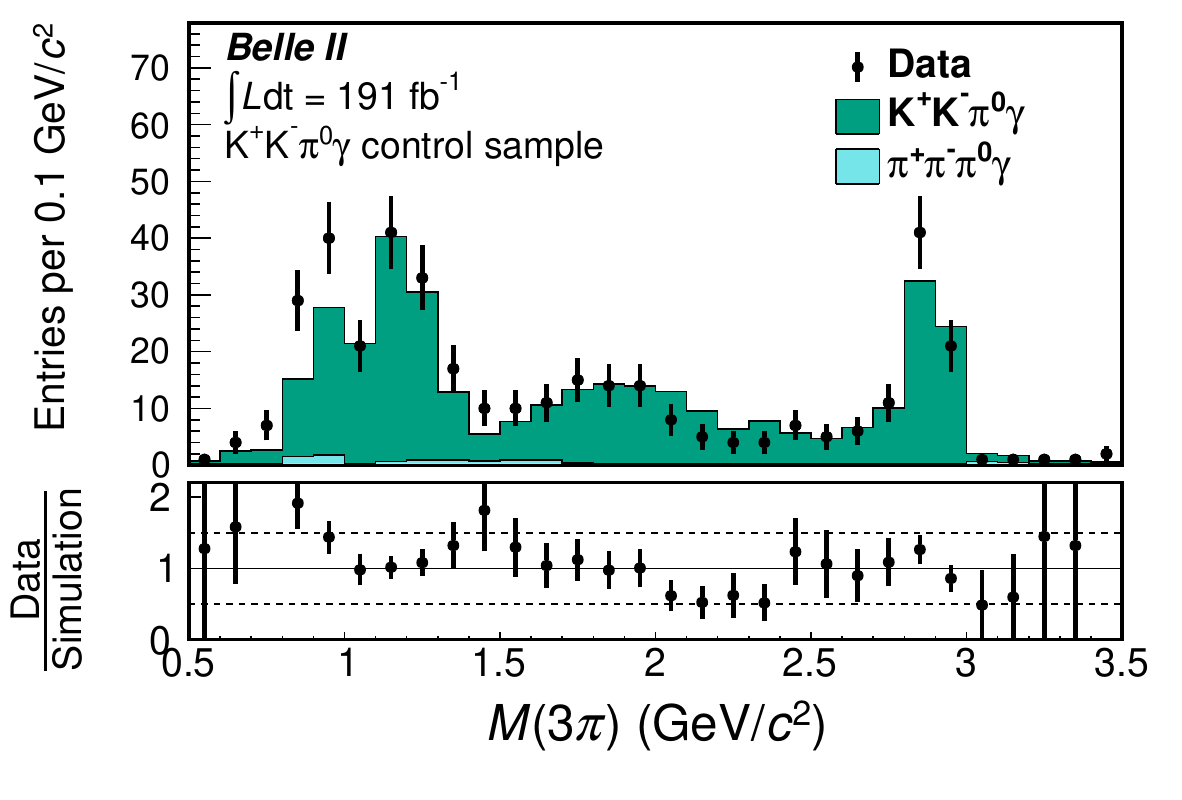}
    \caption{
        Distribution of $3\pi$ mass in the $\epem \to \kkpzg$ control sample, assigning the pion mass to both tracks.
        Points with error bars show the data, and stacked histograms show the simulations.
        Bottom panels show data-to-simulation ratios.
  }
  \label{fig:bkgest_kkpzg_cr}
\end{figure}

A data-to-simulation scale factor for the $\epem \to \kkpzg$ background is evaluated using a control sample that enhances the \kkpzg component, where both charged particles are identified as kaons.
The scale factor as a function of $3\pi$ mass is determined by taking the ratio of the \mpppz distribution of the data to the corresponding simulated sample, which assumes a $K^{*}(892)^{\pm}K^{\mp}$ intermediate state.
Figure~\ref{fig:bkgest_kkpzg_cr} compares the \mpppz distribution of the \kkpzg control sample, where the pion mass hypothesis is assumed for the charged particles.
The $\kkpzg$ model implemented in \texttt{PHOKHARA} and \texttt{EvtGen} reproduces the enhancement of the $\kkpzg$ process with an accuracy of around 50\%.
Contamination from other processes, such as $\pppzg$, is negligible.
The scale factor for the $\kkpzg$ component is obtained in 100-\txtmevcc-wide \mpppz bins.
A 10\%--20\% uncertainty due to the limited size of experimental and simulated samples and an uncertainty in the corrections to the pion-kaon identification efficiency is assigned for the scale factor.
%
%
%
%
%
\par
The non-ISR \qqbar background is negligible in the $\omega$ and $\phi$ mass regions, and is dominant in the high mass region beyond $1.05\gevcc$; the main source is the $\epem \to \pppzpz$ process.
In the latter process, two photons from one of the \piz's, which has a relatively high momentum, are merged into a single cluster in the ECL, resulting in misidentification of the \piz as an ISR photon.
A data-to-simulation scale factor for the non-ISR \qqbar background as a function of $3\pi$ mass is estimated from events in which the ISR photon candidate originates from a merged cluster of \piz decay photons.
A control sample is selected with the requirement that the cluster second moment be above 1.5 and $\chisqfctpg < 50$.
Although the shape of the $3\pi$ mass spectrum is generally consistent, the Monte Carlo simulation overestimates this background compared to data by a factor of two.
Therefore, the scale factor in the range 0.5--1.4\gevcc is 0.5--0.8.
We also test the simulated non-ISR \qqbar background using a control sample selected by requiring that \mggisr is consistent with the \piz mass and $\chisqfctpg < 50$, and obtain consistent results.
\par
For the non-$4\pi$ \qqbar background, we prepare a control sample with $100 < \chisqfctpg < 1000$ and determine the data-to-simulation scale factor using the distributions of the cluster second moment moment and \mggisr.
\par
The background in the $\epem\to \pppzpz$ process is dominated by $\rho\rho$ and $\rho\pi\pi$ production.
However, there is an additional process $\epem \to \omega\piz$ that is not available in the \texttt{KKMC} program. 
We prepare a simulated sample of this process using \texttt{EvtGen}.
To test the correctness of the simulation for the background associated with merged clusters, we prepare a data control sample in which the cluster second-moment requirement in the signal selection is 
reversed from below 1.3 to above 1.5.
This control sample shows that the simulation agrees with the data at the 20\% level.
Using a simulated Monte Carlo sample of the non-ISR $\omega\piz$ process, we estimate the remaining $\omega\piz$ background contamination to be $39 \pm 9$ events, which corresponds to about 0.1\% of the signal.
A systematic uncertainty of 20\% is assigned to the $\omega\piz$ background level based on the largest difference observed when the selection on the cluster second moment is varied from 1.5 to 2.5.
%
%
%
%
\par
\begin{figure}
    \centering
    \includegraphics[width=\userFigureWidth]{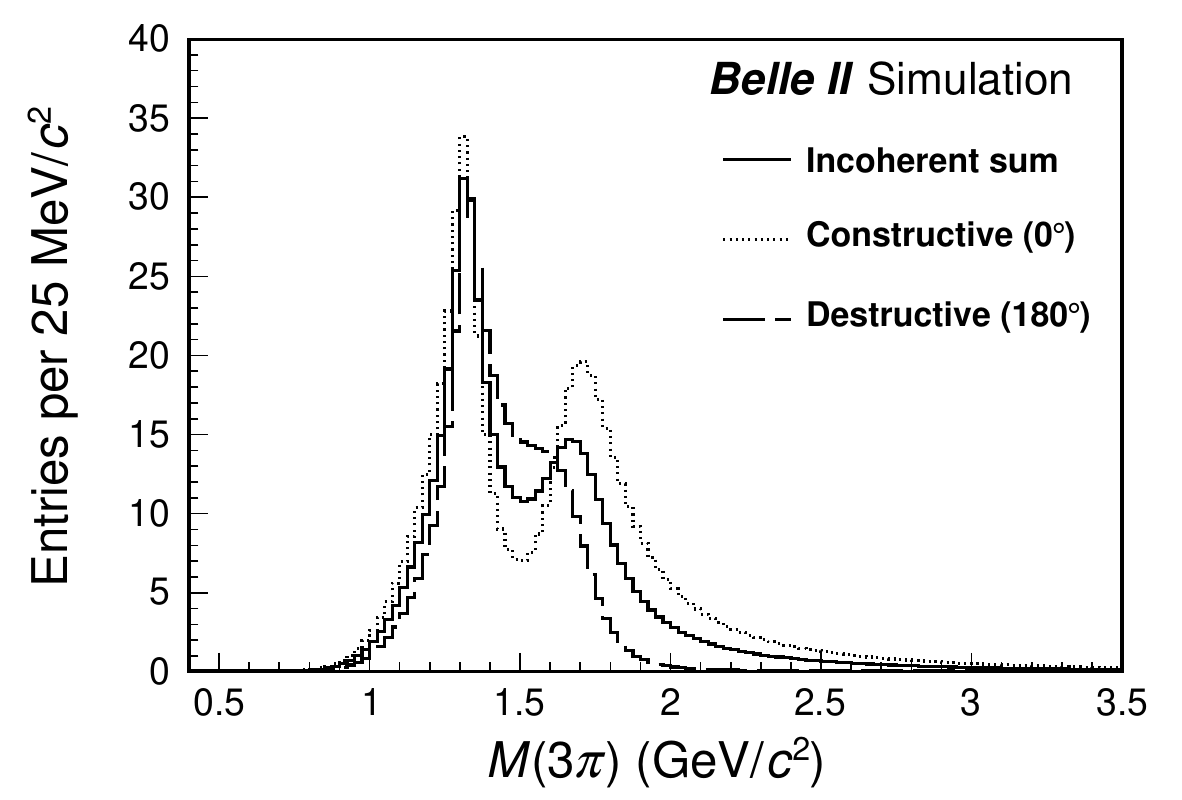}
    \caption{
      Expected \mpppz spectrum of the FSR processes $\epem \to M\g \to \pppzg$, where $M = a_1(1260)$, $ a_2(1320)$, $a_1(1640)$, or $a_2(1700)$.
      The solid histogram is the incoherent sum of the processes.
      The dotted (dashed) histogram is the constructive (destructive) interference of the processes.
    }
    \label{fig:fsr_background}
\end{figure}

There are small potential backgrounds from meson production with final-state radiation~(FSR).
Among the FSR background processes, the cross section for $\epem \to \pppz$ where the charged pion emits a hard FSR photon with $E^{*} > 4\gev$ is negligibly small.
Other FSR processes are from exclusive reactions that involve photon emission from one of the quark legs, $\epem \to M\g$, where $M$ is an intermediate meson.
We follow the discussion of \babar~\cite{BABAR:2021cde} for the estimate of these FSR processes.
The $\epem \to M\g$ processes in which the intermediate meson $M$ is an $a_1(1260)$, $ a_2(1320)$, $a_1(1640)$, or $a_2(1700)$ are taken into account.
These hadronic states are expected to decay to $3\pi$ with branching fractions of 30\%--50\%~\cite{Barnes:1996ff,Pang:2014laa,Chen:2015iqa}.
The FSR background contributes above the 1.2\gevcc region in \mpppz.
The differential cross section as a function of FSR polar angle is evaluated using perturbative QCD, and is given by~\cite{BABAR:2021cde}
\begin{equation}
    \frac{\d \sigma(\epem \to M\g)}{\d \cos{\thetagcms}} = \frac{\pi^{2}\alpha^3}{4} \left|F_{M\gaga}\right| (1+\cos{\thetagcms}^2)
\end{equation}
where $F_{M\gaga}$ is a meson-photon transition form factor.
The total cross sections of the $\epem \to a_1(1260)\g$, $ a_2(1320)\g$, $a_1(1640)\g$, and $a_2(1700)\g$ processes without interference are calculated to be 
6.4~fb,
5.4~fb,
5.1~fb, and
7.2~fb, respectively.
The expected yield is 503 events with a detection efficiency of 19\%, as estimated using a simulated $\epem \to a_2(1320)\g$ sample.
The yield is similar to the background from \pppzpzg in the high $3\pi$ mass region above 1.3\gev.
The baseline model assumes an incoherent sum of resonances, with each resonance parameterized by a Breit-Wigner distribution.
Figure~\ref{fig:fsr_background} shows the expected \mpppz spectrum of the FSR processes for several values of the interference phase.
Distributions taking into account the possible interference between $a_1(1260)$ and $a_1(1640)$, and $a_2(1320)$ and $a_2(1700)$ are calculated.
Differences from the incoherent sum are assigned as systematic uncertainties.
The interference between $a_1$ and $a_2$ states is not included.
A 100\% systematic uncertainty is assigned for the contribution of the poorly established $a_1(1930)$ and $a_2(2030)$ states around 2.0\gevcc.
%
%
%
%
\section{Signal efficiency}~\label{sec:efficiency}
As a first approximation, the signal efficiency including the acceptance, trigger, and selection efficiency is estimated as a function of $3\pi$ mass using a simulated signal sample ten times larger than the data sample.
The branching fraction of $\piz \to \gaga$ is included in the efficiency.
The simulated signal sample is divided into bins of generated \mpppz.
An event with a reconstructed mass outside of the generated mass bin is still treated as signal within the bin.
The event selection and signal-extraction procedure is repeated on the simulated signal sample.
The efficiency in each bin is the ratio of the number of reconstructed events divided by that of generated events.
The efficiency is fitted with a third-order polynomial function over the $3\pi$ mass range 0.7--3.5\gevcc, and is used as the signal efficiency.
\par
\begin{figure}
    \centering
    \includegraphics[width=\userFigureWidth]{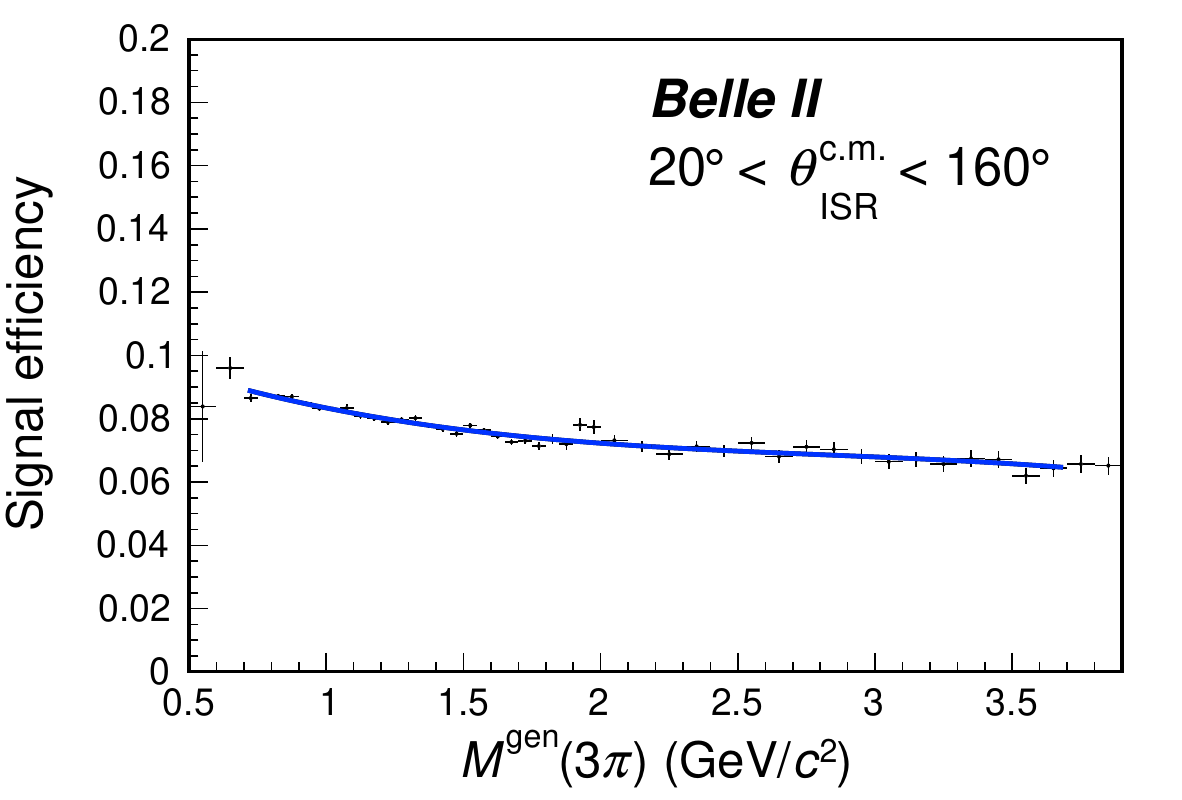}
    \caption{
    Signal efficiency as a function of generated $3\pi$ mass obtained from simulation.
    }
    \label{fig:detection_efficiency}
\end{figure}

The efficiency varies from 8.8\% to 6.6\% in the mass range 0.7--3.5\gevcc as shown in Fig.~\ref{fig:detection_efficiency}.
The main factors that determine the efficiency are the geometrical acceptance of the ISR photon and the \piz efficiency.
The ISR photon is generated in the range $20\degrees < \theta^{*} < 160\degrees$ but is limited to the barrel region $37.3\degrees < \theta^{\mathrm{lab.}}<123.7\degrees$ by the baseline selection.
The efficiency is reduced by 40\% by this requirement.
The \piz efficiency is about 50\%.
The trigger efficiency is close to 100\% (see Sec.~\ref{sec:eff_trg}), but is prescaled by a factor of two, i.e., events meeting the trigger criteria are accepted once every two events, in 47.2\% of the sample to meet data-acquisition bandwidth restrictions, resulting in a 76\% effective trigger efficiency over the entire sample.
\par
We validate the efficiency from Monte Carlo simulation by using data control samples, each specially selected to check an individual contribution to the efficiency, such as
tracking, ISR photon detection, \piz detection, trigger, and background suppression.
The corrected signal efficiency is defined as
\begin{align} \label{eq:efficiency}
  \varepsilon \equiv \varepsilon_{\mathrm{sim}} \prod_{i} (1+\eta_{i}),
\end{align}
where $\varepsilon_{\mathrm{sim}}$ is the efficiency estimated on simulation, $i$ is an index running over the individual corrections, and $(1+\eta_{i})$ are the correction factors.
We determine each value of $\eta_i$ by taking the ratio of the efficiency determined in control data $\varepsilon_{\text{data}, i}$ to the efficiency in simulation $\varepsilon_{\mathrm{sim}, i}$,
as $\eta_{i} \equiv (\varepsilon_{\mathrm{data}, i}/\varepsilon_{\mathrm{sim}, i} - 1)$.
For some effects, the correction factors are calculated by dividing the $3\pi$ mass region into three ranges: below 1.05\gevcc, in 1.05--2.0\gevcc, and above 2.0\gevcc.
In the following subsections, we discuss the evaluation of data-simulation differences for each source.
The total efficiency correction obtained using Eq.~\eqref{eq:efficiency} is $(-4.6 \pm 1.5)\%$ for the region below 1.05\gevcc and $(-4.6 \pm 2.4)\%$ for the region above 1.05\gevcc.
No strong mass dependence is found.
%
%
%
%
\subsection{Tracking efficiency}
The tracking efficiency for pions is studied using $\epem \to \tautau$ data, in which one $\tau$ decays into leptons and the other
decays into three charged pions.
Three good-quality tracks are used to tag tau-pair events and the existence of an additional track is inferred from charge conservation.
The efficiencies in the data and in the simulation are in good agreement, resulting in a systematic uncertainty of 0.3\% per track without any correction factor.
\par
\begin{figure}
  \centering
  \subfloat[][$\Delta\varphi < 0$]{\includegraphics[width=0.48\columnwidth]{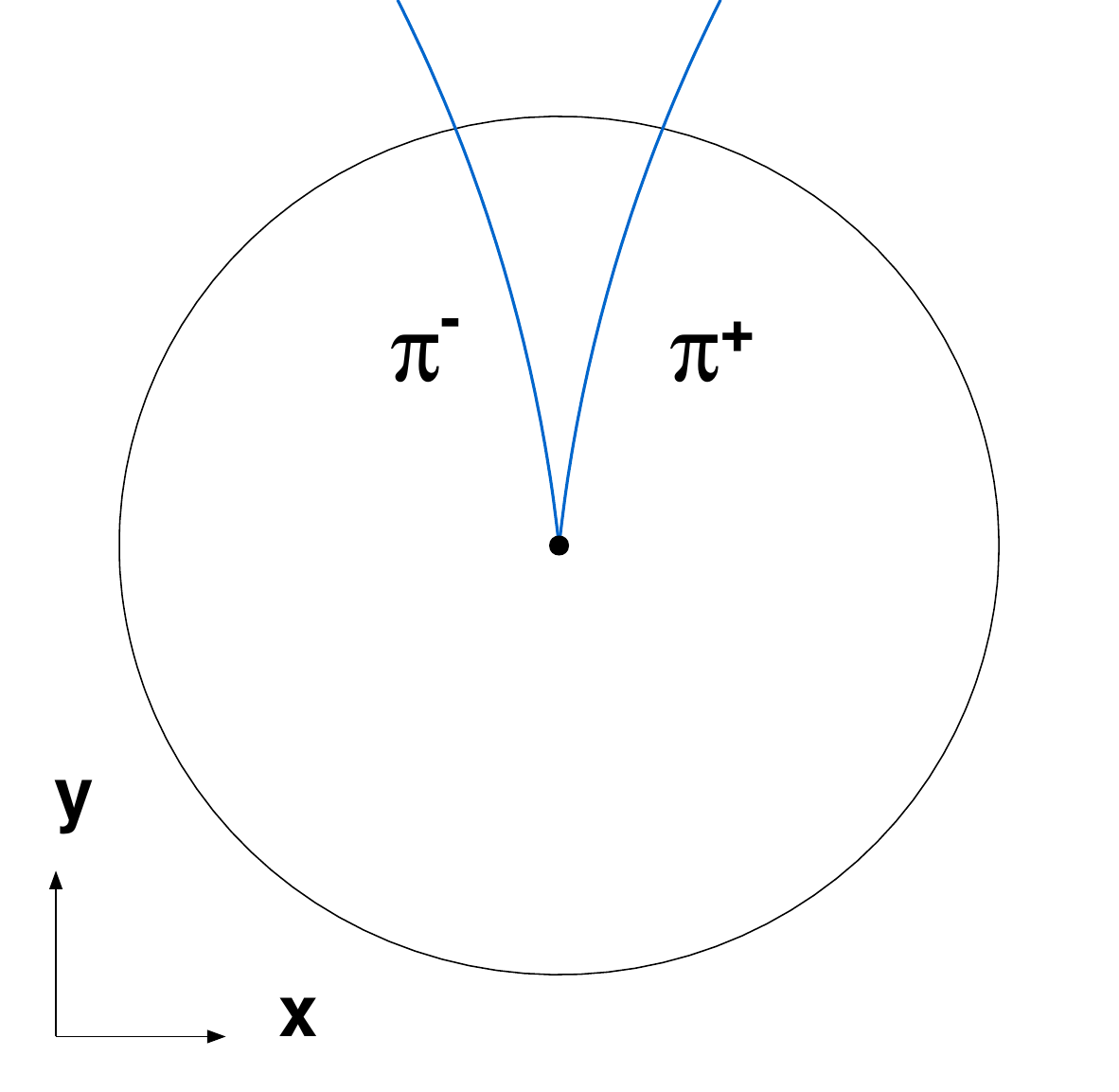}\label{fig:eff_trackloss_ill_a}}
  \subfloat[][$\Delta\varphi > 0$]{\includegraphics[width=0.48\columnwidth]{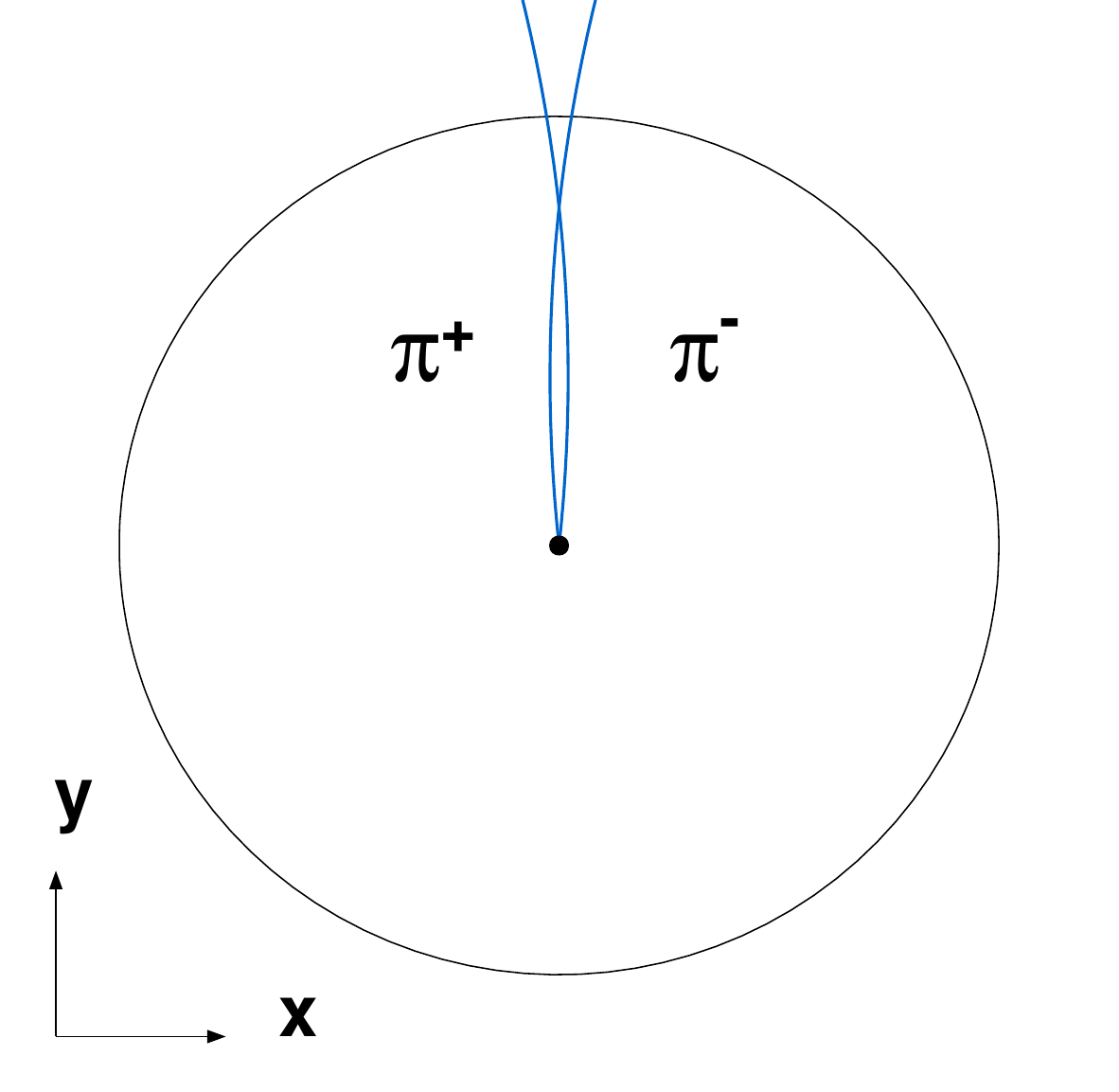}\label{fig:eff_trackloss_ill_b}}
    \caption{
    Visual representation of the two classes of tracks:
    (a) events with tracks having $\Delta\varphi < 0$ and 
    (b) events with tracks having $\Delta\varphi > 0$.
    The magnetic field axis, $+z$ axis, is vertical out of the page.
    The central dot is the IP and the circle is the CDC outer frame.
    The blue arcs are charged tracks.
  }
  \label{fig:eff_trackloss_ill}
\end{figure}
\begin{figure}
  \centering
  \includegraphics[width=\userFigureWidth]{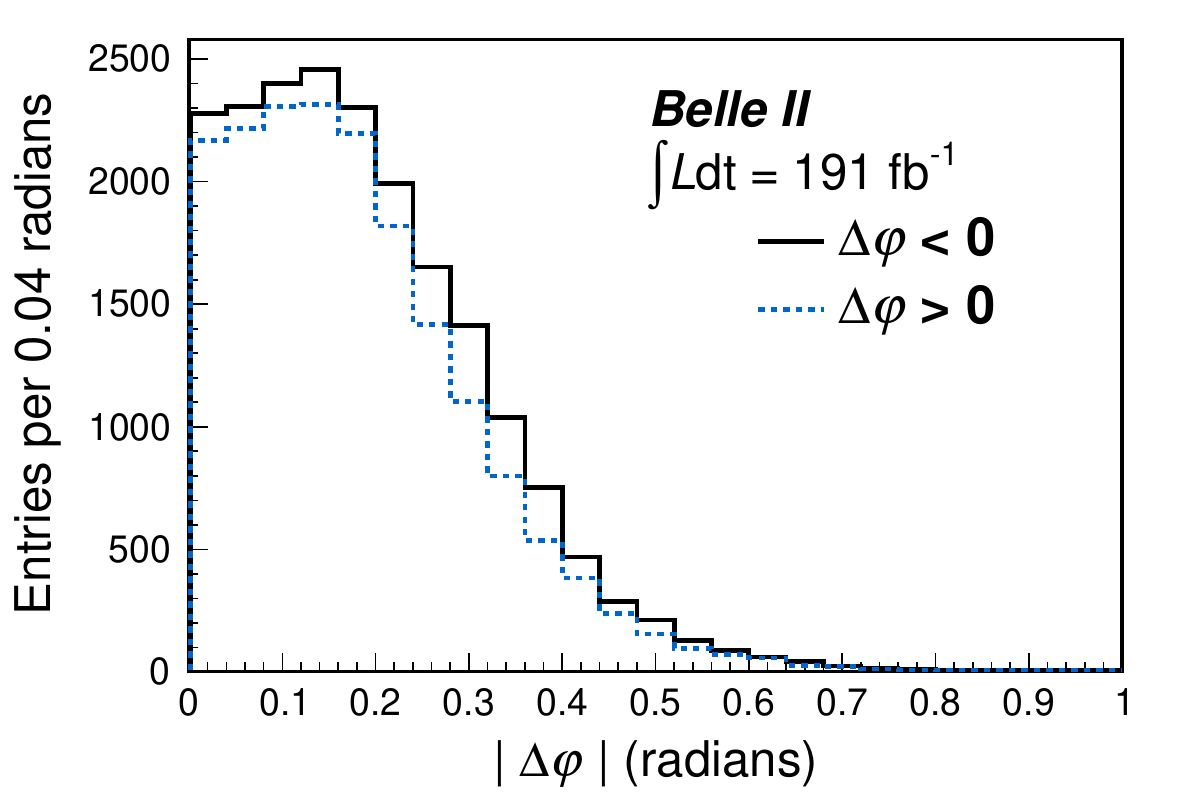}
    \caption{
    Azimuthal difference $|\Delta\varphi|$ in pairs of tracks in data at the $\omega$ and $\phi$ resonances.
    The histograms show  $\Delta\varphi<0$ pairs (solid) and $\Delta\varphi>0$ pairs (dashed).
  }
  \label{fig:eff_trackloss}
\end{figure}

Tracks may be lost when the opening angle between them is small and many CDC wire hits are shared.
The opening angle between two tracks in the azimuthal plane relative to the magnetic field axis is $\Delta\varphi = \varphi^{+} - \varphi^{-}$, where the superscript indicates the charge of the particle.
A sketch of $\Delta\varphi$ is shown in Fig.~\ref{fig:eff_trackloss_ill}.
A negative $\Delta\varphi$ corresponds to fewer shared hits and less track loss,
while a positive $\Delta\varphi$ is more likely to yield track loss.
The azimuth difference $|\Delta\varphi|$ shown in Fig.~\ref{fig:eff_trackloss} indicates a clear difference between positive and negative $\Delta\varphi$ events.
The efficiency due to the track loss effect is calculated as a sum of the number of positive and negative $\Delta\varphi$ events divided by twice the number of negative $\Delta\varphi$ events using a \pppzg signal sample in the $\omega$ and $\phi$ resonance regions, 0.72--0.84\gevcc and 0.98--1.05\gevcc.
The data-to-simulation correction for the track loss efficiency is $(-1.1 \pm 0.5)\%$.
\par
The efficiency for the requirement on the number of CDC wire hits is about 99\%.
To check the dependence of this efficiency on the wire-hits requirements, the fraction of tracks passing the hit requirements is studied using events passing the full signal selection except for the CDC hit criteria.
The difference between the data and simulation is assigned as a correction factor, $(-0.3 \pm 0.1)\%$ for $\mpppz < 1.05\gevcc$ and $(-0.6 \pm 0.2)\%$ for $\mpppz > 1.05\gevcc$.
\par
We examine the fraction of events in which three tracks are required in order to to assess the impact of requiring exactly two tracks. 
The percentage of three-track events is 0.01\% compared to 2-track signal events, which agrees well with the simulation.
\par
By linearly adding these differences, the correction factor associated with the tracking efficiency is estimated to be $\eta_{\mathrm{track}} = (-1.4 \pm 0.8)\%$ for $\mpppz < 1.05\gevcc$ and $(-1.7 \pm 0.8)\%$ for $\mpppz > 1.05\gevcc$.
%
%
%
%
\subsection{ISR photon detection efficiency}
The ISR photon detection efficiency is studied with an $\epem \to \mmg$ control sample.
The expected photon position and energy can be determined from the recoil momentum of the reconstructed muon pair.
A pair of tracks identified as muons with $p>1\gevc$ are kinematically fitted with the recoil mass constraint assuming $\epem \to \mmg$.
Candidates with small recoil mass are used to avoid $\epem \to \tautau$ background.
These muon-pair events are divided into bins of recoil momentum, polar and azimuth angle.
The photon detection efficiency is the fraction of the muon pairs for which a neutral cluster having position and energy inferred from the recoil momentum is reconstructed.
Photon loss occurs primarily due to conversion into \epem pairs in the detector material.
A correction factor for the photon detection efficiency is obtained, $\eta_{\mathrm{photon}} = (0.2 \pm 0.7)\%$.
%
%
%
%
\subsection{\texorpdfstring{\boldmath{$\piz$}}{pi0} detection efficiency}\label{sec:eff_piz}
\begin{figure*}
  \centering
  \subfloat[][]{\includegraphics[width=0.48\textwidth]{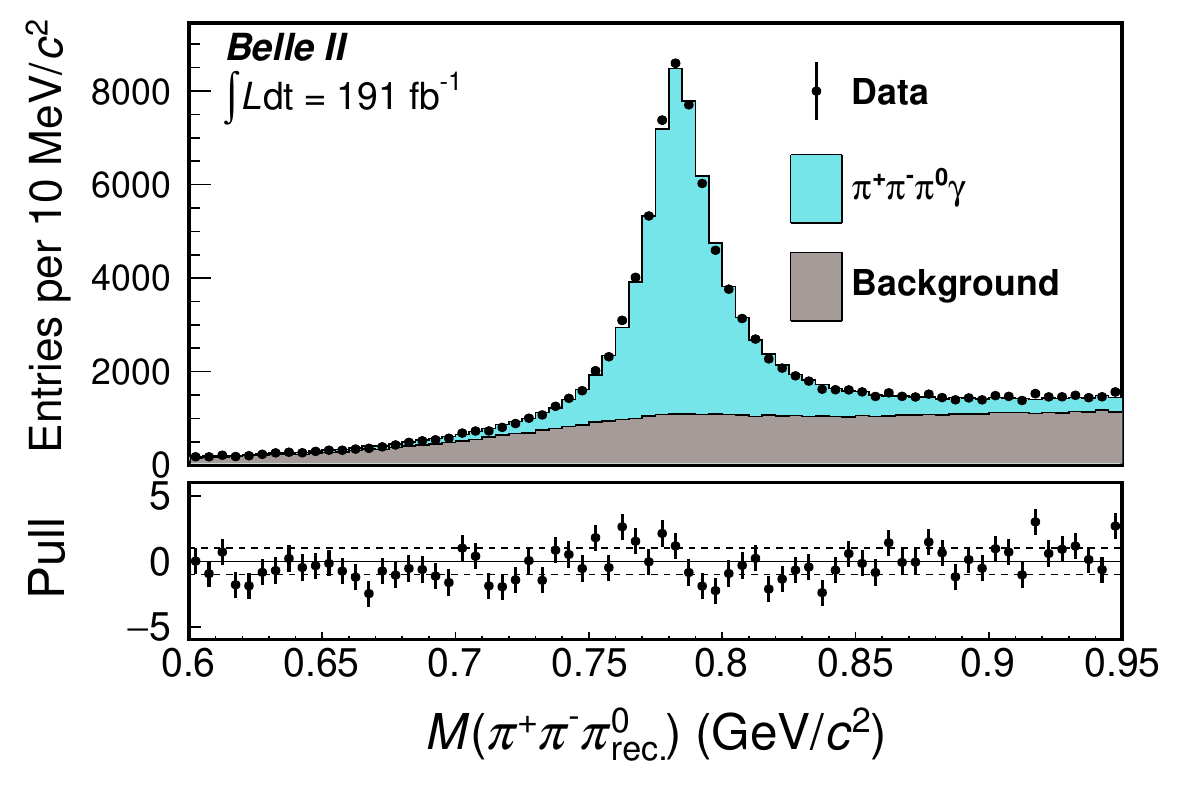}\label{fig:eff_pi0_a}}
  \subfloat[][]{\includegraphics[width=0.48\textwidth]{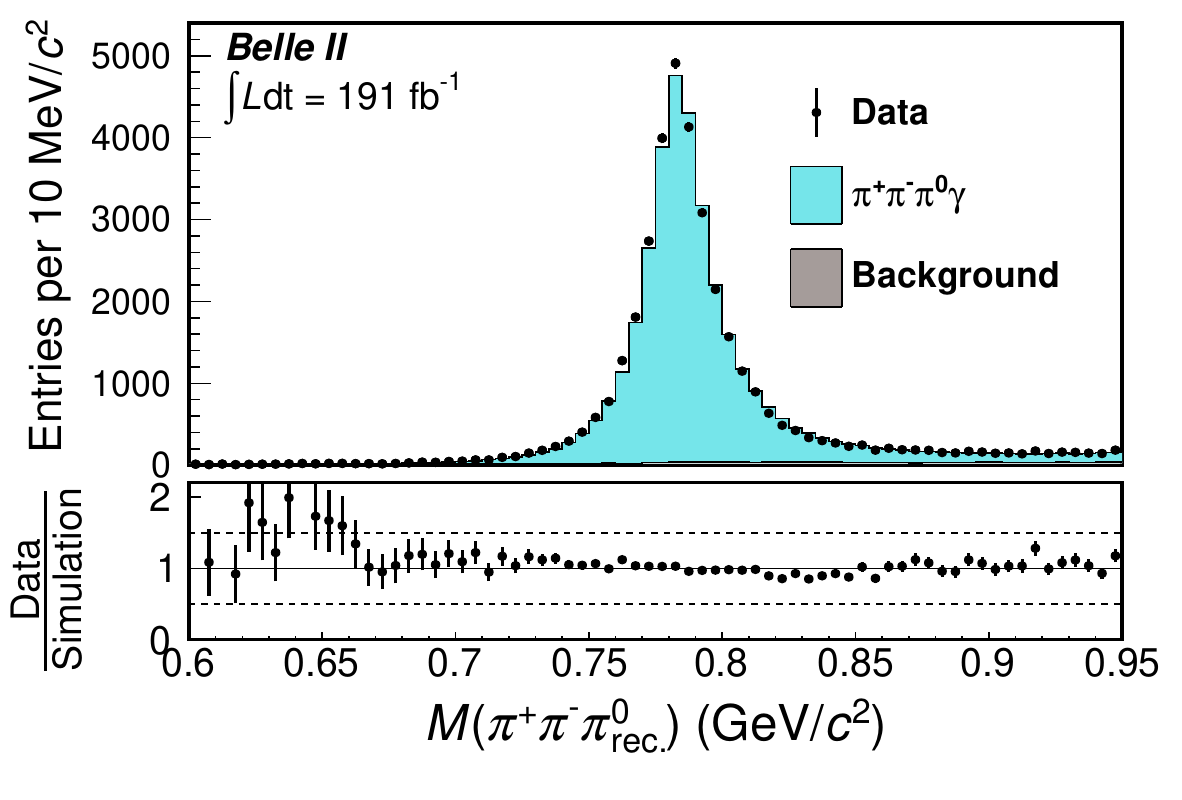}\label{fig:eff_pi0_b}}
  \caption{
    (a) Three-pion mass \mpppzrec distribution for event fully reconstructed using \ppg particles.
    The shaded histograms in the top panel show the result of a \mpppzrec fit to the data with signal and background components.
    The differences between data and fit results divided by the data uncertainties (pull) are shown in the bottom panel.
    (b) The same \mpppzrec distribution for the events fully reconstructed using \pppzg particles.
    The convention of the figure is the same as (a).
    (b) The shaded histograms, which show simulated signal and background, are normalized to the data to check the consistency of the signal model.
    The data-to-simulation ratio is shown in the bottom panel.
  }
  \label{fig:eff_pi0}
\end{figure*}
\def\npiznum{\ensuremath{N_{i,\mathrm{full}}}\xspace}
\def\npizden{\ensuremath{N_{i,\mathrm{part}}}\xspace}
\par
The \piz detection efficiency is obtained using the following reaction in the $\omega$ resonance region
\begin{align}\label{eq:pi0omega}
\epem \to \omega \g \to \pppzg.
\end{align}
We measure the \piz efficiency $\epsilon_{i}(\piz)$ using the formula
\begin{align}
  \label{eq:pi0eff}
  \epsilon_{i}(\piz) = \frac{\npiznum}{\npizden},
\end{align}
where $i$ is an index for data or simulation.
The numerator, \npiznum, is the number of events in which all particles in the reaction, including the \piz, are detected.
We use the same fitting method described in Sec.~\ref{sec:signal_extraction} to determine the \piz yield.
The denominator, \npizden, is the number of events in which the reaction is reconstructed without requiring that the \piz be reconstructed.
Since the process in Eq.~\eqref{eq:pi0omega} is exclusive, one can infer the presence of the \piz from the mass recoiling against the \ppg system, without reconstructing the \piz.
By counting the number of events recorded in this way, the number of \piz's that are needed for the efficiency determination is known.
In addition, the prominent $\omega$ signal is used to determine the relevant yields.
\par
We carry out a one-constraint (1C) kinematic fit with the hypothesis that the mass recoiling against the \ppg system is the known \piz mass~\cite{ParticleDataGroup:2022pth} in order to infer the \piz momentum.
After the kinematic fit, we obtain partially-reconstructed events by requiring good fit quality, $\chisqrec < 10$.
We denote the invariant mass of the $3\pi$ system calculated using the \piz momentum as \mpppzrec.
We obtain fully-reconstructed events from the partially-reconstructed ones with \piz's detected using the same \piz criteria used in the initial event selection.
The criteria of $\chisqfctpg < 50$ is also imposed to ensure signal purity.
\par
The values of \npizden are determined from a fit to the \mpppzrec distribution of the partially-reconstructed \ppg events.
The \mpppzrec distribution, in Fig.~\ref{fig:eff_pi0}\subref{fig:eff_pi0_a}, shows a prominent $\omega$ signal.
The background is from the processes $\epem \to \pppzpzg$, $\epem \to \phi\g \to \ksklg $, and $\epem \to \qqbar (\g)$.
We check the shape of each background predicted by the simulation using data samples specially selected to enhance each contribution.
The signal probability density function~(PDF) for the \mpppzrec distribution is obtained from the simulation,
and the consistency of the PDF shape is confirmed using the data in the fully-reconstructed events, which are shown in Fig.~\ref{fig:eff_pi0}~\subref{fig:eff_pi0_b}.
The values of \npiznum are determined from a fit to the \mgg distribution for the fully-reconstructed events in the same \mpppzrec range as \npizden.
\par
From the data-to-simulation ratio determined in this way, the correction factor for the \piz efficiency is determined to be $\eta_{\pi^{0}} = (-1.4 \pm 1.0)\%$.
The uncertainty in $\eta_{\piz}$ is dominated by the uncertainty in the background contamination for the data in the denominator.
%
%
%
%
%
%
%
\def\chisqfcmmg    {\ensuremath{\chi^{2}_{\mathrm{2\mu\g}}}\xspace} 
\def\cppg    {\ensuremath{\chi^{2}_{\mathrm{2\pi\g}}}\xspace}
\def\cthre {\ensuremath{\chi^2_{\mathrm{thr}}}\xspace}
\def\cpiz {\ensuremath{\chi^2_{\gaga}}\xspace}
\def\effchi {\ensuremath{\epsilon(\cthre)}\xspace}
\subsection{Efficiency for kinematic-fit quality selection}\label{sec:eff_chi2}
The \chisqfctpg of the 4C fit has contributions from the charged particles, the two photons from the \piz, and the ISR photon. 
The dominant uncertainty is from the ISR photon.
To estimate this uncertainty independently from the signal process, we use an $\epem \to \mmg$ control sample, which provides high purity \mmg events without a $\chi^{2}$ requirement.
In addition, the signal and this data control sample have similar kinematic properties as they both include an ISR photon and two oppositely charged particles of similar masses.
We define the efficiency as a function of the $\chi^2$-threshold \cthre as
\begin{equation} \label{eq:eff_chi2}
    \effchi =\frac{N(\chi^2 < \cthre ) }{N_{\mathrm{all}}}, 
\end{equation}
where $N_{\mathrm{all}}$ and $N(\chi^2 < \cthre )$ are the total number of events before and after the $\chi^2$ requirement.
Using the $\chi^2$ distribution function $f(\chi^{2})$, these values are given by
\begin{equation*}
 N_{\mathrm{all}} = \int_{0}^{\infty} f(\chi^{2}) d\chi^{2} , \quad
 N(\chi^2 < \cthre )  = \int_{0}^{\cthre} f(\chi^{2}) d\chi^{2}.
\end{equation*}
\par
\begin{figure}
  \centering
  \includegraphics[width=\userFigureWidth]{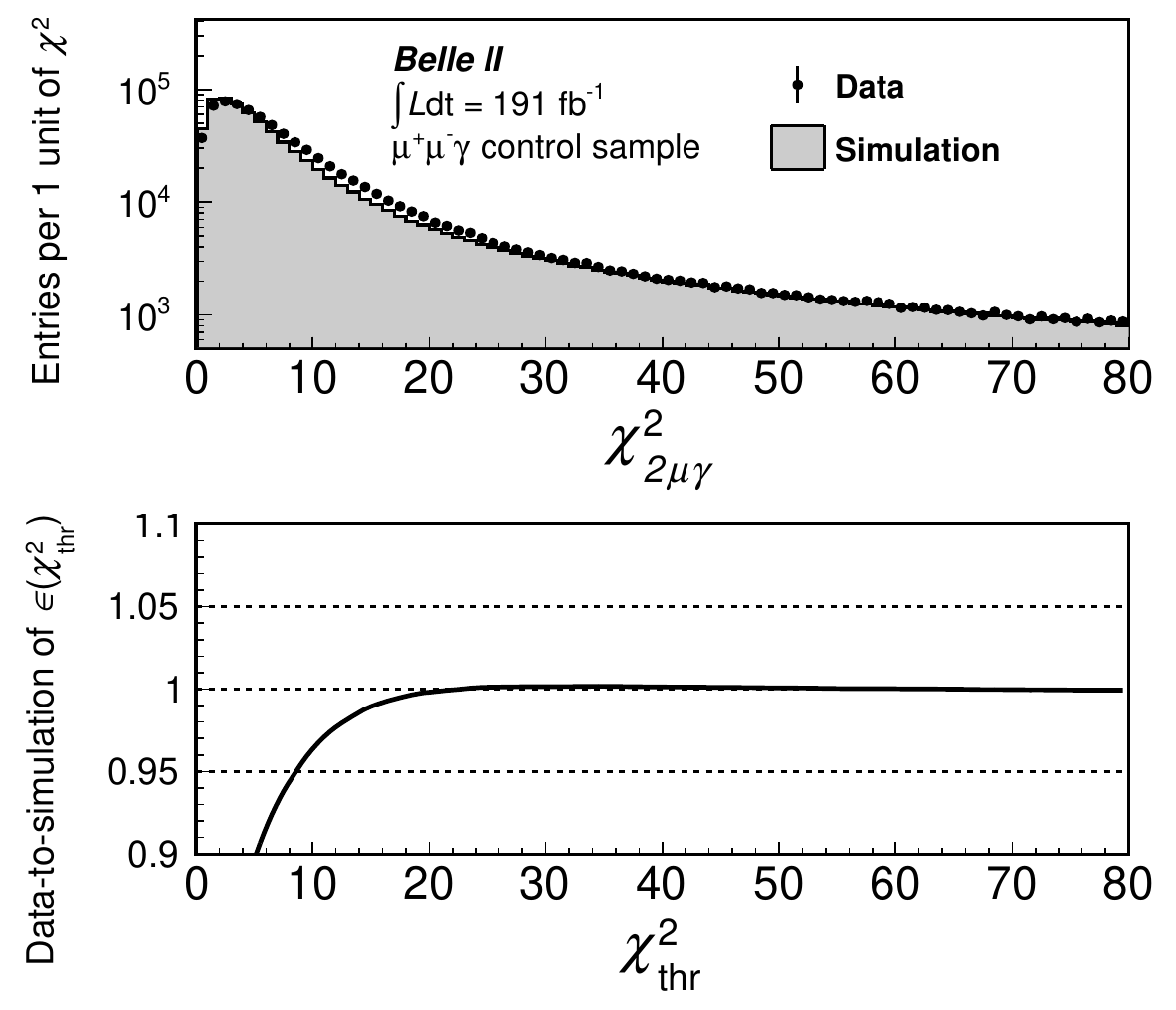}
    \caption{
        (top) Distribution of $\chisqfcmmg$ and (bottom) data-to-simulation ratio of the $\chisqfcmmg$ efficiency $\epsilon(\cthre)$.
        In the top panel, the points with error bars represent the data and the filled histogram represents the simulated sample normalized to the integrated luminosity of data.
  }
  \label{fig:eff_chi2}
\end{figure}

We show the  $\chi^2$ distribution  $f(\chi^{2})$ for $\epem\to \mmg$ events in the dimuon mass range below 1.05\gevcc in the upper figure in Fig.~\ref{fig:eff_chi2}.
The points with error bars are the data and the filled histogram is the simulation. From these distributions, we can obtain the efficiency in Eq.~\eqref{eq:eff_chi2} for the data and the simulation separately.
The lower figure in Fig.~\ref{fig:eff_chi2} shows the data-to-simulation ratio of the efficiency \effchi as a function of \cthre.
For any value of \cthre above $\cthre > 20$, the data-to-simulation ratio of the efficiency for \mmg is close to 1.00 within $0.2\%$.
To address the uncertainty from the difference in the kinematic properties of charged particles between the \mumu and $3\pi$ samples, we examine subsamples divided by muon momentum.
The data-to-simulation ratios are tested in the subsamples, and half of the minimum and maximum ratios are assigned as a systematic uncertainty of 0.6\%.
A correction for the difference in the 4C fit $\chi^2$ originating from the two photons of \piz decay is included in the \piz efficiency correction by imposing the \chisqfctpg requirement on the numerator.
From these results, we determine that the correction factor for the selection on the 4C kinematic fit is $\eta_{\chi^2}= (0.0\pm 0.6)\%$ for $\mpppz <1.05\gevcc$.
A similar test is performed for the mass region above 1.05\gevcc and the results are summarized in Table~\ref{tab:sys_detection}.
%
%
%
\\
\subsection{Monte Carlo generator effects}\label{sec:eff_gen}
The good agreement in $\varepsilon(\cthre)$ between data and simulation samples generated using the \texttt{KKMC} generator indicates that missing higher-order radiative effects,
which could modify the event kinematic properties and degrade the kinematic fit \chisq, are not significant for this generator.
\par
In contrast, a \babar study of the \texttt{PHOKHARA} generator used for signal simulation found a 20\% excess of events with an additional energetic ISR photon along the beam line, in PHOKHARA relative to data~\cite{BaBar:2023xiy}.
The ISR-based analysis at \babar is not impacted as it relied on a different generator, but our analysis, based on \texttt{PHOKHARA}, could be affected.
Since the \babar study used $\epem \to \pipi\g$ and $\epem \to \mumu\g$ processes, we reproduce the effect of these processes using $\epem \to \pppzg$ events in Belle~II data.
A study of $\epem \to \pppz\g$ using a three-constraint kinematic fit, which allows for an additional ISR photon along the beamline, corroborates \babar's findings.
Events with such an additional photon usually fail the \chisqfctpg selection criterion.
Removing this excess in the simulation would increase the measured signal efficiency by $(2.4 \pm 0.7)$\%.
The \babar study also indicates that events with two additional energetic photons, a process not simulated by \texttt{PHOKHARA}, make up $(3.5 \pm 0.4)\%$ of the events in the relevant mass range, which can be compared to the total fraction with a single
photon and two additional photons $(21.8\pm0.4)$\%.
If included in the simulation, such events would be expected to reduce the signal efficiency by an amount comparable to a 1.9\% change in efficiency.
We do not assign a correction to signal efficiency for these generator effects, but instead, assign an additional uncertainty of 1.2\% as a systematic uncertainty for the generator, which is the sum in quadrature of the 0.7\% uncertainty due to a single additional photon and 0.95\%, corresponding to half of the uncertainty due to two additional photons.
%
%
%
%
%
%
\subsection{Trigger efficiency}\label{sec:eff_trg}
\par
To evaluate the efficiency of the ECL energy trigger, $\epem \to \mmg$ events triggered by a tracking trigger based on CDC and KLM signals are selected and used as a reference.
The track trigger matches CDC tracks with KLM hits, allowing for high trigger efficiency for events with one or more barrel muons.
This provides a data sample of 12 million $\epem \to \mmg$ events independent of the ECL trigger.
The trigger efficiency tested in this way using the data is close to 100\%.
Taking the ratio of the trigger efficiency between the data and the simulation,
we obtain a data-to-simulation correction factor, $\eta_{\mathrm{trig}} = (-0.09 \pm 0.08)\%$ for the 0.62--1.05\gevcc in the three-pion mass range that includes the $\omega$ and $\phi$ resonances, $(-0.08 \pm 0.08)\%$ for the 1.05--2.0\gevcc range, and $(-0.06 \pm 0.08)\%$ for the 2.0--3.5\gevcc range.
%
%
%
%
%
\subsection{Efficiency for background-suppression criteria}
Several background suppression criteria are applied in Sec.~\ref{sec:sel_bkgsup}.
We evaluate the net signal efficiency by comparing the ratio of the signal yield before and after applying these criteria for the data and the simulation separately.
The criteria include particle identification and $\ppg$, $\mmg$, $\pppzpzg$, and non-ISR \qqbar suppression.
The efficiency in data and simulation is the fraction of background-subtracted events passing the selection.
Even without these selections the signal purity near the $\omega$ and $\phi$ resonances is 97\%.
The correction factor for the three-pion mass region below 1.05\gevcc is estimated to be $\eta_{\mathrm{sel}} = (-1.9 \pm 0.2)\%$ using events in the $\omega$ and $\phi$ resonance regions.
The correction factor in the region above 1.05\gevcc is evaluated from a fit to the \jpsi resonance to be $(-1.8 \pm 1.9)\%$.
%
%
%
%
\section{Signal extraction and unfolding}\label{sec:unfold}
%
\subsection{Measured three-pion mass spectrum}\label{sec:spectrum}
\begin{figure*}
  \centering
  \subfloat[][]{\includegraphics[width=\userFigureWidth]{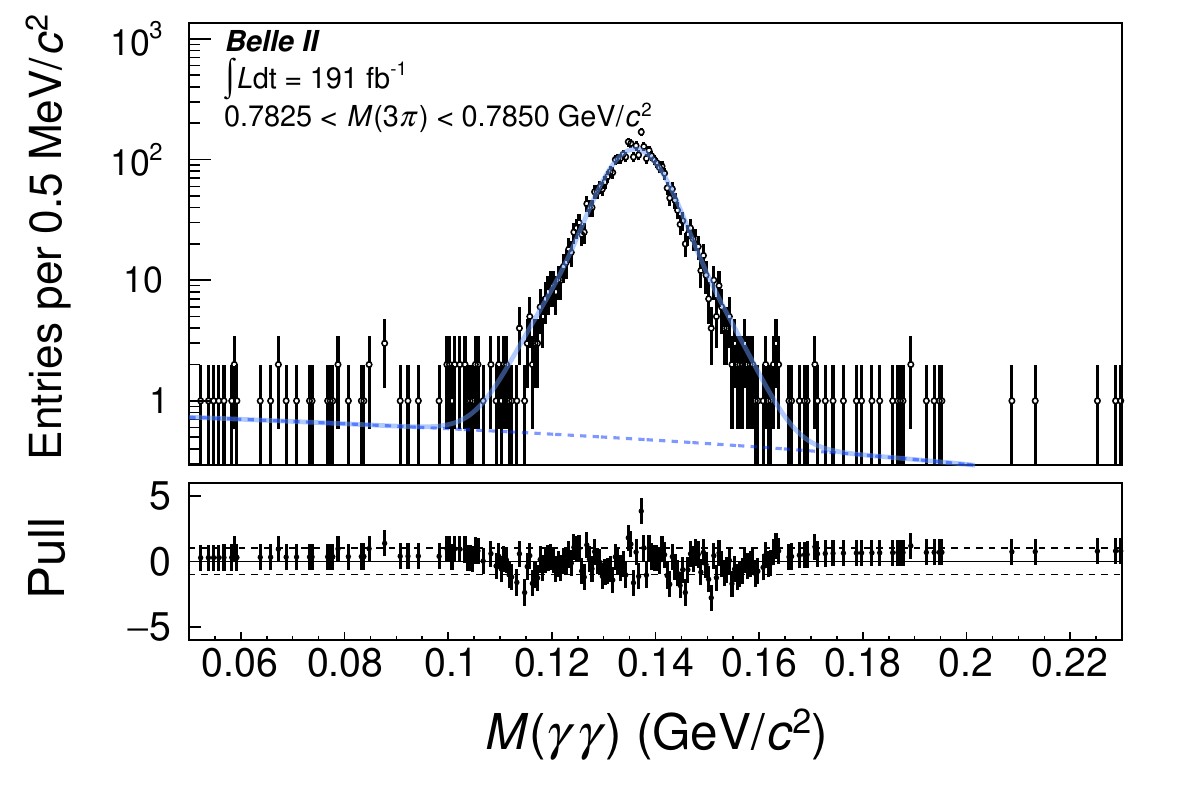}\label{fig:mggibin_a}}
  \subfloat[][]{\includegraphics[width=\userFigureWidth]{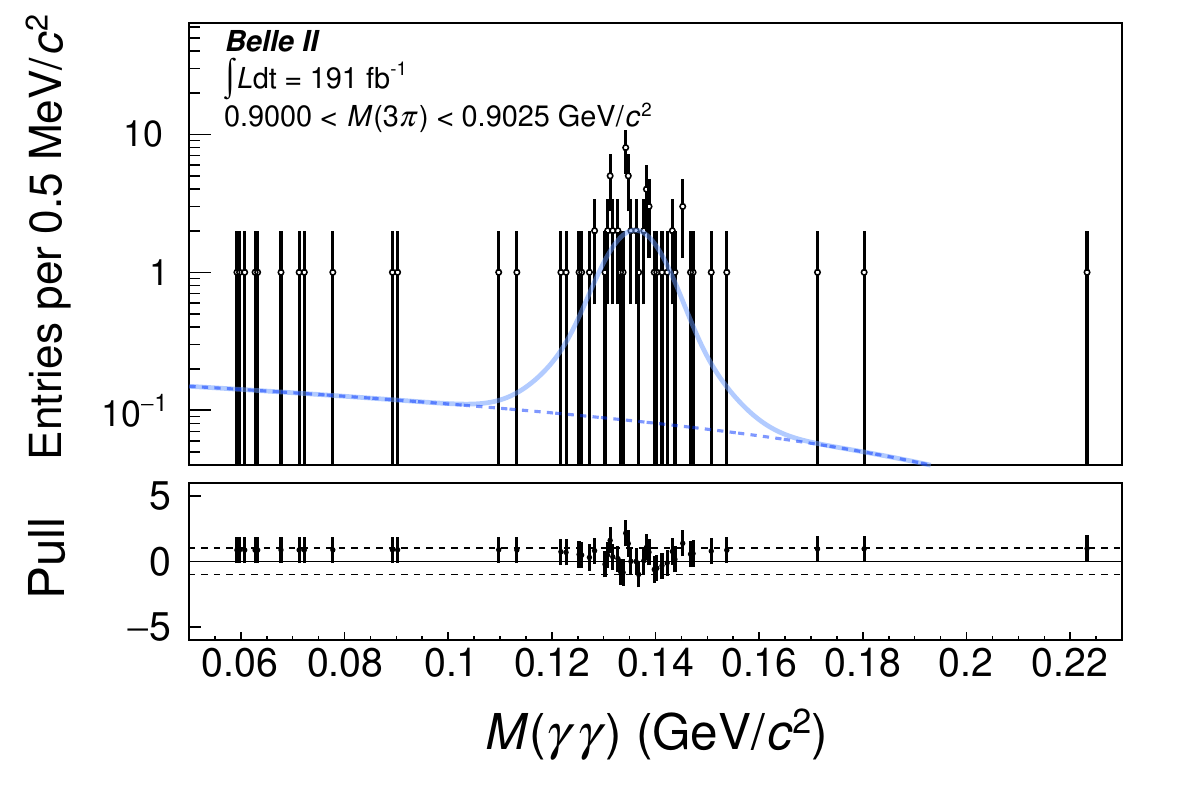}\label{fig:mggibin_b}}
    \caption{
      Diphoton mass distributions (a) for events in the $3\pi$ mass range 0.7825--0.7850\gevcc and (b) for events in the $3\pi$ mass range 0.9000--0.9025\gevcc.
      The convention in the figure is the same as in Fig.~\ref{fig:final_sample_pi0mass}.
    }
  \label{fig:mggibin}
\end{figure*}

\begin{figure*}
  \centering
  \includegraphics[width=0.48\textwidth]{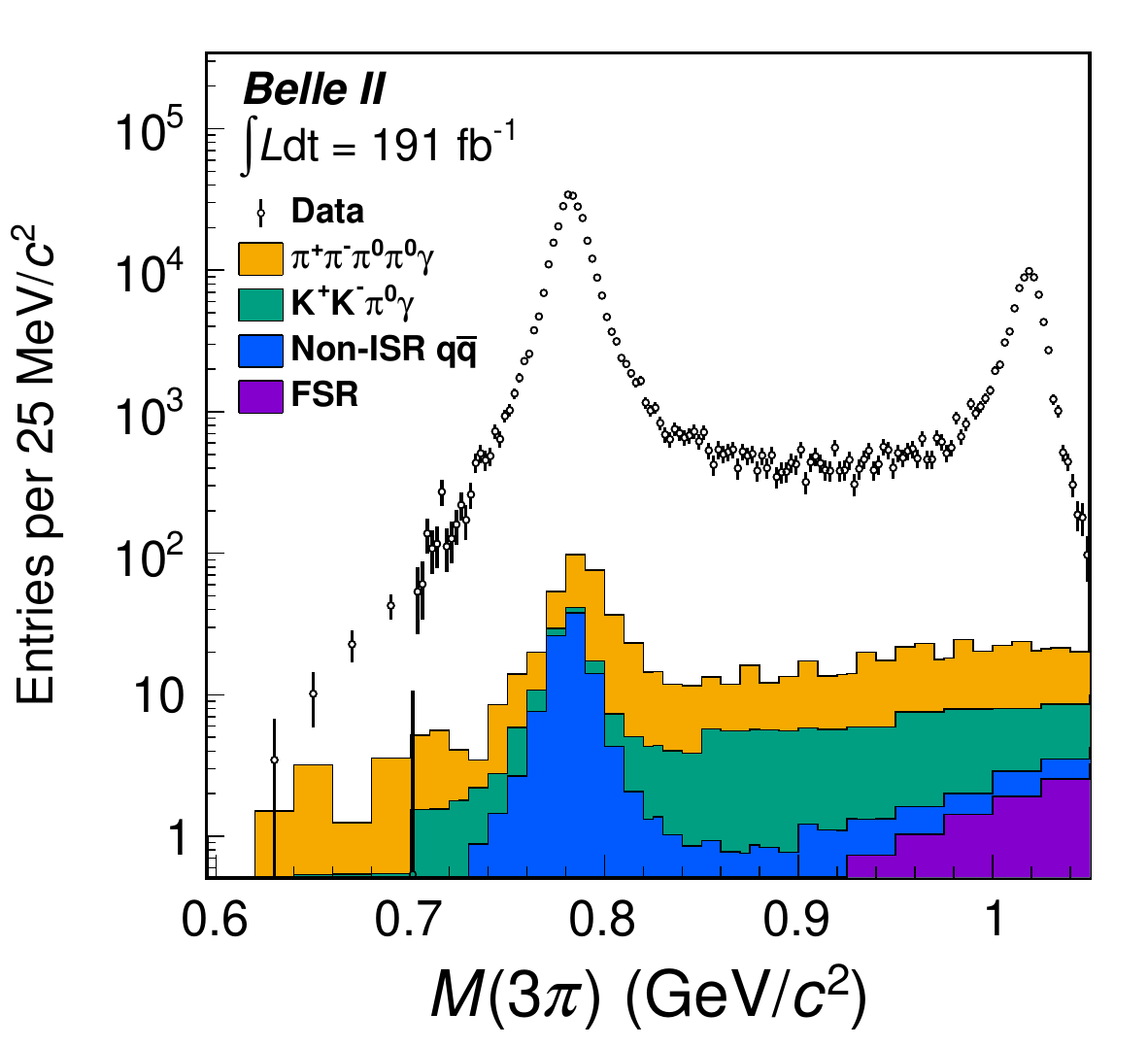}
  \includegraphics[width=0.48\textwidth]{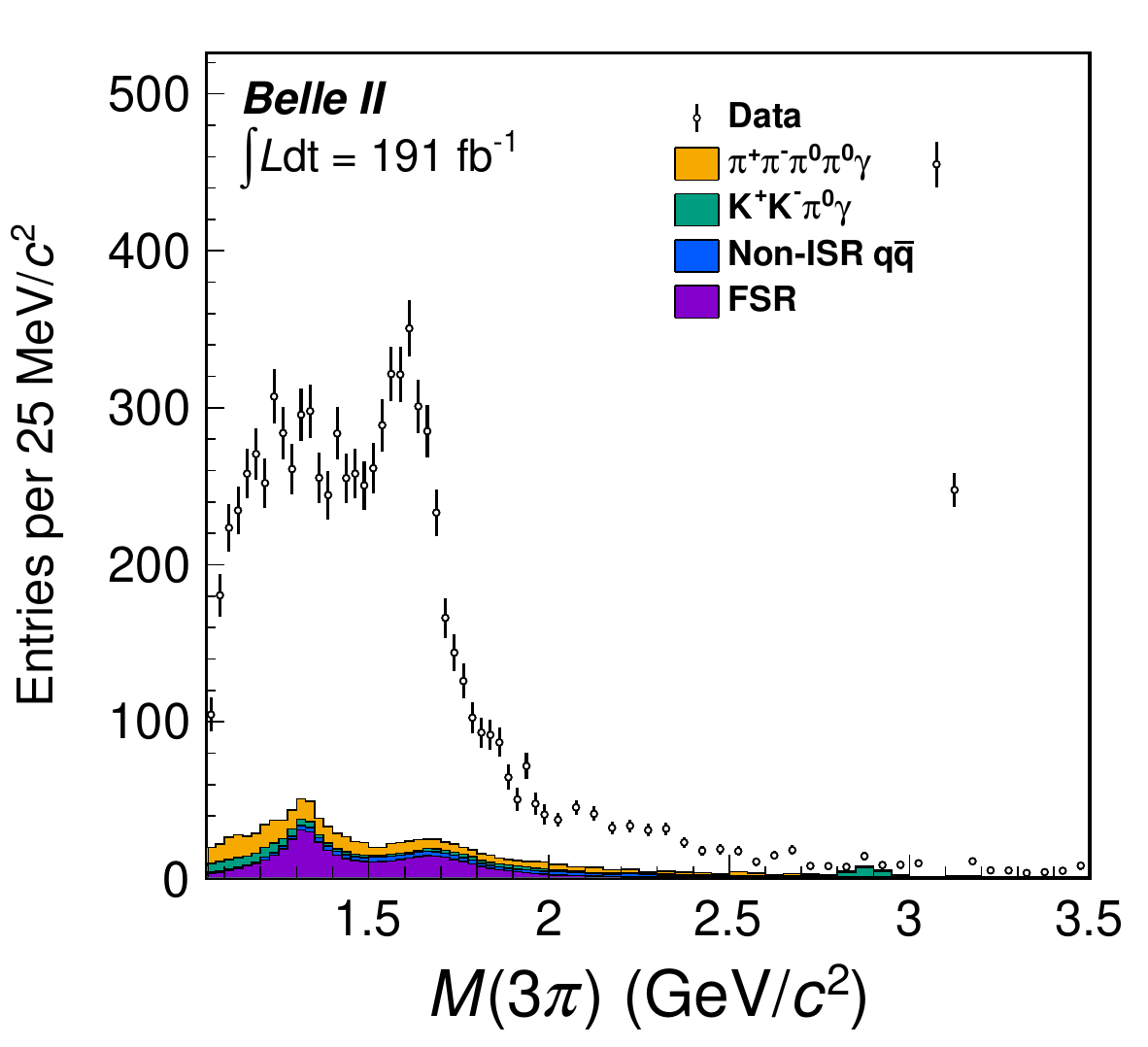}
    \caption{
      Distribution of $3\pi$ mass spectrum (left) for \mpppz less than 1.05\gevcc and (right) for \mpppz greater than 1.05\gevcc.
      The points with error bars are determined from diphoton mass fits in each \mpppz bin.
      The filled stacked histograms are the estimated contributions of residual backgrounds.
      The number of events measured in each \mpppz bin is scaled to the 25\mevcc bin width.
    }
  \label{fig:final_sample_spectrum}
\end{figure*}

To better estimate the three-pion mass, we re-evaluate the $\pppz$ mass using the momenta determined from five-constraint kinematic fits, which constrain the diphoton mass to match the known \piz mass, in addition to the constraints of four-momentum conservation.
The \mpppz bin width is varied depending on the $3\pi$ invariant mass: 2.5\mevcc in the $\omega$ and $\phi$ resonance regions, 20\mevcc in the region below 0.7\gevcc, and 25--50\mevcc in the region above 1.05\gevcc.
As noted in Sec.~\ref{sec:signal_extraction}, we fit the \mgg distribution in each \mpppz bin to extract the \piz yield.
Before performing the fit for each bin, the six line-shape parameters of the signal PDF, three parameters for the Novosibirsk function, and two parameters for a Gaussisan, and their ratio, are determined by fitting events in the \mpppz range below and above 1.05\gevcc (Fig.~\ref{fig:final_sample_pi0mass}).
To extract the signal yield in each \mpppz bin, a binned maximum-likelihood fit is carried out with the signal PDF described above and a linear function to describe the combinatorial background.
The linear function is intended to stabilize the fit in \pppz bins with small amounts of combinatorial background.
In the fit, the signal normalization and background parameters are allowed to float.
Figure~\ref{fig:mggibin} shows the \mgg distributions for events with \mpppz in the $\omega$ resonance region and around 900\mevcc, respectively.
These are typical examples of signal-extraction fits in cases of large or small \mpppz bin populations.
Figures~\ref{fig:final_sample_spectrum} (a) and (b) show the signal yields resulting from the \mgg fits as functions of \mpppz.%
The histograms show the expected background from other processes, where the yields are obtained by fitting the \mgg distribution for each background process and the correction factors discussed in Sec.~\ref{sec:bkg} are applied.
We find prominent $\omega$, $\phi$, and $\jpsi$ signal peaks.
In the 1.1--1.8\gevcc region, broad enhancements due to the $\omega(1420)$ and $\omega(1650)$ resonances are also visible.
\begin{figure*}
  \centering
  \includegraphics[width=\userFigureWidth]{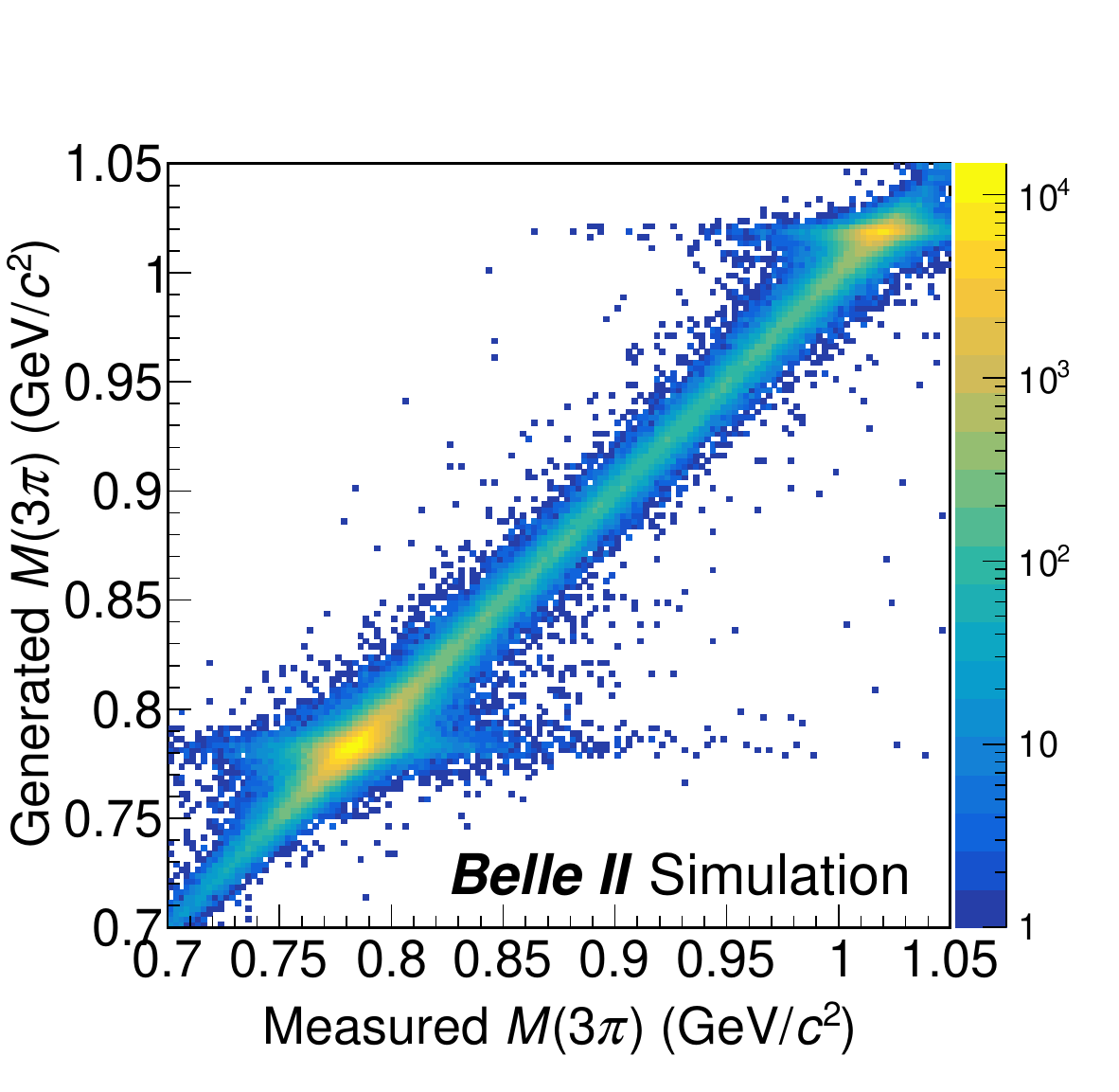}
  \includegraphics[width=\userFigureWidth]{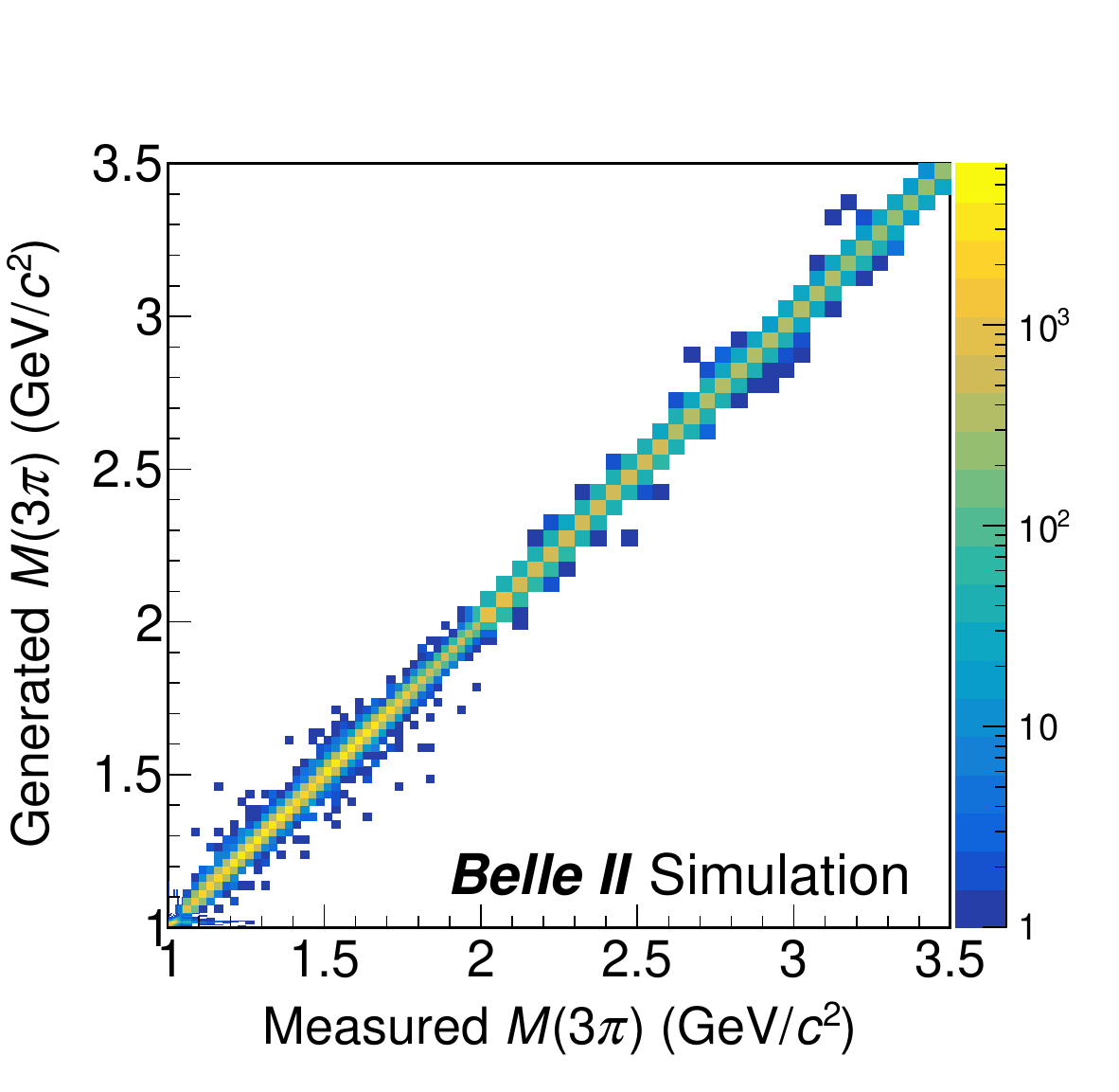}
  \caption{
    Transfer matrix obtained from simulation (left) for $\mpppz < 1.05\gevcc$ and (right) for $\mpppz > 1.05\gevcc$.
    The horizontal axis is the measured \mpppz value and the vertical axis is the generated \mpppz value.
  }
  \label{fig:transfer_matrix}
\end{figure*}
%
%
%
%
\subsection{Unfolding procedure}
The signal-only $3\pi$ mass spectrum resulting from the signal extraction fits is unfolded to account for the migration of events between bins due to the effect of detector response and FSR. 
An iterative dynamic stable unfolding method (IDS)~\cite{Malaescu:2011yg} is used to unfold the original signal yield.
The typical \mpppz mass resolution based on simulation is 6.5\mevcc at the $\omega$ resonance.
The detector resolution is comparable to the width of the $\omega$ and $\phi$ resonances.
\par
Unfolding transforms a measured spectrum into a generated spectrum based on a transfer matrix $A_{ij}$.
The matrix $A_{ij}$, which describes the number of events generated in the $j$th \mpppz bin and reconstructed in the $i$th \mpppz bin, is obtained from the simulated sample shown in Fig.~\ref{fig:transfer_matrix}.
The IDS method allows the unfolding of structures that are not modeled in the simulation and avoids fluctuations from the background subtraction.
\par
\begin{figure}
  \centering
  \includegraphics[width=\userFigureWidth]{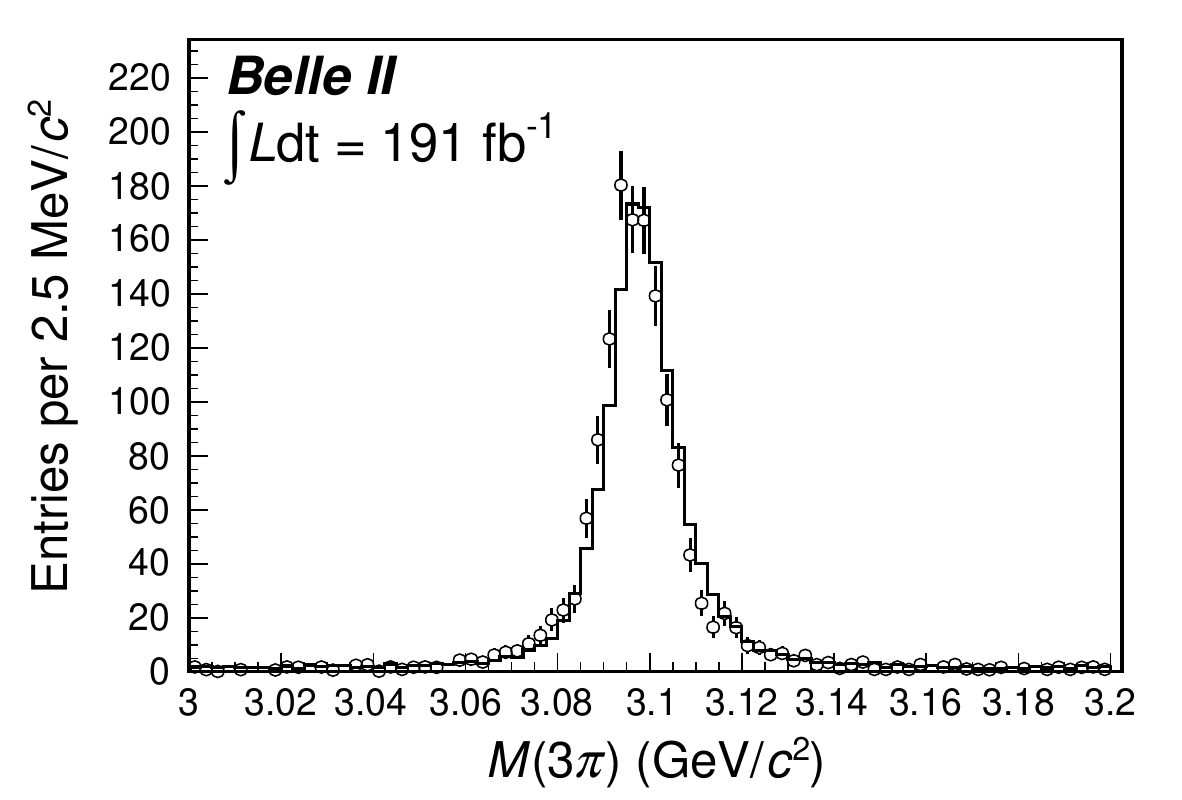}
  \caption{
    Three-pion mass distribution at the \jpsi resonance region.
    The points with error bars are the data and the solid histogram is the simulation normalized by the signal yields in data.
  }
  \label{fig:mass_resolution_jpsi}
\end{figure}

Before performing the unfolding procedure, we evaluate potential data-simulation differences in the transfer matrix resulting from an incomplete simulation of the mass resolution and momentum or energy scale since the IDS method does not compensate for these differences.
We assess these differences by fitting the signal-only $3\pi$ mass spectrum in data using a model that includes the transfer matrix, the simulated distribution, and a Gaussian smearing term to represent a possible shift in the measured mass and a degradation of the resolution in data.
In the $\omega$ and $\phi$ resonance regions, the simulated spectrum convolved with a single Gaussian function serves as the fit function, where the parameters of resonance shapes rely on \texttt{PHOKHARA}.
The fit function can be written as
\begin{align} \label{eq:resolution}
 \left( \frac{\d N}{\d \sqrt{s'}} \right)^{\mathrm{meas}}_{i} = \sum_{j} A_{ij} \left[\left( \frac{\d N}{\d \sqrt{s'}} \right)^{\mathrm{gen}} * G \right]_{j},
\end{align}
where $i$ is the index of the \mpppz mass bin where the event is reconstructed,
$j$ is that of the generated one,
$( \d N/\d \sqrt{s'} )^{\mathrm{meas/gen}}_{i/j}$ is the simulated signal yield observed in bin $i$ or generated in bin $j$,
$G$ is a Gaussian function with mean and width left free in the fit,
and the operator $*$ represents the convolution integral.
In the $\omega$ resonance region, the mass bias, represented by the mean of the convolved Gaussian function, is $-0.46 \pm 0.06$\mevcc and the width is $1.1 \pm 0.2$\mevcc.
For the $\phi$ resonance, the mass bias is $-0.78 \pm 0.01$\mevcc and the width is consistent with zero.
Consistency in the high invariant mass region is confirmed using the \jpsi resonance as shown in Fig.~\ref{fig:mass_resolution_jpsi}.
The intrinsic width of the \jpsi is much smaller than the expected detector resolution of 11\mevcc.
Therefore, the shape of the \jpsi resonance is a good probe of the resolution.
The mass bias and resolution at the \jpsi resonance are obtained by fitting a Voigt function with resolution taken from simulation.
The mass bias is $-1.4 \pm 0.1$\mevcc, and the resolution for the $\jpsi$ signal is 11\mevcc (Fig.~\ref{fig:mass_resolution_jpsi}).
No difference in the resolution is observed between data and simulation.
\par
\babar reports that there are differences between data and simulation in the long tails of the resolution function that can be described by a Lorentzian function~\cite{BABAR:2021cde}.
To check for the presence of the Lorentzian term, we use the fitting function given in Eq.~(16) of \babar's paper.
In our case, we observe that, in Belle~II data, this Lorentzian term is consistent with zero.
\par
\begin{figure}
  \centering
  \includegraphics[width=\userFigureWidth]{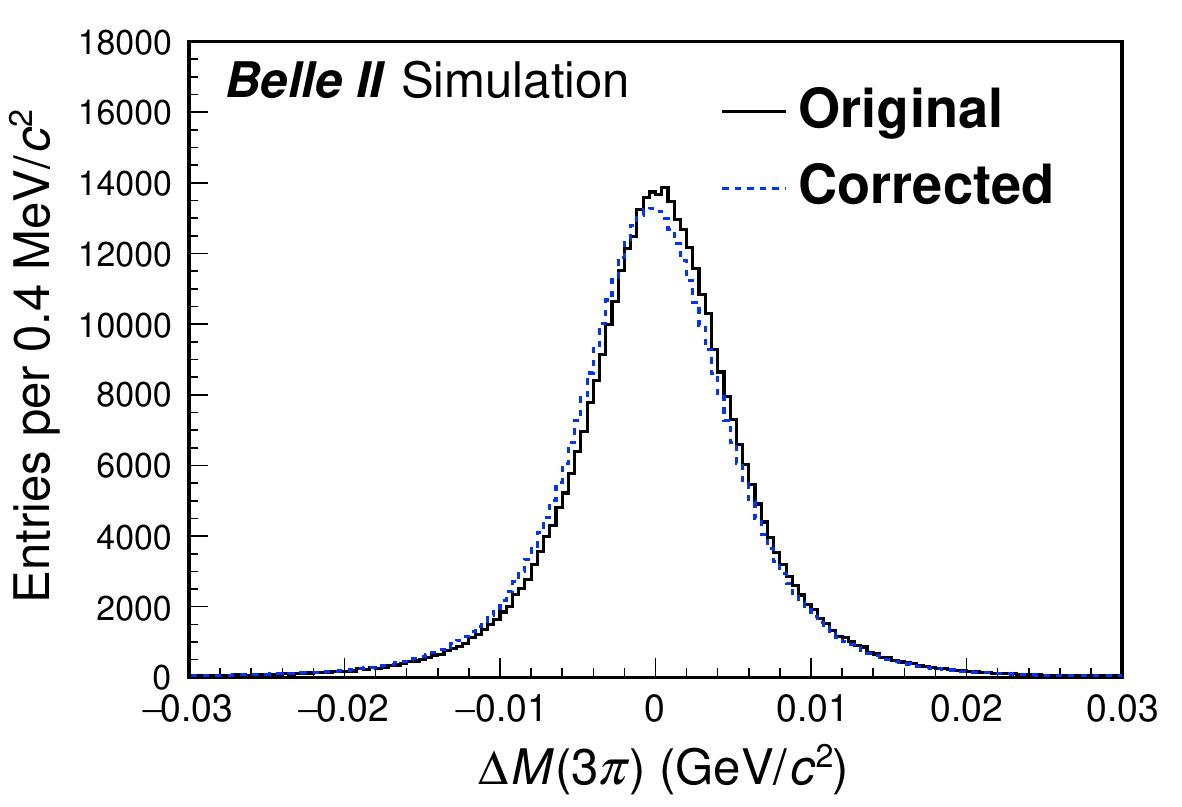}
  \caption{
    Mass resolution in the range $0.72 < \mpppz < 0.84\gevcc$.
    The solid histogram shows the resolution of the simulation and the dashed histogram shows the simulated resolution convolved with the Gaussian correction function obtained from data.
  }
  \label{fig:mass_resolution}
\end{figure}

Figure~\ref{fig:mass_resolution} compares the simulated \mpppz resolution with that from the simulation convolved with the Gaussian function that includes the data-to-simulation differences determined using the data fits in the $\omega$ resonance region.
A slight difference is observed and used to correct for the unfolding matrix with a Gaussian smearing.
The typical mass resolution after correction has a width of 6.6\mevcc at the $\omega$ resonance, 7.5\mevcc at $\phi$ resonance, 8\mevcc at 2\gevcc, and 11\mevcc at 3\gevcc.

%
%
%
%
%
\section{Cross-section measurement and systematic uncertainty}
\begin{table*}
  \centering
  \caption{
    Summary of fractional systematic uncertainty (\%)
    in the $\epem \to \pppz$ cross section for three different energy regions.
    The values in parentheses are the efficiency correction factors $\eta_i$ defined in Eq.~\eqref{eq:efficiency}.
    The systematic uncertainties due to the background subtraction and unfolding, which depend on the $3\pi$ mass bin, show the range from minimum to maximum in the energy regions of \mpppz near the $\omega$ resonance, 1.1--1.8\gevcc and 2.0--3.0\gevcc.
    }
    \label{tab:sys_detection}
  \begin{adjustbox}{width=\textwidth}{
\begin{tabular}{l@{\hspace*{5mm}}|@{\hspace*{8mm}}c@{\hspace*{4mm}}l@{\hspace*{20mm}}c@{\hspace*{4mm}}l@{\hspace*{20mm}}c@{\hspace*{4mm}}l@{\hspace*{8mm}}}
\hline \hline
\multirow{2}{*}{Source} & \multicolumn{6}{c}{Systematic uncertainty (efficiency correction factor $\eta_i$) \%} \\
                        & \multicolumn{2}{l}{0.62--1.05\gevcc} &\multicolumn{2}{l}{1.05--2.00\gevcc} & \multicolumn{2}{l}{2.0--3.5\gevcc}\\
\hline
Tracking                             &  0.8      &$(-1.35)$ & 0.8         &$(-1.71)$&  0.8      &$(-1.71)$\\
ISR photon detection                 &  0.7      &$(+0.15)$ & 0.7         &$(+0.15)$&  0.7      &$(+0.15)$\\
\piz detection                       &  1.0      &$(-1.43)$ & 1.0         &$(-1.43)$&  1.0      &$(-1.43)$\\
Kinematic fit ($\chi^{2}$)           &  0.6      &$(+0.0)$  & 0.3         &$(+0.30)$&  0.3      &$(+0.30)$\\
Trigger                              &  0.1      &$(-0.09)$ & 0.1         &$(-0.08)$&  0.1      &$(-0.06)$\\
Background suppression               &  0.2      &$(-1.90)$ & 1.9         &$(-1.78)$&  1.9      &$(-1.78)$\\
Monte Carlo generator                &  1.2      &         & 1.2         &       &  1.2      &   \\
Integrated luminosity                &  0.6      &         & 0.6         &       &  0.6      &   \\
Radiative corrections                &  0.5      &         & 0.5         &       &  0.5      &   \\
Simulated sample size                &  0.2      &         & 0.2--0.5    &       &  0.5--1.6 &   \\
Background subtraction               &  0.2--2.3 &         & 0.4--7.2    &       &  4.4--44  &   \\
Unfolding                            &  0.7--25  &         & 0.2--5.1    &       &  0.3--11     &   \\
\hline
Total uncertainty                    &  2.3-25 &           & 2.9--8.8    &        & 6.4-44   &        \\
 (Total correction $\varepsilon/\varepsilon_{\text{sim}}-1$)&       & $(-4.61)$  &        & $(-4.55)$  &       &$(-4.53)$ \\
\hline \hline
\end{tabular}
 }
 \end{adjustbox}
\end{table*}

The dressed cross section is calculated from the unfolded $3\pi$ mass spectrum using Eq.~\eqref{eq:spectrum}, where $\d N_{\mathrm{vis}}/\d\sqrt{s'}$ corresponds to the unfolded spectrum.
The correction factor for the higher-order ISR processes, \rrad, is evaluated using the \texttt{PHOKHARA} generator.
We generate events using the leading order radiator function with single ISR photon emission and a next-to-leading order contribution with one or two ISR photon emissions.
FSR emission is not simulated.
An ISR photon is generated within the range 20\degrees--160\degrees, and the invariant mass of the $3\pi$ system and the ISR photon is required to be greater than 8\gevcc.
The ratio of the yields of next-to-leading-order to leading-order events is calculated as a function of $3\pi$ invariant mass.
The radiative correction \rrad is 1.0080 for the mass range 0.5--1.05\gevcc, 1.0078 for the mass range 1.05--2\gevcc, and 1.0125 for the mass range 2.0--3.5\gevcc with a systematic uncertainty of 0.5\%~\cite{Rodrigo:2001kf}.
\par
A summary of the contributions to the systematic uncertainty for the cross section measurement is given in Table~\ref{tab:sys_detection}, for three different energy regions.
In the table, we also give the efficiency correction factors $\eta_i$ in parentheses.
We discuss each contribution following the order shown in the table.
%
%
\par
Uncertainties from the trigger, ISR photon detection, tracking, and \piz detection efficiencies are discussed in Sec.~\ref{sec:efficiency}, and are mainly driven by the size of the data sample in the control region.
The uncertainty in the event selection using 4C fits as well as other special requirements for the background suppression are also checked by comparing the data-to-simulation difference with and without the corresponding requirement.
In the region below 1.05\gevcc, the main systematic uncertainty is from the correction for tracking and \piz detection efficiency.
In the region above 1.05\gevcc, the main systematic uncertainty is from the correction for background suppression.
This originates from the statistical uncertainty of the $\jpsi \to \pppz$ data sample used for the evaluation.
%
%
\par
The systematic uncertainty due to mismodeling of the Monte Carlo generator is 1.2\%.
The uncertainty about how the extra ISR photon simulation in \texttt{PHOKHARA} will change the event selection efficiency, as discussed in Sec.~\ref{sec:eff_gen}.
%
\par
The integrated luminosity obtained using Bhabha and \gaga processes is $\intlum = 190.6 \pm 1.2 \invfb$, resulting in a 0.63\% systematic uncertainty~\cite{Belle-II:2019usr}.
A drift of the beam energy of at most 6\mev is observed during data taking.
The impact of this shift on the effective luminosity is 0.1\%, which is accounted for by the systematic uncertainty of the integrated luminosity.
%
%
\par
The systematic uncertainty for radiative corrections is 0.5\%, which arises from higher-order ISR processes in the \texttt{PHOKHARA} generator~\cite{Rodrigo:2001kf}.
%
%
\par
The size of the simulated sample used for the first approximation of the signal efficiency gives a systematic uncertainty, which depends on the $3\pi$ mass and is typically 0.2\% at the $\omega$ resonance, increasing to about 1.6\% as the mass increases.
%
%
\par
The uncertainty in the background subtraction is determined by the statistical and systematic uncertainties of the control sample.
The systematic uncertainty is 0.2\%--2.3\% at the $\omega$ resonance, and 0.4\%--7.2\% for the range 1.1--1.8\gevcc.
%
%
\par
Several checks on the uncertainty of the unfolding are carried out.
We check the dependence on the four regularization parameters in the IDS method by varying them from their optimum values and observe no significant change.
In addition, a possible bias in the unfolding procedure is checked by a simulation study with at least ten times more events than in data.
A simulated spectrum is generated by fluctuating the population of each bin in the generated spectrum by its statistical uncertainty, and the corrected transfer matrix is used to produce a measured spectrum.
The differences between the unfolded and true spectrum are assigned as a systematic uncertainty of about 0.3\% near the $\omega$ and $\phi$ resonances, and 0.9\%--6\% above 1.05\gevcc.
In the mass range below 0.7\gevcc, the systematic uncertainty for the unfolding is greater than 10\% and is the main source of systematic uncertainty.
%
%
\par
The total systematic uncertainty, listed in Table~\ref{tab:sys_detection}, is obtained by taking the quadrature sum of all contributions in the table.
The total systematic uncertainty in the $\omega$ resonance region is 2.1\%, which is dominated by the uncertainty of efficiency correction factors.
%
%
%
%
\vspace{1em}
\section{Cross-section result}\label{sec:xsec}
\begin{figure*}
  \centering
    \includegraphics[width=\textwidth]{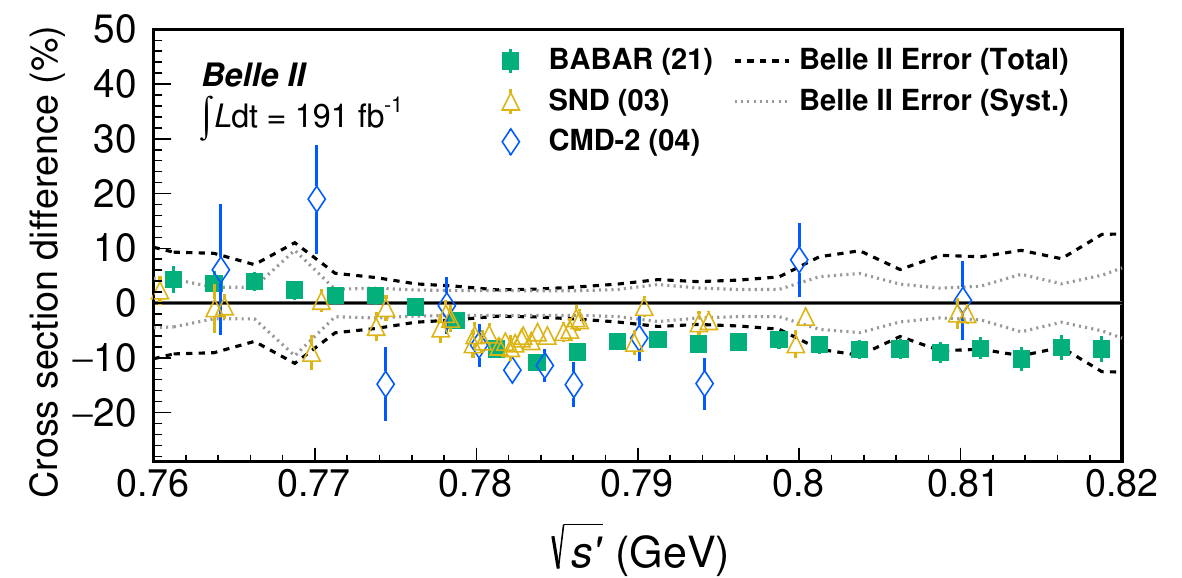}
    \caption{
    Differences between $\epem \to \pppz$ cross-section results from previous measurements and results of this work, as functions of energy (markers with error bars).
    Belle~II results are taken as the reference at zero, with dashed (dotted) lines corresponding to the total (systematic) uncertainties.
    Squares are the \babar results~\cite{BABAR:2021cde},
    triangles are the SND results~\cite{Achasov:2003ir},
    and diamonds are the CMD-2 results~\cite{CMD-2:2003gqi}.
    }
  \label{fig:xsec_comp}
\end{figure*}

The resulting dressed cross section for the $\epem \to \pppz$ process is given in Fig.~\ref{fig:dressed_cross_section_1} and summarized in Appendix~\ref{app:xsectable}.
The corresponding covariance matrices, for both statistical and systematic components, are provided in the Supplemental Material~\cite{supple}.
Other systematic uncertainties that are largely common in wide energy regions are listed in Table~\ref{tab:sys_detection}.
The total correlated systematic uncertainty is 2.15\% below 1.05\gev and is 2.80\% above 1.05\gev.
Figures~\ref{fig:dressed_cross_section_1}~{\subref{fig:dressed_cross_section_a}}--{\subref{fig:dressed_cross_section_f}} show comparisons of the dressed cross section to previous measurements.
In the energy range $3.0 < \sqrt{s'} < 3.2\gev$, the nonresonant cross sections are obtained after subtracting the $\jpsi \to \pppz$ contribution from the spectrum.
A comparison of previous measurements with the results of this work is shown in Fig.~\ref{fig:xsec_comp}, where the dotted and dashed lines show our systematic and total uncertainties.
\par
The differences across the measurements are visible in the cross section at the $\omega$ resonance.
Below 0.78\gev, our results are consistent with those of \babar due to large statistical uncertainties.
In the energy region 0.78--0.80\gev, \babar reports a smaller cross section than Belle~II.
Above 0.8\gev, \babar results are also smaller than ours but the difference is about the same size as our total uncertainties.
\begin{figure*}[p]
  \centering
  \subfloat[][]{\includegraphics[width=0.48\textwidth]{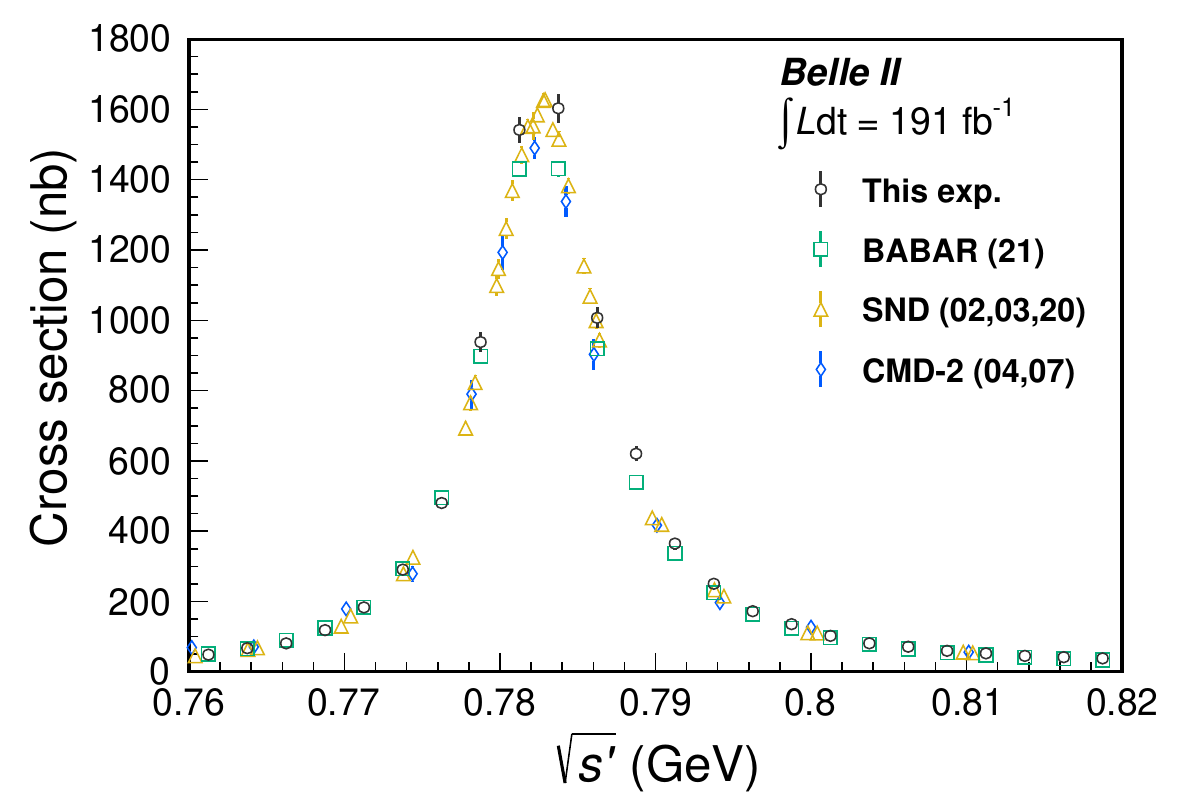}\label{fig:dressed_cross_section_a}}
  \subfloat[][]{\includegraphics[width=0.48\textwidth]{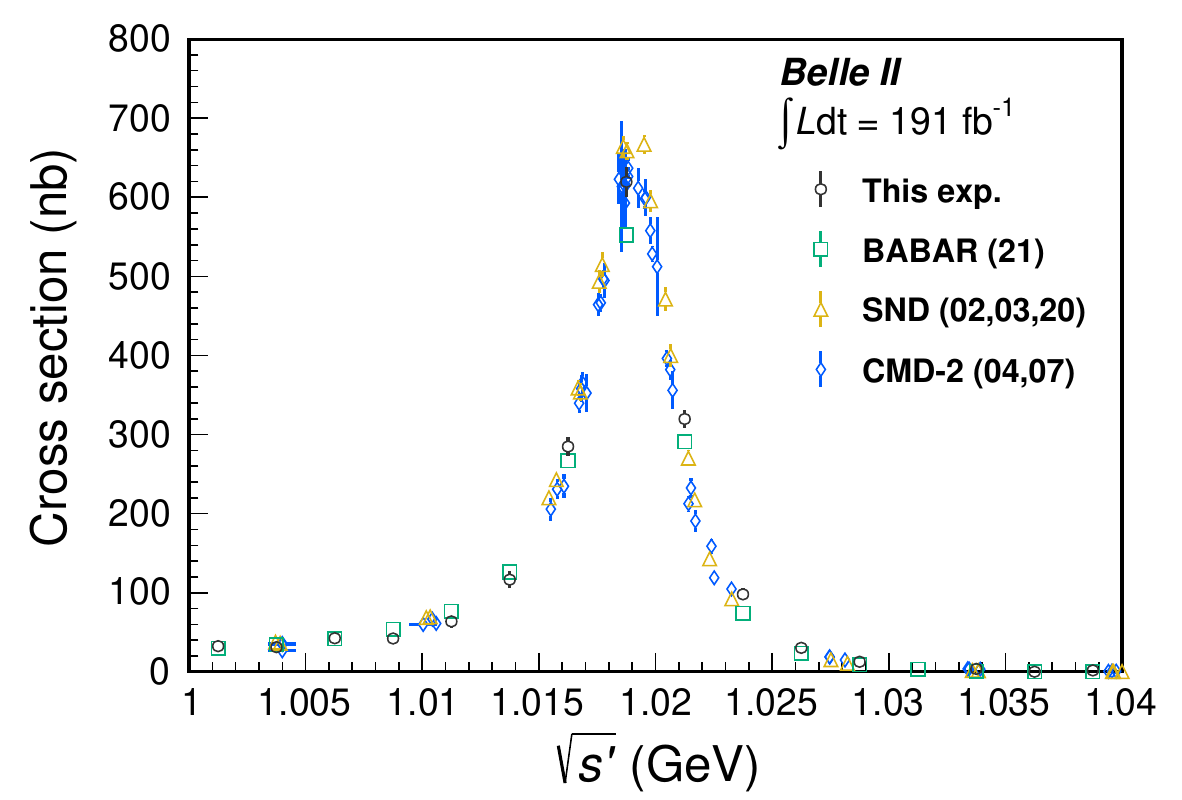}\label{fig:dressed_cross_section_b}}\\
  \subfloat[][]{\includegraphics[width=0.48\textwidth]{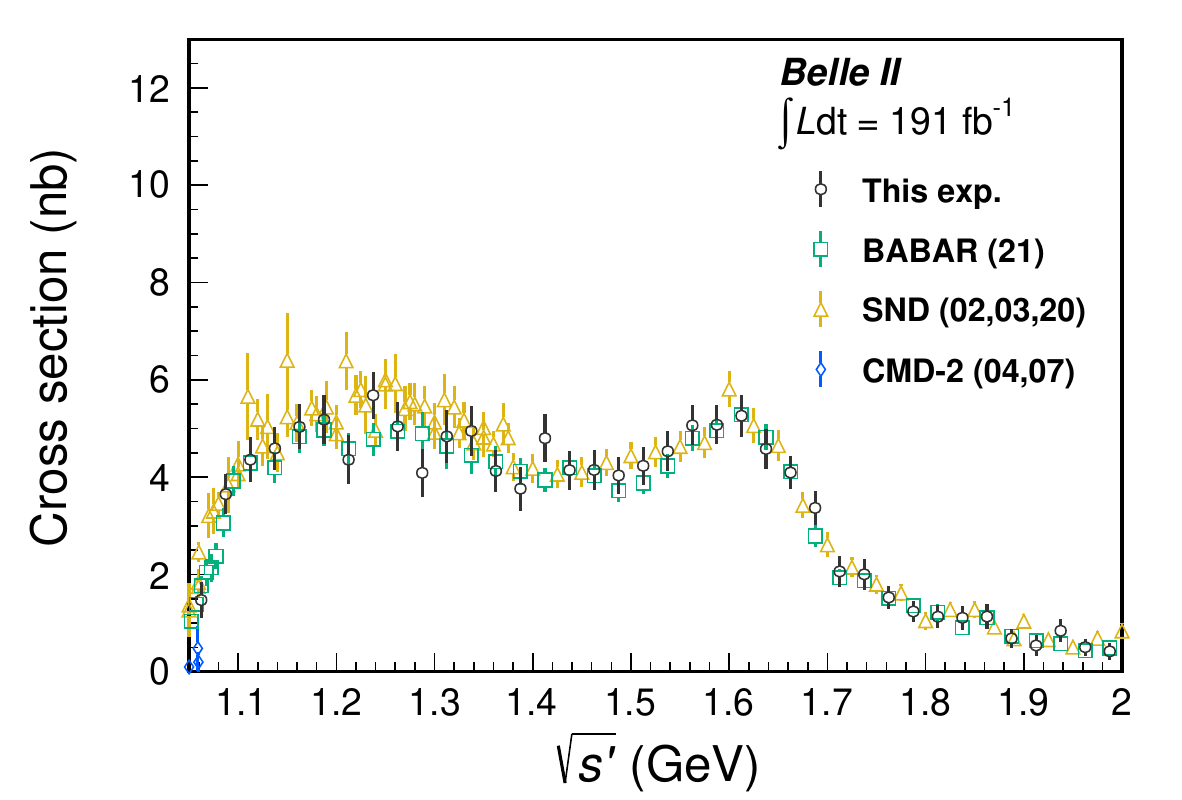} \label{fig:dressed_cross_section_c}}
  \subfloat[][]{\includegraphics[width=0.48\textwidth]{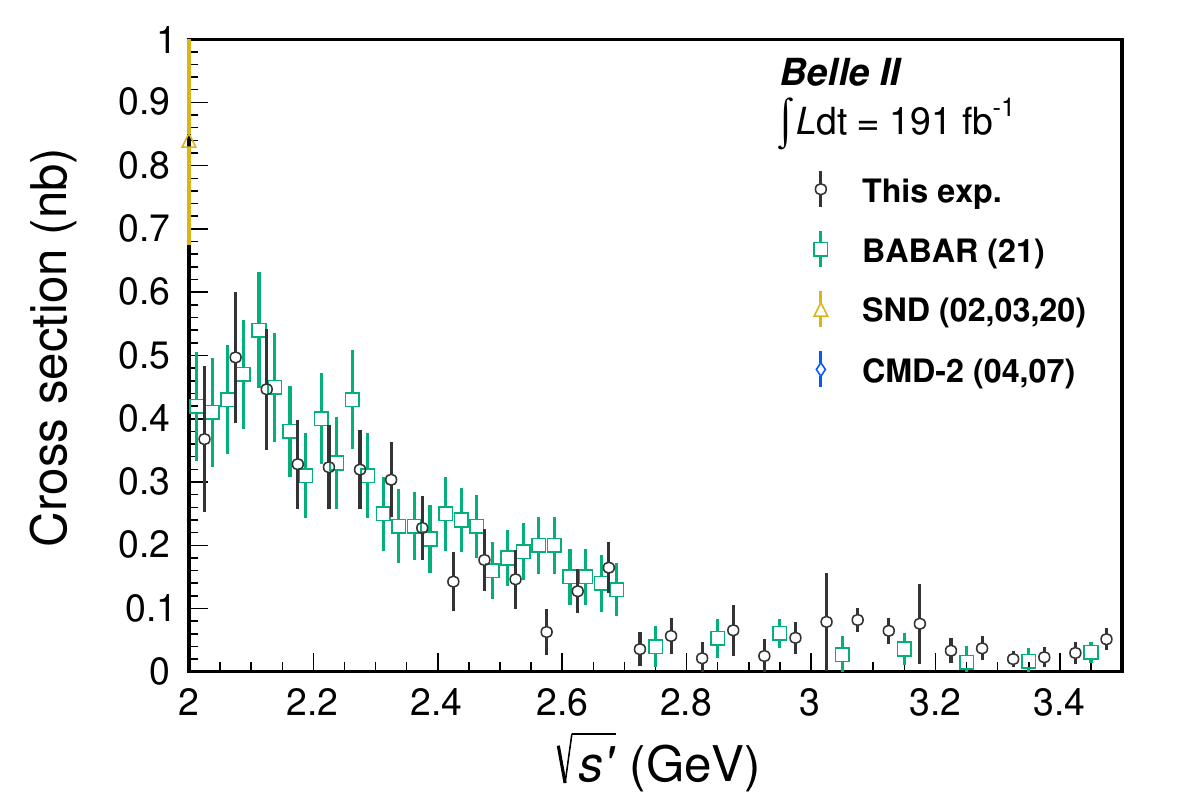} \label{fig:dressed_cross_section_d}}\\
  \subfloat[][]{\includegraphics[width=0.48\textwidth]{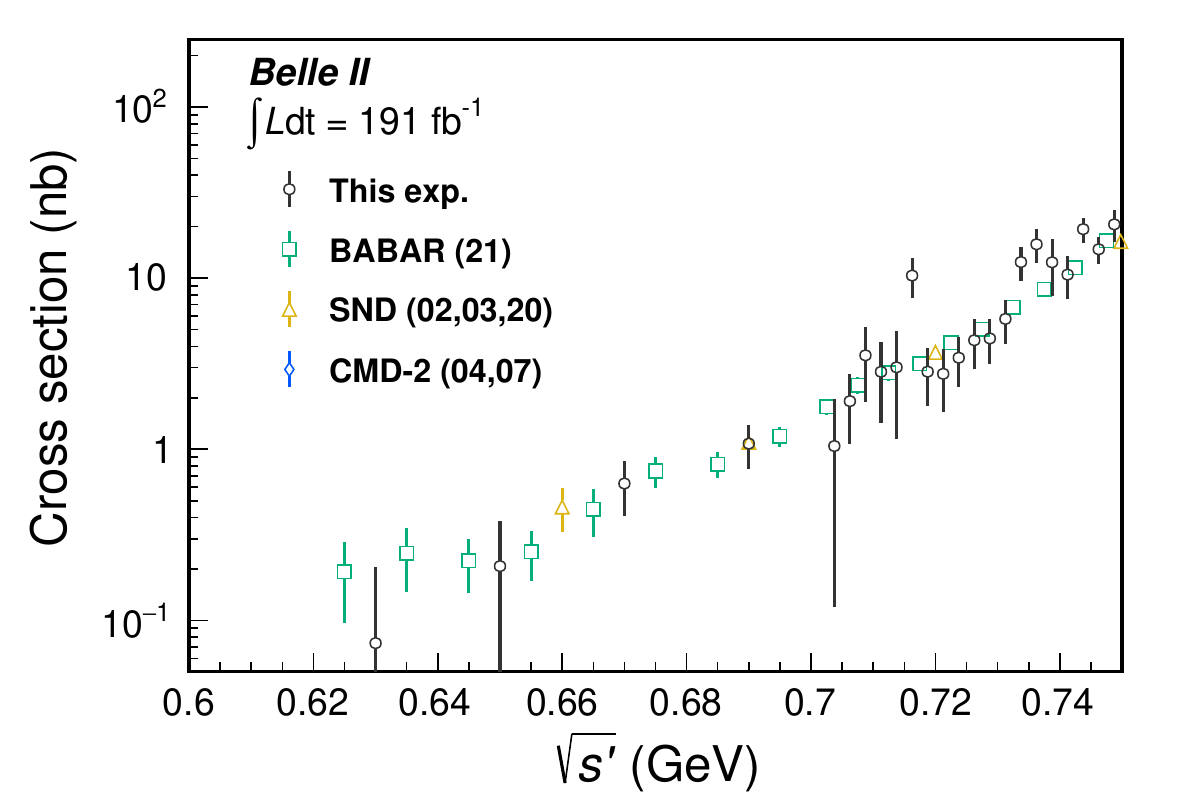}\label{fig:dressed_cross_section_e}} 
  \subfloat[][]{\includegraphics[width=0.48\textwidth]{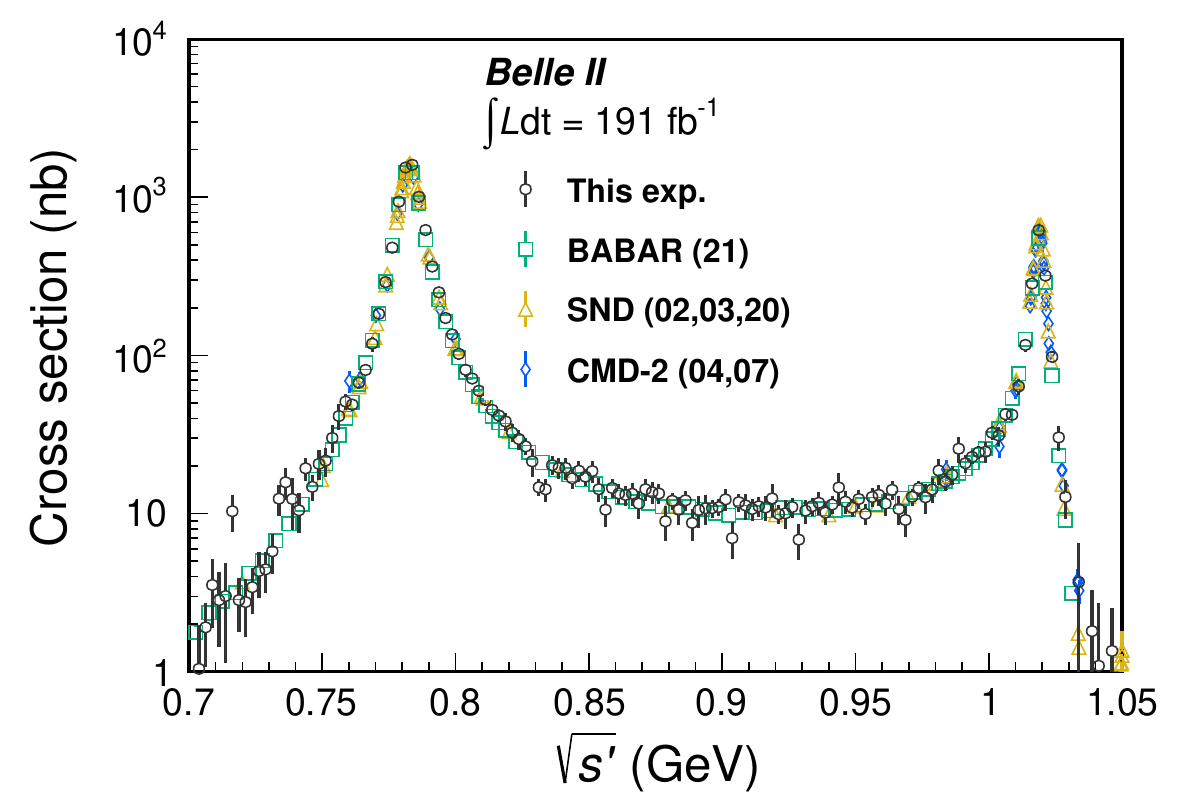} \label{fig:dressed_cross_section_f}} 
  \caption{
    Observed $\epem \to \pppz$ cross section as a function of energy compared with previous results.
    Each panel covers a different energy range: 
    (a) 0.76--0.82\gev ($\omega$ resonance),
    (b) 1.00--1.05\gev ($\phi$ resonance),
    (c) 1.05--2.00\gev,
    and (d) 2.0--3.5\gev regions with a linear scale.
    A logarithmic scale for (e) a threshold region ($<0.75$\gev) and (f) $\omega$ and $\phi$ region (0.7--1.05 GeV).
    Circles with error bars are the Belle~II results, 
    squares are the \babar results~\cite{BABAR:2021cde},
    triangles are the SND results~\cite{Achasov:2002ud,Achasov:2003ir,Aulchenko:2015mwt},
    and diamonds are the CMD-2 results~\cite{CMD-2:2003gqi,Akhmetshin:2006sc}.
  }
  \label{fig:dressed_cross_section_1}
\end{figure*}
%
%
%
%
%
%
\section{\texorpdfstring{Contribution to \boldmath{\ammhvp}}{Contribution to muon anomalous magnetic moment}} \label{sec:amm}
\begin{figure}
  \centering
  \includegraphics[width=\userFigureWidth]{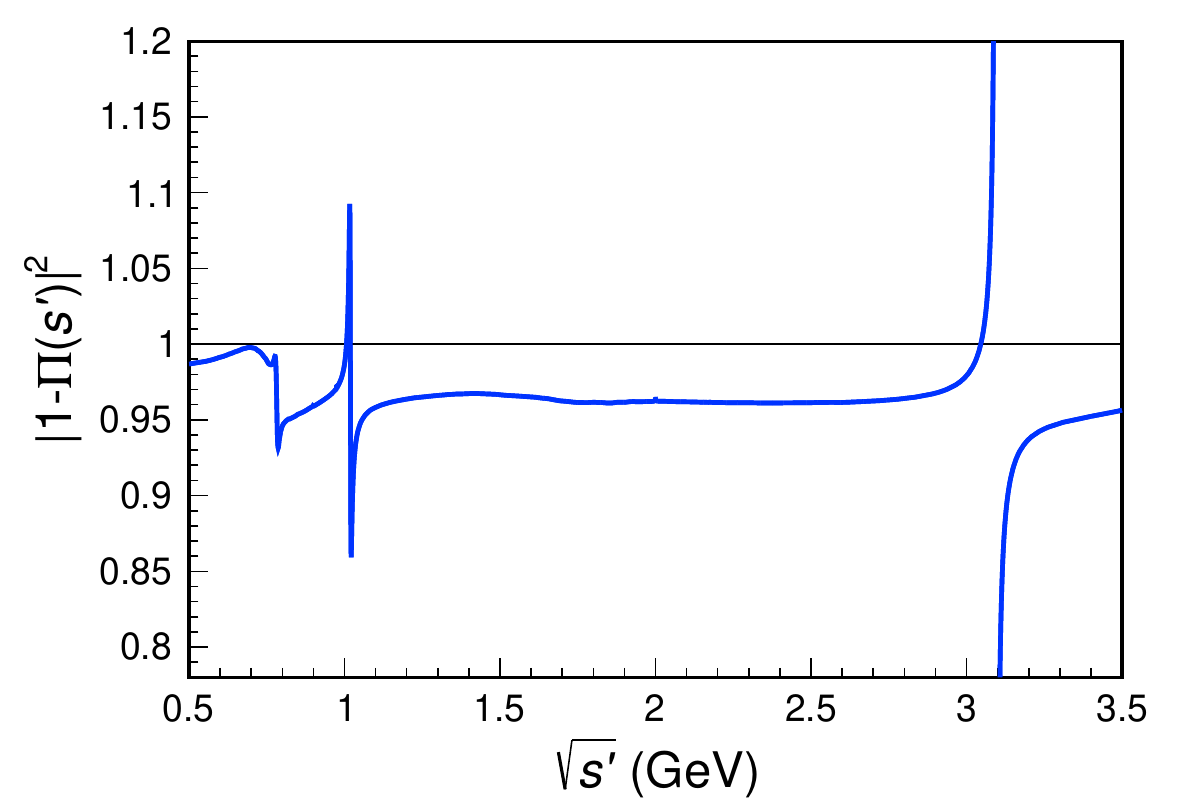}
  \caption{
  Energy dependence of the vacuum polarization corrections $|1-\Pi(s')|^2$ reproduced from Ref.~\cite{F.Ignatov:2019}.
  }
  \label{fig:vpc}
\end{figure}

The $3\pi$ contribution to the leading-order HVP term in \amm is given by
\begin{align}
    \ammhvplo = \frac{\alpha}{3\pi^2} \int^{\infty}_{m^{2}_{\pi}} \frac{K(s)}{s} R_\mathrm{had}(s) ds,
\end{align} where $K (s)$ is the QED kernel function~\cite{Aoyama:2020ynm}.
The hadronic $R$-ratio is expressed in terms of the ratio of bare hadron- and muon-pair cross sections, 
\begin{align}
R_{\mathrm{had}}(s) = \frac{\sigma_{0}(\epem \to \mathrm{hadrons})}{\sigma_{\text{pt}}(\epem\to\mumu)},
\end{align}
where the point-like cross section $\sigma_{\text{pt}}(\epem\to\mumu) = 4\pi\alpha^2/3s$.
The bare cross section $\sigma_0$ is obtained from the dressed cross section, given in Eq.~\eqref{eq:spectrum}, by removing the vacuum polarization effects
\begin{align}
 \sigma_0  = \sigma^{\mathrm{dressed}} |1 -  \Pi(s')|^2.
\end{align}
The value and uncertainty of the vacuum polarization correction $|1-\Pi(s')|^2$ are given in Ref.~\cite{F.Ignatov:2019} and displayed in Fig.~\ref{fig:vpc}.
The uncertainty is 0.06\% below 1.05\gev and 0.02\% above 1.05\gev.
\par
Using these values and integrating over the $3\pi$ cross section measured by Belle~II from 0.62 to 1.8\gev, we obtain
\begin{equation}
    \ammpppz = (48.91 \pm 0.23 \pm 1.07) \times 10^{-10},
\end{equation}
where the first uncertainty is statistical and the second is systematic.
The value of \ammpppz is determined with 2.2\% accuracy.
The contributions to the systematic uncertainty for \ammpppz are summarized in Table~\ref{tab:sys_amm}.
The main sources are the uncertainty due to efficiency corrections and Monte Carlo generator.
The results can be compared to those obtained by the \babar experiment~\cite{BABAR:2021cde},
\begin{equation}\notag
  \ammpppz (0.62\textrm{--}2.0\gev) = (45.86 \pm 0.14 \pm 0.58) \times 10^{-10}
\end{equation}
and the global fit of Ref.~\cite{Hoferichter:2023bjm}, which includes the \babar result,
\begin{equation}\notag
  \ammpppz (0.62\textrm{--}1.8\gev) = (45.91 \pm  0.37 \pm  0.38) \times 10^{-10}.
\end{equation}
The Belle~II cross section is 6.9\% higher than the cross section observed by \babar and 6.5\% higher than the result of the global fit.
The compatibility with either is 2.5$\sigma$.
The values of \ammpppz are calculated separately for the energy ranges below 1.05\gev and 1.05--2.0\gev to compare with \babar, and in both regions, the differences are 7\%.
%

\section{Discussion\label{sec:discuss}}
Although similar analysis procedures are used by \babar~\cite{BABAR:2021cde} and Belle~II measurements, there are several differences.
The data size used by Belle~II ($191\invfb$) is 2.4 times smaller than that of \babar ($469\invfb$).
The generator used for the signal simulation is \texttt{AfkQed}~\cite{Czyz:2000wh,Caffo:1997yy,Caffo:1994dm} in \babar and is \texttt{PHOKHARA}~\cite{Rodrigo:2001jr,Rodrigo:2001kf,Czyz:2005as} in Belle~II.
There is a difference in the ISR QED simulation between the two programs.
Both experiments use kinematical 4C fits for the signal selection.
However, \babar uses only the measured direction for the ISR photon keeping the energy as a free parameter of the fit while Belle~II uses the measured ISR photon energy in their 4C fit.
\babar selects \piz's by counting the number of events in a mass window in \mgg, while Belle~II determines the \piz yield by fitting the \mgg distribution.
Although the size of the background in the $\omega$ region is less than 1\% in both experiments, these differences affect the size of the remaining background.
\par
The systematic uncertainty of the cross section in the $\omega$ resonance region is 1.3\% for BABAR and is 2.2\% in Belle~II.
\babar's systematic uncertainty is dominated by detector effects (1.2\%), which are mainly due to the uncertainty in \piz detection and tracking.
Belle~II's uncertainty is also dominated by the uncertainty on the \piz efficiency (1.0\%) and tracking efficiency (0.8\%).
In addition, Belle~II takes into account 1.2\% due to the uncertainty in ISR photon simulation according to the recent observation in Ref.~\cite{BaBar:2023xiy}.
\begin{table}
\centering
\caption{Summary of contributions to the systematic uncertainty in \ammpppz (\%).}
\label{tab:sys_amm}
\begin{tabular}{lS}
\hline
\hline
{Source} &  {Systematic uncertainty (\%)}\\
\hline
{Efficiency corrections}            & 1.63     \\
{Monte Carlo generator}             & 1.20    \\
{Integrated luminosity}             & 0.64     \\
{Simulated sample size}             & 0.15     \\
{Background subtraction}            & 0.02     \\
{Unfolding}                         & 0.12    \\
{Radiative corrections}             & 0.50    \\
{Vacuum polarization corrections}   & 0.04     \\
\hline
{Total}                             & 2.19     \\
\hline
\hline
\end{tabular}
\end{table}
%
%
%
%
\section{Summary}\label{sec:summary}
In summary, we have measured the cross section for the process $\epem \to \pppz$ in an energy range from 0.62 to 3.5\gev, using the ISR technique.
We use a 191\invfb $\epem$ data sample collected by Belle~II at an \epem c.m.~energy at or near the \Y4S resonance.
The systematic uncertainty of the cross section is about 2.2\% at  the $\omega$ and $\phi$ resonances, where the cross section is large.
At other energies, the precision is limited by the statistical uncertainty.
The resulting contribution, at leading order in HVP, to the muon anomalous magnetic moment is
$\ammpppz = (48.91 \pm 0.23 \pm 1.07) \times 10^{-10}$
in the 0.62--1.8\gev energy range.
The Belle~II result differs by $2.5\sigma$ from the current most precise measurement~\cite{BABAR:2021cde} and global fits~\cite{Hoferichter:2023bjm}.

\acknowledgements
This work, based on data collected using the Belle II detector, which was built and commissioned prior to March 2019,
was supported by
Higher Education and Science Committee of the Republic of Armenia Grant No.~23LCG-1C011;
Australian Research Council and Research Grants
No.~DP200101792, 
No.~DP210101900, 
No.~DP210102831, 
No.~DE220100462, 
No.~LE210100098, 
and
No.~LE230100085; 
Austrian Federal Ministry of Education, Science and Research,
Austrian Science Fund
No.~P~31361-N36
and
No.~J4625-N,
and
Horizon 2020 ERC Starting Grant No.~947006 ``InterLeptons'';
Natural Sciences and Engineering Research Council of Canada, Compute Canada and CANARIE;
National Key R\&D Program of China under Contract No.~2022YFA1601903,
National Natural Science Foundation of China and Research Grants
No.~11575017,
No.~11761141009,
No.~11705209,
No.~11975076,
No.~12135005,
No.~12150004,
No.~12161141008,
and
No.~12175041,
and Shandong Provincial Natural Science Foundation Project~ZR2022JQ02;
the Czech Science Foundation Grant No.~22-18469S;
European Research Council, Seventh Framework PIEF-GA-2013-622527,
Horizon 2020 ERC-Advanced Grants No.~267104 and No.~884719,
Horizon 2020 ERC-Consolidator Grant No.~819127,
Horizon 2020 Marie Sklodowska-Curie Grant Agreement No.~700525 ``NIOBE''
and
No.~101026516,
and
Horizon 2020 Marie Sklodowska-Curie RISE project JENNIFER2 Grant Agreement No.~822070 (European grants);
L'Institut National de Physique Nucl\'{e}aire et de Physique des Particules (IN2P3) du CNRS
and
L'Agence Nationale de la Recherche (ANR) under grant ANR-21-CE31-0009 (France);
BMBF, DFG, HGF, MPG, and AvH Foundation (Germany);
Department of Atomic Energy under Project Identification No.~RTI 4002,
Department of Science and Technology,
and
UPES SEED funding programs
No.~UPES/R\&D-SEED-INFRA/17052023/01 and
No.~UPES/R\&D-SOE/20062022/06 (India);
Israel Science Foundation Grant No.~2476/17,
U.S.-Israel Binational Science Foundation Grant No.~2016113, and
Israel Ministry of Science Grant No.~3-16543;
Istituto Nazionale di Fisica Nucleare and the Research Grants BELLE2;
Japan Society for the Promotion of Science, Grant-in-Aid for Scientific Research Grants
No.~16H03968,
No.~16H03993,
No.~16H06492,
No.~16K05323,
No.~17H01133,
No.~17H05405,
No.~18K03621,
No.~18H03710,
No.~18H05226,
No.~19H00682, 
No.~20H05850,
No.~20H05858,
No.~22H00144,
No.~22K14056,
No.~22K21347,
No.~23H05433,
No.~26220706,
and
No.~26400255,
and
the Ministry of Education, Culture, Sports, Science, and Technology (MEXT) of Japan;  
National Research Foundation (NRF) of Korea Grants
No.~2016R1\-D1A1B\-02012900,
No.~2018R1\-A2B\-3003643,
No.~2018R1\-A6A1A\-06024970,
No.~2019R1\-I1A3A\-01058933,
No.~2021R1\-A6A1A\-03043957,
No.~2021R1\-F1A\-1060423,
No.~2021R1\-F1A\-1064008,
No.~2022R1\-A2C\-1003993,
and
No.~RS-2022-00197659,
Radiation Science Research Institute,
Foreign Large-Size Research Facility Application Supporting project,
the Global Science Experimental Data Hub Center of the Korea Institute of Science and Technology Information
and
KREONET/GLORIAD;
Universiti Malaya RU grant, Akademi Sains Malaysia, and Ministry of Education Malaysia;
Frontiers of Science Program Contracts
No.~FOINS-296,
No.~CB-221329,
No.~CB-236394,
No.~CB-254409,
and
No.~CB-180023, and SEP-CINVESTAV Research Grant No.~237 (Mexico);
the Polish Ministry of Science and Higher Education and the National Science Center;
the Ministry of Science and Higher Education of the Russian Federation
and
the HSE University Basic Research Program, Moscow;
University of Tabuk Research Grants
No.~S-0256-1438 and No.~S-0280-1439 (Saudi Arabia);
Slovenian Research Agency and Research Grants
No.~J1-9124
and
No.~P1-0135;
Agencia Estatal de Investigacion, Spain
Grant No.~RYC2020-029875-I
and
Generalitat Valenciana, Spain
Grant No.~CIDEGENT/2018/020;
National Science and Technology Council,
and
Ministry of Education (Taiwan);
Thailand Center of Excellence in Physics;
TUBITAK ULAKBIM (Turkey);
National Research Foundation of Ukraine, Project No.~2020.02/0257,
and
Ministry of Education and Science of Ukraine;
the U.S. National Science Foundation and Research Grants
No.~PHY-1913789 
and
No.~PHY-2111604, 
and the U.S. Department of Energy and Research Awards
No.~DE-AC06-76RLO1830, 
No.~DE-SC0007983, 
No.~DE-SC0009824, 
No.~DE-SC0009973, 
No.~DE-SC0010007, 
No.~DE-SC0010073, 
No.~DE-SC0010118, 
No.~DE-SC0010504, 
No.~DE-SC0011784, 
No.~DE-SC0012704, 
No.~DE-SC0019230, 
No.~DE-SC0021274, 
No.~DE-SC0021616, 
No.~DE-SC0022350, 
No.~DE-SC0023470; 
and
the Vietnam Academy of Science and Technology (VAST) under Grants
No.~NVCC.05.12/22-23
and
No.~DL0000.02/24-25.

These acknowledgements are not to be interpreted as an endorsement of any statement made
by any of our institutes, funding agencies, governments, or their representatives.

We thank the SuperKEKB team for delivering high-luminosity collisions;
the KEK cryogenics group for the efficient operation of the detector solenoid magnet;
the KEK Computer Research Center for on-site computing support; the NII for SINET6 network support;
and the raw-data centers hosted by BNL, DESY, GridKa, IN2P3, INFN, 
and the University of Victoria.

\appendix
\section{CROSS SECTION RESULTS FOR DIFFERENT ENERGIES}\label{app:xsectable}
Tables~\ref{tab:xsec_lt1050} and \ref{tab:xsec_gt1050} list the energy range, the number of events after unfolding, the signal efficiency, and the dressed cross section for $\epem \to \pppz$.

\newcommand{\xsecspace}{~~~~~~~}
\begin{table*}[p]
  \centering
  \caption{
       Energy bin range ($\sqrt{s'}$), number of events after unfolding ($N_{\mathrm{unf}}$), corrected efficiency ($\varepsilon$), and cross section ($\sigma_{3\pi}$) for $\epem \to \pppz$ in energy range 0.62--1.05\gev.
      The two uncertainties in the cross section are the statistical and systematic contributions.
      The statistical uncertainties for the unfolding and cross section are square roots of the diagonal components of the unfolding covariance matrix.
  }
  \label{tab:xsec_lt1050}
  \begin{adjustbox}{width=0.98\textwidth}
  { \scriptsize
    \begin{tabularx}{\linewidth}{lccrc|clccr}
  \hline\hline
  $\sqrt{s'}$ (\gev) & \xsecspace$N_{\mathrm{unf}}$\xsecspace& \xsecspace$\varepsilon$ (\%)\xsecspace& $\xsecspace\sigma$ (nb) & &&  $\sqrt{s'}$ (\gev) & \xsecspace$N_{\mathrm{unf}}$\xsecspace& \xsecspace$\varepsilon$ (\%)\xsecspace& $\xsecspace\sigma$ (nb) \\
  \hline
 0.6200--0.6400& $     2 \pm      3$& $9.16 \pm 0.15$& $    0.07 \pm   0.13 \pm   0.02$ & &&0.8700--0.8725& $    52 \pm      8$& $8.18 \pm 0.13$& $   14.26 \pm   2.12 \pm   0.89$ \\
 0.6400--0.6600& $     5 \pm      4$& $9.16 \pm 0.15$& $    0.21 \pm   0.17 \pm   0.02$ & &&0.8725--0.8750& $    50 \pm      7$& $8.18 \pm 0.13$& $   13.65 \pm   1.81 \pm   1.07$ \\
 0.6600--0.6800& $    16 \pm      5$& $9.16 \pm 0.15$& $    0.63 \pm   0.22 \pm   0.03$ & &&0.8750--0.8775& $    49 \pm      6$& $8.17 \pm 0.13$& $   13.40 \pm   1.64 \pm   1.11$ \\
 0.6800--0.7000& $    28 \pm      8$& $9.16 \pm 0.15$& $    1.08 \pm   0.31 \pm   0.05$ & &&0.8775--0.8800& $    33 \pm      7$& $8.17 \pm 0.13$& $    8.95 \pm   1.92 \pm   1.12$ \\
 0.7000--0.7025& $    -2 \pm      2$& $8.52 \pm 0.14$& $   -2.06 \pm   1.82 \pm   1.82$ & &&0.8800--0.8825& $    44 \pm      5$& $8.16 \pm 0.13$& $   12.06 \pm   1.38 \pm   0.62$ \\
 0.7025--0.7050& $     3 \pm      2$& $8.51 \pm 0.14$& $    1.04 \pm   0.77 \pm   0.52$ & &&0.8825--0.8850& $    39 \pm      7$& $8.16 \pm 0.13$& $   10.57 \pm   1.97 \pm   0.70$ \\
 0.7050--0.7075& $     6 \pm      2$& $8.51 \pm 0.14$& $    1.91 \pm   0.75 \pm   0.36$ & &&0.8850--0.8875& $    45 \pm      6$& $8.15 \pm 0.13$& $   12.14 \pm   1.56 \pm   0.66$ \\
 0.7075--0.7100& $    11 \pm      4$& $8.50 \pm 0.14$& $    3.54 \pm   1.40 \pm   0.89$ & &&0.8875--0.8900& $    32 \pm      7$& $8.15 \pm 0.13$& $    8.74 \pm   1.88 \pm   0.63$ \\
 0.7100--0.7125& $     9 \pm      4$& $8.50 \pm 0.14$& $    2.84 \pm   1.30 \pm   0.55$ & &&0.8900--0.8925& $    39 \pm      8$& $8.14 \pm 0.13$& $   10.55 \pm   2.08 \pm   1.19$ \\
 0.7125--0.7150& $     9 \pm      6$& $8.49 \pm 0.14$& $    3.01 \pm   1.80 \pm   0.51$ & &&0.8925--0.8950& $    40 \pm      6$& $8.14 \pm 0.13$& $   10.82 \pm   1.64 \pm   1.62$ \\
 0.7150--0.7175& $    32 \pm      8$& $8.49 \pm 0.14$& $   10.36 \pm   2.43 \pm   1.24$ & &&0.8950--0.8975& $    40 \pm      5$& $8.14 \pm 0.13$& $   10.84 \pm   1.31 \pm   0.60$ \\
 0.7175--0.7200& $     9 \pm      3$& $8.48 \pm 0.14$& $    2.84 \pm   0.92 \pm   0.52$ & &&0.8975--0.9000& $    41 \pm      4$& $8.13 \pm 0.13$& $   11.01 \pm   1.18 \pm   0.69$ \\
 0.7200--0.7225& $     9 \pm      3$& $8.48 \pm 0.14$& $    2.76 \pm   0.98 \pm   0.50$ & &&0.9000--0.9025& $    46 \pm      7$& $8.13 \pm 0.13$& $   12.30 \pm   1.86 \pm   0.55$ \\
 0.7225--0.7250& $    11 \pm      2$& $8.47 \pm 0.14$& $    3.43 \pm   0.79 \pm   0.78$ & &&0.9025--0.9050& $    26 \pm      7$& $8.12 \pm 0.13$& $    7.00 \pm   1.74 \pm   0.44$ \\
 0.7250--0.7275& $    14 \pm      4$& $8.46 \pm 0.14$& $    4.34 \pm   1.23 \pm   0.69$ & &&0.9050--0.9075& $    44 \pm      6$& $8.12 \pm 0.13$& $   11.82 \pm   1.48 \pm   0.34$ \\
 0.7275--0.7300& $    14 \pm      3$& $8.46 \pm 0.14$& $    4.45 \pm   1.09 \pm   0.69$ & &&0.9075--0.9100& $    42 \pm      7$& $8.11 \pm 0.13$& $   11.28 \pm   1.87 \pm   0.99$ \\
 0.7300--0.7325& $    18 \pm      5$& $8.45 \pm 0.14$& $    5.78 \pm   1.58 \pm   0.49$ & &&0.9100--0.9125& $    40 \pm      5$& $8.11 \pm 0.13$& $   10.70 \pm   1.35 \pm   1.06$ \\
 0.7325--0.7350& $    39 \pm      8$& $8.45 \pm 0.14$& $   12.43 \pm   2.56 \pm   1.00$ & &&0.9125--0.9150& $    42 \pm      6$& $8.10 \pm 0.13$& $   11.13 \pm   1.51 \pm   0.46$ \\
 0.7350--0.7375& $    50 \pm      9$& $8.44 \pm 0.14$& $   15.79 \pm   3.00 \pm   1.94$ & &&0.9150--0.9175& $    41 \pm      5$& $8.10 \pm 0.13$& $   10.94 \pm   1.42 \pm   1.39$ \\
 0.7375--0.7400& $    39 \pm     10$& $8.44 \pm 0.14$& $   12.39 \pm   3.17 \pm   3.17$ & &&0.9175--0.9200& $    47 \pm     10$& $8.10 \pm 0.13$& $   12.48 \pm   2.70 \pm   1.22$ \\
 0.7400--0.7425& $    33 \pm      9$& $8.43 \pm 0.14$& $   10.50 \pm   2.88 \pm   0.72$ & &&0.9200--0.9225& $    38 \pm      5$& $8.09 \pm 0.13$& $    9.93 \pm   1.31 \pm   0.28$ \\
 0.7425--0.7450& $    62 \pm      9$& $8.43 \pm 0.13$& $   19.35 \pm   2.93 \pm   1.30$ & &&0.9225--0.9250& $    38 \pm      7$& $8.09 \pm 0.13$& $   10.06 \pm   1.78 \pm   1.06$ \\
 0.7450--0.7475& $    47 \pm      9$& $8.42 \pm 0.13$& $   14.77 \pm   2.66 \pm   0.47$ & &&0.9250--0.9275& $    42 \pm      5$& $8.08 \pm 0.13$& $   11.01 \pm   1.35 \pm   0.79$ \\
 0.7475--0.7500& $    66 \pm     12$& $8.42 \pm 0.13$& $   20.64 \pm   3.76 \pm   2.35$ & &&0.9275--0.9300& $    26 \pm      6$& $8.08 \pm 0.13$& $    6.85 \pm   1.65 \pm   0.49$ \\
 0.7500--0.7525& $    70 \pm     10$& $8.41 \pm 0.13$& $   21.63 \pm   3.17 \pm   3.16$ & &&0.9300--0.9325& $    40 \pm      7$& $8.07 \pm 0.13$& $   10.45 \pm   1.70 \pm   0.63$ \\
 0.7525--0.7550& $    97 \pm     14$& $8.41 \pm 0.13$& $   30.10 \pm   4.26 \pm   4.31$ & &&0.9325--0.9350& $    43 \pm      5$& $8.07 \pm 0.13$& $   11.14 \pm   1.32 \pm   0.64$ \\
 0.7550--0.7575& $   134 \pm     18$& $8.40 \pm 0.13$& $   41.39 \pm   5.54 \pm   5.27$ & &&0.9350--0.9375& $    45 \pm      7$& $8.06 \pm 0.13$& $   11.71 \pm   1.86 \pm   0.54$ \\
 0.7575--0.7600& $   166 \pm     17$& $8.40 \pm 0.13$& $   51.12 \pm   5.14 \pm   2.39$ & &&0.9375--0.9400& $    39 \pm      6$& $8.06 \pm 0.13$& $   10.16 \pm   1.47 \pm   0.88$ \\
 0.7600--0.7625& $   159 \pm     13$& $8.39 \pm 0.13$& $   48.88 \pm   4.01 \pm   2.13$ & &&0.9400--0.9425& $    44 \pm      5$& $8.06 \pm 0.13$& $   11.42 \pm   1.39 \pm   0.58$ \\
 0.7625--0.7650& $   220 \pm     19$& $8.39 \pm 0.13$& $   67.36 \pm   5.80 \pm   1.93$ & &&0.9425--0.9450& $    57 \pm     11$& $8.05 \pm 0.13$& $   14.63 \pm   2.90 \pm   1.75$ \\
 0.7650--0.7675& $   265 \pm     17$& $8.38 \pm 0.13$& $   80.95 \pm   5.17 \pm   2.34$ & &&0.9450--0.9475& $    46 \pm      7$& $8.05 \pm 0.13$& $   11.89 \pm   1.80 \pm   0.72$ \\
 0.7675--0.7700& $   389 \pm     21$& $8.38 \pm 0.13$& $  118.63 \pm   6.35 \pm  11.43$ & &&0.9475--0.9500& $    43 \pm      5$& $8.04 \pm 0.13$& $   10.92 \pm   1.40 \pm   0.26$ \\
 0.7700--0.7725& $   603 \pm     29$& $8.37 \pm 0.13$& $  183.30 \pm   8.86 \pm   4.57$ & &&0.9500--0.9525& $    48 \pm      7$& $8.04 \pm 0.13$& $   12.37 \pm   1.68 \pm   0.34$ \\
 0.7725--0.7750& $   959 \pm     36$& $8.37 \pm 0.13$& $  290.79 \pm  10.98 \pm   7.94$ & &&0.9525--0.9550& $    39 \pm      5$& $8.03 \pm 0.13$& $    9.97 \pm   1.39 \pm   0.57$ \\
 0.7750--0.7775& $  1588 \pm     41$& $8.36 \pm 0.13$& $  480.21 \pm  12.35 \pm  11.41$ & &&0.9550--0.9575& $    50 \pm      7$& $8.03 \pm 0.13$& $   12.75 \pm   1.73 \pm   0.31$ \\
 0.7775--0.7800& $  3110 \pm     64$& $8.36 \pm 0.13$& $  938.09 \pm  19.25 \pm  21.37$ & &&0.9575--0.9600& $    52 \pm      8$& $8.03 \pm 0.13$& $   13.15 \pm   1.93 \pm   0.31$ \\
 0.7800--0.7825& $  5124 \pm     43$& $8.35 \pm 0.13$& $ 1541.62 \pm  12.86 \pm  34.87$ & &&0.9600--0.9625& $    46 \pm      6$& $8.02 \pm 0.13$& $   11.57 \pm   1.63 \pm   1.04$ \\
 0.7825--0.7850& $  5342 \pm     63$& $8.35 \pm 0.13$& $ 1603.13 \pm  19.00 \pm  36.07$ & &&0.9625--0.9650& $    56 \pm      8$& $8.02 \pm 0.13$& $   14.14 \pm   1.95 \pm   0.38$ \\
 0.7850--0.7875& $  3365 \pm     63$& $8.34 \pm 0.13$& $ 1007.06 \pm  18.93 \pm  22.62$ & &&0.9650--0.9675& $    42 \pm      8$& $8.01 \pm 0.13$& $   10.70 \pm   1.99 \pm   1.06$ \\
 0.7875--0.7900& $  2079 \pm     50$& $8.34 \pm 0.13$& $  620.45 \pm  15.02 \pm  15.34$ & &&0.9675--0.9700& $    36 \pm      8$& $8.01 \pm 0.13$& $    9.13 \pm   1.96 \pm   0.41$ \\
 0.7900--0.7925& $  1226 \pm     32$& $8.33 \pm 0.13$& $  364.90 \pm   9.58 \pm  12.33$ & &&0.9700--0.9725& $    51 \pm      5$& $8.00 \pm 0.13$& $   12.78 \pm   1.36 \pm   0.88$ \\
 0.7925--0.7950& $   844 \pm     24$& $8.33 \pm 0.13$& $  250.54 \pm   6.98 \pm   6.88$ & &&0.9725--0.9750& $    58 \pm      6$& $8.00 \pm 0.13$& $   14.48 \pm   1.58 \pm   0.59$ \\
 0.7950--0.7975& $   583 \pm     20$& $8.32 \pm 0.13$& $  172.52 \pm   5.85 \pm   4.25$ & &&0.9750--0.9775& $    51 \pm      8$& $8.00 \pm 0.13$& $   12.69 \pm   2.03 \pm   1.79$ \\
 0.7975--0.8000& $   459 \pm     19$& $8.32 \pm 0.13$& $  135.69 \pm   5.53 \pm   3.37$ & &&0.9775--0.9800& $    57 \pm      8$& $7.99 \pm 0.13$& $   14.22 \pm   2.08 \pm   1.56$ \\
 0.8000--0.8025& $   348 \pm     24$& $8.31 \pm 0.13$& $  102.39 \pm   7.11 \pm   4.93$ & &&0.9800--0.9825& $    75 \pm     11$& $7.99 \pm 0.13$& $   18.72 \pm   2.73 \pm   2.17$ \\
 0.8025--0.8050& $   276 \pm     22$& $8.31 \pm 0.13$& $   81.14 \pm   6.39 \pm   4.39$ & &&0.9825--0.9850& $    64 \pm      7$& $7.98 \pm 0.13$& $   15.93 \pm   1.79 \pm   1.02$ \\
 0.8050--0.8075& $   244 \pm     12$& $8.30 \pm 0.13$& $   71.49 \pm   3.61 \pm   2.47$ & &&0.9850--0.9875& $    71 \pm      9$& $7.98 \pm 0.13$& $   17.62 \pm   2.21 \pm   1.15$ \\
 0.8075--0.8100& $   205 \pm     17$& $8.30 \pm 0.13$& $   59.85 \pm   4.96 \pm   1.62$ & &&0.9875--0.9900& $   104 \pm     15$& $7.98 \pm 0.13$& $   25.73 \pm   3.83 \pm   2.72$ \\
 0.8100--0.8125& $   180 \pm     14$& $8.29 \pm 0.13$& $   52.57 \pm   4.13 \pm   1.62$ & &&0.9900--0.9925& $    84 \pm     10$& $7.97 \pm 0.13$& $   20.83 \pm   2.37 \pm   1.32$ \\
 0.8125--0.8150& $   156 \pm     13$& $8.29 \pm 0.13$& $   45.31 \pm   3.65 \pm   2.38$ & &&0.9925--0.9950& $    93 \pm      9$& $7.97 \pm 0.13$& $   22.84 \pm   2.20 \pm   0.68$ \\
 0.8150--0.8175& $   144 \pm     10$& $8.28 \pm 0.13$& $   41.71 \pm   3.04 \pm   1.46$ & &&0.9950--0.9975& $    99 \pm      9$& $7.96 \pm 0.13$& $   24.47 \pm   2.20 \pm   0.67$ \\
 0.8175--0.8200& $   132 \pm     15$& $8.28 \pm 0.13$& $   38.26 \pm   4.39 \pm   1.94$ & &&0.9975--1.0000& $   100 \pm     11$& $7.96 \pm 0.13$& $   24.60 \pm   2.65 \pm   2.51$ \\
 0.8200--0.8225& $   113 \pm     11$& $8.28 \pm 0.13$& $   32.49 \pm   3.27 \pm   2.52$ & &&1.0000--1.0025& $   133 \pm     12$& $7.95 \pm 0.13$& $   32.47 \pm   3.01 \pm   1.42$ \\
 0.8225--0.8250& $   103 \pm     12$& $8.27 \pm 0.13$& $   29.65 \pm   3.57 \pm   2.31$ & &&1.0025--1.0050& $   128 \pm     16$& $7.95 \pm 0.13$& $   31.20 \pm   4.00 \pm   1.53$ \\
 0.8250--0.8275& $    93 \pm     11$& $8.27 \pm 0.13$& $   26.59 \pm   3.15 \pm   0.88$ & &&1.0050--1.0075& $   174 \pm     14$& $7.95 \pm 0.13$& $   42.48 \pm   3.33 \pm   2.46$ \\
 0.8275--0.8300& $    75 \pm     12$& $8.26 \pm 0.13$& $   21.42 \pm   3.47 \pm   2.47$ & &&1.0075--1.0100& $   173 \pm      8$& $7.94 \pm 0.13$& $   42.16 \pm   2.03 \pm   1.58$ \\
 0.8300--0.8325& $    51 \pm      6$& $8.26 \pm 0.13$& $   14.68 \pm   1.66 \pm   0.79$ & &&1.0100--1.0125& $   262 \pm     25$& $7.94 \pm 0.13$& $   63.75 \pm   5.98 \pm   3.89$ \\
 0.8325--0.8350& $    50 \pm      8$& $8.25 \pm 0.13$& $   14.25 \pm   2.24 \pm   0.89$ & &&1.0125--1.0150& $   481 \pm     35$& $7.93 \pm 0.13$& $  116.54 \pm   8.42 \pm   6.60$ \\
 0.8350--0.8375& $    71 \pm      8$& $8.25 \pm 0.13$& $   20.24 \pm   2.23 \pm   1.65$ & &&1.0150--1.0175& $  1177 \pm     37$& $7.93 \pm 0.13$& $  284.82 \pm   8.94 \pm   7.24$ \\
 0.8375--0.8400& $    69 \pm      9$& $8.24 \pm 0.13$& $   19.50 \pm   2.65 \pm   1.09$ & &&1.0175--1.0200& $  2565 \pm     37$& $7.93 \pm 0.13$& $  619.46 \pm   8.87 \pm  17.02$ \\
 0.8400--0.8425& $    69 \pm      8$& $8.24 \pm 0.13$& $   19.46 \pm   2.29 \pm   2.22$ & &&1.0200--1.0225& $  1326 \pm     34$& $7.92 \pm 0.13$& $  319.66 \pm   8.30 \pm   7.09$ \\
 0.8425--0.8450& $    59 \pm      8$& $8.23 \pm 0.13$& $   16.73 \pm   2.25 \pm   0.86$ & &&1.0225--1.0250& $   407 \pm     30$& $7.92 \pm 0.13$& $   97.99 \pm   7.26 \pm   2.16$ \\
 0.8450--0.8475& $    66 \pm      9$& $8.23 \pm 0.13$& $   18.68 \pm   2.57 \pm   1.03$ & &&1.0250--1.0275& $   126 \pm     17$& $7.91 \pm 0.13$& $   30.34 \pm   4.10 \pm   3.42$ \\
 0.8475--0.8500& $    61 \pm      6$& $8.22 \pm 0.13$& $   17.13 \pm   1.73 \pm   1.30$ & &&1.0275--1.0300& $    53 \pm     15$& $7.91 \pm 0.13$& $   12.75 \pm   3.48 \pm   0.67$ \\
 0.8500--0.8525& $    66 \pm     10$& $8.22 \pm 0.13$& $   18.58 \pm   2.79 \pm   0.65$ & &&1.0300--1.0325& $    -3 \pm      9$& $7.91 \pm 0.13$& $   -3.37 \pm   9.40 \pm   9.40$ \\
 0.8525--0.8550& $    51 \pm      8$& $8.21 \pm 0.13$& $   14.27 \pm   2.28 \pm   0.99$ & &&1.0325--1.0350& $    15 \pm      9$& $7.90 \pm 0.13$& $    3.70 \pm   2.12 \pm   1.89$ \\
 0.8550--0.8575& $    38 \pm      8$& $8.21 \pm 0.13$& $   10.59 \pm   2.21 \pm   0.94$ & &&1.0350--1.0375& $     1 \pm      6$& $7.90 \pm 0.13$& $    0.15 \pm   1.49 \pm   0.18$ \\
 0.8575--0.8600& $    52 \pm      7$& $8.20 \pm 0.13$& $   14.53 \pm   1.99 \pm   0.44$ & &&1.0375--1.0400& $     8 \pm      6$& $7.89 \pm 0.13$& $    1.81 \pm   1.43 \pm   0.32$ \\
 0.8600--0.8625& $    48 \pm      7$& $8.20 \pm 0.13$& $   13.41 \pm   1.90 \pm   0.49$ & &&1.0400--1.0425& $     5 \pm      7$& $7.89 \pm 0.13$& $    1.09 \pm   1.58 \pm   0.35$ \\
 0.8625--0.8650& $    47 \pm      7$& $8.20 \pm 0.13$& $   13.14 \pm   1.85 \pm   0.47$ & &&1.0425--1.0450& $     4 \pm      5$& $7.89 \pm 0.13$& $    0.95 \pm   1.13 \pm   0.52$ \\
 0.8650--0.8675& $    48 \pm      7$& $8.19 \pm 0.13$& $   13.36 \pm   1.88 \pm   1.08$ & &&1.0450--1.0475& $     6 \pm      4$& $7.88 \pm 0.13$& $    1.36 \pm   1.05 \pm   0.51$ \\
 0.8675--0.8700& $    42 \pm      7$& $8.19 \pm 0.13$& $   11.58 \pm   1.81 \pm   1.55$ & &&1.0475--1.0500& $     3 \pm      3$& $7.88 \pm 0.13$& $    0.68 \pm   0.82 \pm   0.22$ \\
  \hline\hline
\end{tabularx}

  }
  \end{adjustbox}
\end{table*}

\renewcommand{\xsecspace}{~~~~~~~~}
\begin{table*}
  \centering
  \caption{
    Cross section for $\epem \to \pppz$ in the energy range 1.05--3.50\gev.
    The conventions are the same as in Table~\ref{tab:xsec_lt1050}.
  }
  \label{tab:xsec_gt1050}
  \begin{adjustbox}{width=0.98\textwidth}
  {\scriptsize
    \begin{tabularx}{\linewidth}{lccrc|clccr}
  \hline\hline
  $\sqrt{s'}$ (\gev) & \xsecspace$N_{\mathrm{unf}}$\xsecspace& \xsecspace$\varepsilon$ (\%)\xsecspace& $\xsecspace\sigma$ (nb) & &&  $\sqrt{s'}$ (\gev) & \xsecspace$N_{\mathrm{unf}}$\xsecspace& \xsecspace$\varepsilon$ (\%)\xsecspace& $\xsecspace\sigma$ (nb) \\
  \hline
 1.050--1.075& $    63 \pm     15$& $7.86 \pm 0.13$& $    1.48 \pm   0.35 \pm   0.10$ & &&1.900--1.925& $    38 \pm     11$& $6.95 \pm 0.17$& $    0.54 \pm   0.16 \pm   0.14$ \\
 1.075--1.100& $   159 \pm     17$& $7.82 \pm 0.13$& $    3.65 \pm   0.39 \pm   0.13$ & &&1.925--1.950& $    61 \pm     13$& $6.94 \pm 0.17$& $    0.84 \pm   0.18 \pm   0.15$ \\
 1.100--1.125& $   194 \pm     20$& $7.78 \pm 0.19$& $    4.36 \pm   0.44 \pm   0.15$ & &&1.950--1.975& $    37 \pm      9$& $6.92 \pm 0.17$& $    0.51 \pm   0.13 \pm   0.11$ \\
 1.125--1.150& $   208 \pm     19$& $7.75 \pm 0.19$& $    4.59 \pm   0.41 \pm   0.17$ & &&1.975--2.000& $    31 \pm     10$& $6.91 \pm 0.17$& $    0.42 \pm   0.13 \pm   0.12$ \\
 1.150--1.175& $   232 \pm     20$& $7.71 \pm 0.18$& $    5.03 \pm   0.44 \pm   0.15$ & &&2.000--2.050& $    55 \pm     12$& $6.88 \pm 0.17$& $    0.37 \pm   0.08 \pm   0.08$ \\
 1.175--1.200& $   244 \pm     22$& $7.68 \pm 0.18$& $    5.18 \pm   0.47 \pm   0.16$ & &&2.050--2.100& $    76 \pm     12$& $6.85 \pm 0.17$& $    0.50 \pm   0.08 \pm   0.07$ \\
 1.200--1.225& $   208 \pm     23$& $7.64 \pm 0.18$& $    4.36 \pm   0.48 \pm   0.16$ & &&2.100--2.150& $    70 \pm     13$& $6.83 \pm 0.17$& $    0.45 \pm   0.08 \pm   0.05$ \\
 1.225--1.250& $   276 \pm     22$& $7.61 \pm 0.18$& $    5.68 \pm   0.45 \pm   0.18$ & &&2.150--2.200& $    53 \pm     10$& $6.80 \pm 0.17$& $    0.33 \pm   0.06 \pm   0.03$ \\
 1.250--1.275& $   249 \pm     23$& $7.58 \pm 0.18$& $    5.04 \pm   0.47 \pm   0.19$ & &&2.200--2.250& $    54 \pm     10$& $6.78 \pm 0.17$& $    0.32 \pm   0.06 \pm   0.02$ \\
 1.275--1.300& $   206 \pm     24$& $7.54 \pm 0.18$& $    4.09 \pm   0.47 \pm   0.16$ & &&2.250--2.300& $    54 \pm     10$& $6.75 \pm 0.17$& $    0.32 \pm   0.06 \pm   0.02$ \\
 1.300--1.325& $   247 \pm     26$& $7.51 \pm 0.18$& $    4.84 \pm   0.52 \pm   0.17$ & &&2.300--2.350& $    53 \pm     10$& $6.73 \pm 0.17$& $    0.30 \pm   0.06 \pm   0.02$ \\
 1.325--1.350& $   257 \pm     25$& $7.48 \pm 0.18$& $    4.95 \pm   0.48 \pm   0.17$ & &&2.350--2.400& $    40 \pm      9$& $6.71 \pm 0.17$& $    0.23 \pm   0.05 \pm   0.02$ \\
 1.350--1.375& $   218 \pm     22$& $7.45 \pm 0.18$& $    4.13 \pm   0.41 \pm   0.13$ & &&2.400--2.450& $    26 \pm      8$& $6.69 \pm 0.17$& $    0.14 \pm   0.05 \pm   0.01$ \\
 1.375--1.400& $   201 \pm     22$& $7.42 \pm 0.18$& $    3.76 \pm   0.41 \pm   0.17$ & &&2.450--2.500& $    33 \pm      8$& $6.67 \pm 0.17$& $    0.18 \pm   0.05 \pm   0.02$ \\
 1.400--1.425& $   261 \pm     25$& $7.39 \pm 0.18$& $    4.80 \pm   0.45 \pm   0.19$ & &&2.500--2.550& $    28 \pm      8$& $6.65 \pm 0.17$& $    0.15 \pm   0.04 \pm   0.01$ \\
 1.425--1.450& $   229 \pm     20$& $7.37 \pm 0.18$& $    4.14 \pm   0.36 \pm   0.16$ & &&2.550--2.600& $    12 \pm      7$& $6.63 \pm 0.17$& $    0.06 \pm   0.03 \pm   0.01$ \\
 1.450--1.475& $   232 \pm     21$& $7.34 \pm 0.18$& $    4.15 \pm   0.37 \pm   0.16$ & &&2.600--2.650& $    25 \pm      7$& $6.61 \pm 0.17$& $    0.13 \pm   0.03 \pm   0.01$ \\
 1.475--1.500& $   229 \pm     20$& $7.31 \pm 0.18$& $    4.03 \pm   0.35 \pm   0.15$ & &&2.650--2.700& $    33 \pm      8$& $6.60 \pm 0.17$& $    0.16 \pm   0.04 \pm   0.01$ \\
 1.500--1.525& $   244 \pm     21$& $7.29 \pm 0.17$& $    4.23 \pm   0.36 \pm   0.15$ & &&2.700--2.750& $     7 \pm      5$& $6.58 \pm 0.17$& $    0.04 \pm   0.03 \pm   0.01$ \\
 1.525--1.550& $   265 \pm     22$& $7.26 \pm 0.17$& $    4.53 \pm   0.38 \pm   0.16$ & &&2.750--2.800& $    12 \pm      6$& $6.56 \pm 0.17$& $    0.06 \pm   0.03 \pm   0.01$ \\
 1.550--1.575& $   300 \pm     22$& $7.24 \pm 0.17$& $    5.06 \pm   0.38 \pm   0.16$ & &&2.800--2.850& $     5 \pm      5$& $6.55 \pm 0.17$& $    0.02 \pm   0.02 \pm   0.01$ \\
 1.575--1.600& $   305 \pm     22$& $7.21 \pm 0.17$& $    5.08 \pm   0.37 \pm   0.15$ & &&2.850--2.900& $    14 \pm      8$& $6.53 \pm 0.17$& $    0.07 \pm   0.04 \pm   0.02$ \\
 1.600--1.625& $   321 \pm     24$& $7.19 \pm 0.17$& $    5.26 \pm   0.39 \pm   0.17$ & &&2.900--2.950& $     6 \pm      6$& $6.51 \pm 0.17$& $    0.02 \pm   0.03 \pm   0.01$ \\
 1.625--1.650& $   283 \pm     24$& $7.17 \pm 0.17$& $    4.58 \pm   0.38 \pm   0.16$ & &&2.950--3.000& $    12 \pm      6$& $6.49 \pm 0.17$& $    0.05 \pm   0.02 \pm   0.01$ \\
 1.650--1.675& $   256 \pm     19$& $7.14 \pm 0.17$& $    4.09 \pm   0.31 \pm   0.15$ & &&3.000--3.050& $    18 \pm      4$& $6.48 \pm 0.17$& $    0.08 \pm   0.02 \pm   0.08$ \\
 1.675--1.700& $   214 \pm     20$& $7.12 \pm 0.17$& $    3.37 \pm   0.32 \pm   0.14$ & &&3.050--3.100& $    19 \pm      4$& $6.46 \pm 0.17$& $   0.082 \pm  0.018 \pm  0.004$ \\
 1.700--1.725& $   133 \pm     19$& $7.10 \pm 0.17$& $    2.06 \pm   0.29 \pm   0.15$ & &&3.100--3.150& $    16 \pm      5$& $6.44 \pm 0.17$& $   0.065 \pm  0.020 \pm  0.004$ \\
 1.725--1.750& $   130 \pm     17$& $7.08 \pm 0.17$& $    2.00 \pm   0.26 \pm   0.18$ & &&3.150--3.200& $    19 \pm      5$& $6.42 \pm 0.17$& $    0.08 \pm   0.02 \pm   0.06$ \\
 1.750--1.775& $   101 \pm     14$& $7.06 \pm 0.17$& $    1.53 \pm   0.21 \pm   0.11$ & &&3.200--3.250& $     8 \pm      5$& $6.40 \pm 0.17$& $    0.03 \pm   0.02 \pm   0.01$ \\
 1.775--1.800& $    83 \pm     12$& $7.04 \pm 0.17$& $    1.24 \pm   0.18 \pm   0.10$ & &&3.250--3.300& $     9 \pm      5$& $6.38 \pm 0.17$& $   0.037 \pm  0.019 \pm  0.003$ \\
 1.800--1.825& $    77 \pm     14$& $7.02 \pm 0.17$& $    1.13 \pm   0.20 \pm   0.12$ & &&3.300--3.350& $     5 \pm      3$& $6.36 \pm 0.17$& $   0.020 \pm  0.012 \pm  0.004$ \\
 1.825--1.850& $    76 \pm     14$& $7.01 \pm 0.17$& $    1.11 \pm   0.21 \pm   0.13$ & &&3.350--3.400& $     6 \pm      4$& $6.34 \pm 0.17$& $   0.023 \pm  0.015 \pm  0.005$ \\
 1.850--1.875& $    79 \pm     15$& $6.99 \pm 0.17$& $    1.14 \pm   0.21 \pm   0.15$ & &&3.400--3.450& $     8 \pm      4$& $6.31 \pm 0.18$& $   0.030 \pm  0.017 \pm  0.005$ \\
 1.875--1.900& $    48 \pm     11$& $6.97 \pm 0.17$& $    0.69 \pm   0.16 \pm   0.11$ & &&3.450--3.500& $    14 \pm      4$& $6.29 \pm 0.18$& $    0.05 \pm   0.02 \pm   0.01$ \\
  \hline\hline
\end{tabularx}

   }
   \end{adjustbox}
\end{table*}

\bibliographystyle{apsrev4-1}
\bibliography{references}

\end{document}